\documentclass[11pt,a4paper,oneside,british,headsepline,idxtotoc, bibtotoc,liststotoc]{scrbook}
\usepackage[T1]{fontenc}
\usepackage[latin1]{inputenc}
\usepackage{amsmath}
\usepackage{makeidx}
\makeindex
\usepackage{graphicx}
\usepackage{setspace}
\usepackage{amssymb}

\makeatletter

\providecommand{\tabularnewline}{\\}

 \input{db.std}

\typearea[20mm]{12}
\usepackage{cite}
\providecommand{\openone}{\leavevmode\hbox{\small1\kern-3.8pt\normalsize1}}
\setcapindent{0cm}

\usepackage{babel}
\makeatother
\begin{document}
\begin{center}\thispagestyle{plain}\par\end{center}

\vspace{-4cm}

\hspace{-4.63cm}\includegraphics[width=21cm]{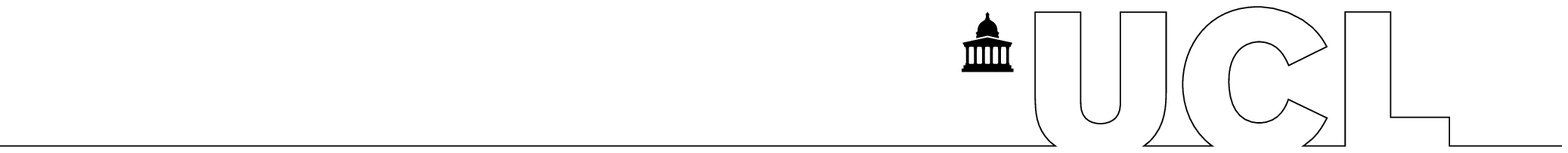}\vspace{2cm}

\begin{center}\textsf{\textbf{\LARGE Quantum State Transfer with Spin
Chains}} \par\end{center}

\begin{center}\vspace{1cm}
\par\end{center}

\begin{center}\textsf{\textbf{\LARGE Daniel Klaus Burgarth}}\vspace{1cm}
\par\end{center}

\begin{center}A thesis submitted to the University of London\\
 for the degree of Doctor of Philosophy\par\end{center}

\begin{center}\vspace{1cm}
\par\end{center}

\begin{center}Department of Physics and Astronomy\\
University College London\vspace{1cm}
\par\end{center}

\begin{center}December 2006\par\end{center}

\chapter*{Declaration}

I, Daniel Klaus Burgarth, confirm that the work presented in this
thesis is my own. Where information has been derived from other sources,
I confirm that this has been indicated in the thesis.

\chapter*{Abstract}

In the last few decades the idea came up that by making use of the
superposition principle from Quantum Mechanics, one can process information
in a new and much faster way. Hence a new field of information technology,
QIT (Quantum Information Technology), has emerged. From a physics
point of view it is important to find ways of implementing these new
methods in real systems. One of the most basic tasks required for
QIT is the ability to connect different components of a Quantum Computer
by \emph{quantum wires} that obey the superposition principle. Since
superpositions can be very sensitive to noise this turns out to be
already quite difficult. Recently, it was suggested to use chains
of permanently coupled spin-1/2 particles (\emph{quantum chains})
for this purpose. They have the advantage that no external control
along the wire is required during the transport of information, which
makes it possible to isolate the wire from sources of noise. The purpose
of this thesis is to develop and investigate advanced schemes for
using quantum chains as wires. We first give an introduction to basic
quantum state transfer and review existing advanced schemes by other
authors. We then introduce two new methods which were created as a
part of this thesis. First, we show how the fidelity of transfer can
be made perfect by performing measurements at the receiving end of
the chain. Then we introduce a scheme which is based on performing
unitary operations at the end of the chain. We generalise both methods
and discuss them from the more fundamental point of view of mixing
properties of a quantum channel. Finally, we study the effects of
a non-Markovian environment on quantum state transfer.

\chapter*{Acknowledgements}

Most of all, I would like to thank my supervisor Sougato Bose for
much inspiration and advice. I am very grateful for many inspiring
and fruitful discussions and collaborations with Vittorio Giovannetti,
and with Floor Paauw, Christoph Bruder, Jason Twamley, Andreas Buchleitner
and Vladimir Korepin. Furthermore I would like to thank all my teachers
and those who have guided and motivated me along my journey through
physics, including Heinz-Peter Breuer, Francesco Petruccione, Lewis
Ryder, John Strange, Werner Riegler, Carsten Schuldt and Rolf Bussmann.
I acknowledge financial support by the UK Engineering and Physical
Sciences Research Council through the grant GR/S62796/01. Finally
I would like to thank my parents for their loving support.

\chapter*{Notation}


\begin{quotation}
X,Y,Z\hfill{}Pauli matrices

$X_{n},Y_{n},Z_{n}$\hfill{}Pauli matrices acting on the Hilbert-space
of qubit $n$

$|0\rangle,|1\rangle$ \hfill{}Single qubit state in the canonical
basis

$|\boldsymbol{0}\rangle$\hfill{}Quantum chain in the product state
$|0\rangle\otimes\cdots\otimes|0\rangle$

$|\boldsymbol{n}\rangle$\hfill{}''Single excitation'' state $X_{n}|\boldsymbol{0}\rangle$

$\mbox{Tr}_{X}$\hfill{}Partial trace over subsystem $X$

$||\ldots||$\hfill{}Euclidean vector norm

$||\ldots||_{1}$\hfill{}Trace norm

$||\ldots||_{2}$\hfill{}Euclidean matrix norm 
\end{quotation}
\begin{flushleft}We also use the following graphical representation:\vspace{5mm}
\includegraphics[width=0.663\columnwidth]{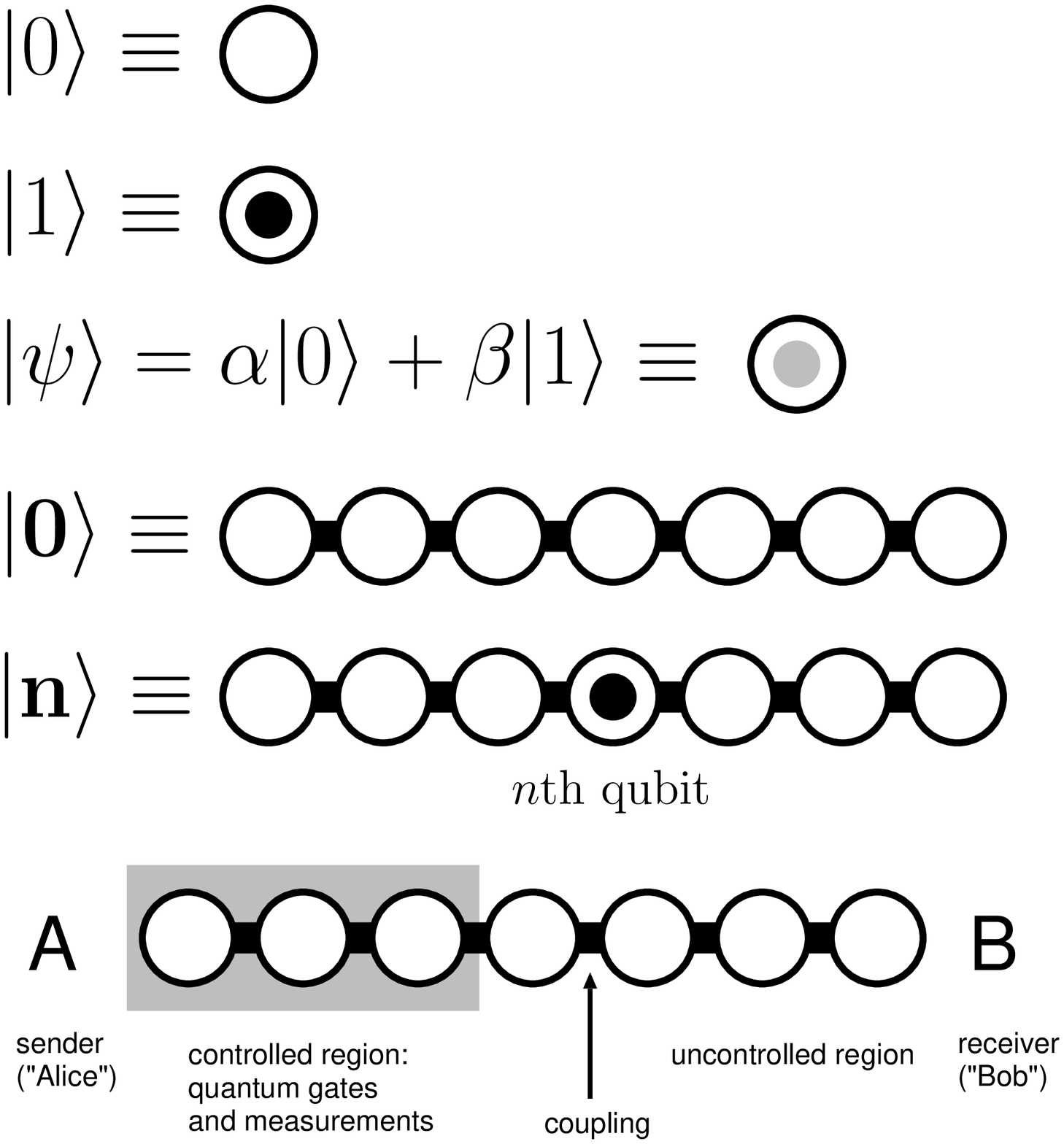}\par\end{flushleft}

\tableofcontents{}

\chapter{Introduction}

The Hilbert space that contains the states of quantum mechanical objects
is huge, scaling exponentially with the number of particles described.
In 1982, Richard Feynman suggested to make use of this as a resource
for \emph{simulating} quantum mechanics in a \emph{quantum computer,}
i.e. a device where the physical interaction could be {}``programmed''
to yield a specific Hamiltonian. This has led to the new fields of
Quantum Computation and Quantum Information. A quantum computer can
solve questions one could never imagine to solve using an ordinary
computer. For example, it can factorise large numbers into primes
efficiently, a task of greatest importance for cryptography. It may
thus be a surprise that more than twenty years after the initial ideas,
these devices still haven't been built or only in ridiculously small
size. The largest quantum computer so far can only solve problems
that any child could solve within seconds. A closer look reveals that
the main problem in the realisation of quantum computers is the {}``programming'',
i.e. the design of a specific (time-dependent) Hamiltonian, usually
described as a set of discrete unitary gates. This turns out to be
extremely difficult because we need to connect microscopic objects
(those behaving quantum mechanically) with macroscopic devices that
\emph{control} the microscopic behaviour. Even if one manages to find
a link between the micro- and the macroscopic world, such as laser
pulses and electric or magnetic fields, then the connection introduces
not only control but also noise (dissipation and decoherence) to the
microscopic system, and its quantum behaviour is diminished.

The vision of this thesis is to develop theoretical methods narrowing
the gap between what is imagined theoretically and what can be done
experimentally. As a method we consider chains (or more general graphs)
of \emph{permanently coupled} quantum systems. This idea has been
originally put forward by S. Bose for the specific task of quantum
communication~\cite{Bose2003}. Due to the permanent coupling, these
devices can in principle be built in such a way that they don't require
external control to perform their tasks, just like a mechanical clockwork.
This also overcomes the problem of decoherence as they can be separated
from any source of noise. Unfortunately, most schemes that have been
developed so far still require external control, though much less
than an {}``ordinary'' quantum computer. Furthermore, internal dispersion
in these devices is leading to a decrease of their fidelity. A third
problem is, that for building these devices the permanent couplings
still need to be realised, although only once, and experimental constraints
such as resolution and errors need to be considered. We are thus left
with the following questions: which is the best way to perform quantum
state transfer using a permanently coupled graph? How much control
do we need, and how difficult will it be to implement the couplings?
How do errors and noise affect the scheme? All these points are highly
related and it cannot be expected to find an \emph{absolute,} i.e.
system independent answer. The purpose of this research is to develop
advanced schemes for the transfer of quantum information, to improve
and generalise existing ideas, to relate them to each other and to
investigate their stability and efficiency.

\section{Quantum Computation and Quantum Information\label{sec:Quantum-Computation}}

In this Section we review some of the basic concepts of Quantum Computation.
We will be very brief and only focus on those aspects that we require
later on in the thesis. A more detailed introduction can be found
in~\cite{NIELSEN}.

In information science, an algorithm is a list of instructions that
a computer performs on a given input to achieve a specific task. For
instance, a \emph{factoring algorithm} has an arbitrary integer as
its input, and gives its prime factors as an output. A \emph{quantum
factoring algorithm} can be thought of in a similar way, i.e. it has
an integer as input, and its prime factors as an output. \emph{In-between}
however it encodes information in a quantum mechanical system. Due
to the superposition principle, the information of a quantum system
cannot be represented as \emph{bits.} The valid generalisation of
the bit to the quantum case is called \emph{qubit.} The possible states
of a qubit are written as\begin{equation}
\alpha|0\rangle+\beta|1\rangle,\end{equation}
where $\alpha,\beta$ are normalised complex coefficients, and $|0\rangle$
and $|1\rangle$ are vectors of a two-dimensional complex vector space.
Peter W. Shor has shown in a famous paper~\cite{Shor1997} that the
\emph{detour} of representing the intermediate part of a factoring
algorithm in a quantum system (as well as using quantum gates, see
below) can be very beneficial: it runs \emph{much} faster. This is
important, because many cryptographic methods rely on factoring algorithms
being slow. \index{Shor's algorithm}Shor's algorithm is definitely
not the only reason why it would be very nice to have a \emph{quantum
computer\index{quantum computer},} i.e. a machine that represents
information in a quantum way and can perform instructions on it, and
many more details can be found in the textbook mentioned above.

Algorithms on a computer can be represented as list of logical operations
on bits. Likewise, a (standard) quantum algorithm can be represented
as a list of \emph{quantum logical operations}, or \emph{\index{quantum gates}quantum
gates,} acting on qubits. The most general quantum algorithm is given
by an arbitrary unitary operator. A \emph{universal set of gates}
is a set such that any quantum algorithm (i.e. unitary operator) can
be decomposed into a sequence of gates belonging to this set. In the
\emph{standard model} of quantum computation, one assumes that such
a set is available on the machine~\cite{DiVincenzo2000}. Also the
ability to perform measurements is assumed. We refer to this as the
\emph{full control} case.

From a information theoretic point of view, qubits are not only \emph{useful}
objects to perform algorithms with, but also very interesting from
a fundamental point of view. To give a (too simple) analogy consider
the following. If you read the word ''chocolate'', you can associate
a positive/negative or neutral feeling of whether you would like to
eat some chocolate now. However, what was the state of your mind concerning
chocolate \emph{before you read} the word? \emph{}Unless you were
already craving for chocolate\index{chocolate} beforehand, or you
have just eaten a lot, your mind was probably \emph{undecided}. Moreover,
it would have been very difficult - if not impossible - to describe
to someone in plain language which opinion you had about the chocolate
before you read the word.

In a similar manner, the quantum information contained in a single
\emph{arbitrary and unknown} qubit\index{arbitrary and unknown qubit}
cannot be described by classical information. When it is measured,
it behaves like a normal bit in the sense that the outcome is only
$0$ or $1,$ but when it is not measured, it behaves in some way
as if it was undecided between $0$ and 1. Of course one has to be
very careful with these analogies. But for the purpose of this thesis
it is important to stress that quantum information cannot be transported
by any classical methods~\cite{WERNER}. This is why it is so important
and also so difficult to develop new wires, dubbed \emph{quantum wires},
that are capable of doing this.

\section{Quantum state transfer along short distances\label{sec:Quantum-state-transfer}}

\begin{figure}[htbp]
\begin{centering}\includegraphics[width=0.7\columnwidth]{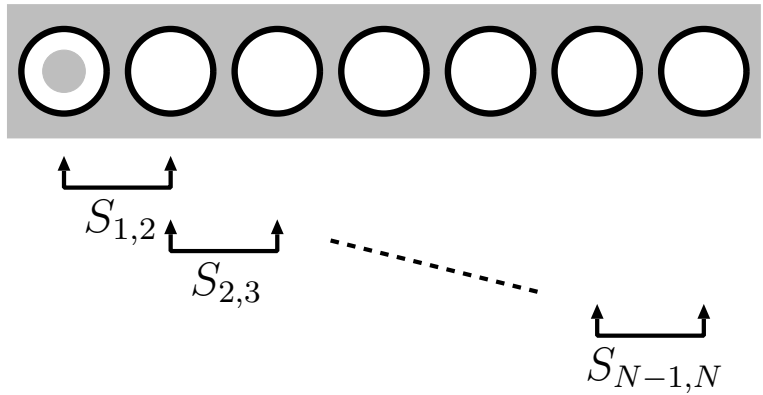}\par\end{centering}

\caption{\label{fig:swapping0}In areas of universal control, quantum states
can easily be transferred by sequences of unitary swap gates $S_{j,k}$
between nearest neighbours.}
\end{figure}

In theory, additional devices for the transfer of unknown quantum
states are not required for building a quantum computer, unless it
is being used for typical quantum communication purposes, such as
secret key distribution \cite{DiVincenzo2000}. This is because the
\emph{universal set of gates} on the quantum computer can be used
to transfer quantum states by applying sequences of two-qubit swap
gates (Fig.~\ref{fig:swapping0}). %

However in practice it is crucial to minimise the required number
of quantum gates, as each gate typically introduces \emph{errors}.
In this light it appears costly to perform $N-1$ swap gates between
nearest neighbours to just move a qubit state over a distance of $N$
sites. For example, Shor's algorithm on $N$ qubits can be implemented
by only $\log N$ quantum gating operations~\cite{Cleve2000} if
long distant qubit gates are available. These long distant gates could
consist of local gates followed by a quantum state transfer. If however
the quantum state transfer is implemented as a sequence of local gates,
then the number of operations blows up to the order of $N$ gates.
The quantum state transfer can even be thought of as the source of
the \emph{power of quantum computation}, as any quantum circuit with
$\log N$ gates and \emph{local gates only} can be efficiently simulated
on a classical computer~\cite{Josza2006,Short2006}.

\begin{figure}[htbp]
\begin{centering}\includegraphics[width=0.7\columnwidth]{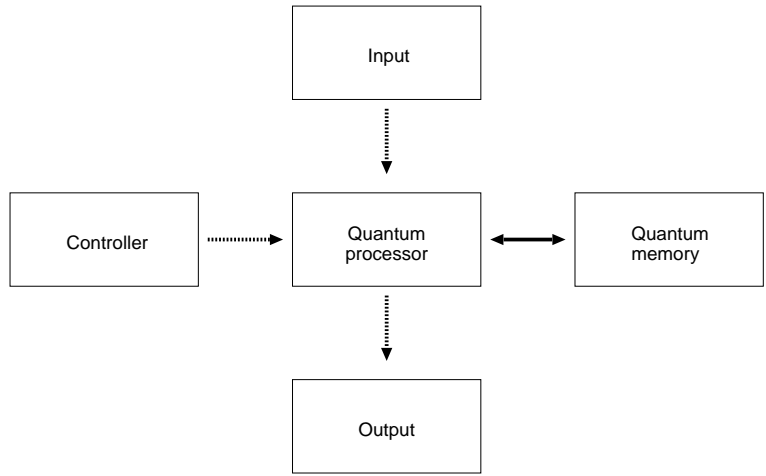}\par\end{centering}

\caption{\label{fig:quantumcomputer}Schematic layout of a quantum computer.
The solid arrows represent the flow of quantum information, and the
dashed arrows the flow of classical information.}
\end{figure}

\begin{figure}[htbp]
\begin{centering}\includegraphics[width=0.7\columnwidth]{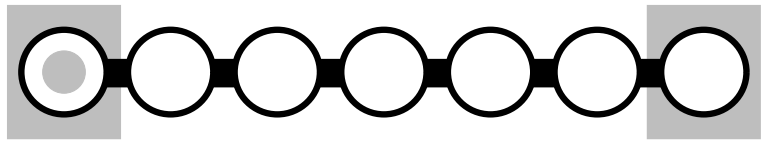}\par\end{centering}

\caption{\label{fig:qchain}Permanently coupled quantum chains can transfer
quantum states without control along the line. Note that the ends
still need to be controllable to initialise and read out quantum states.}
\end{figure}

A second reason to consider devices for quantum state transfer is
related to \emph{scalability}\index{scalability}. While small quantum
computers have already been built~\cite{Haffner2005}, it is very
difficult to build large arrays of fully controllable qubits. A \emph{black
box}\index{black box} that transports unknown quantum states could
be used to build larger quantum computers out of small components
by connecting them. Likewise, quantum state transfer can be used to
connect \emph{different} components of a quantum computer, such as
the processor and the memory (see also Fig.~\ref{fig:quantumcomputer}).
On larger distances, flying qubits such as photons, ballistic electrons
and guided atoms/ions are considered for this purpose~\cite{Skinner2003,Kielpinski2002}.
However, converting back and forth between stationary qubits and mobile
carriers of quantum information and interfacing between different
physical implementations of qubits is very difficult and worthwhile
only for short communication distances. This is the typical situation
one has to face in solid state systems, where quantum information
is usually contained in the states of \emph{fixed objects} such as
quantum dots or Josephson junctions. In this case permanently coupled
\emph{quantum chains} have recently been proposed as prototypes of
reliable quantum communication lines~\cite{Bose2003,LLOYD}. A quantum
chain\index{quantum chain} (also referred to as \emph{\index{spin chain}spin
chain}) is a one-dimensional array of qubits which are coupled by
some Hamiltonian (cf. Fig.~\ref{fig:qchain}). These couplings can
transfer states \emph{without external classical control.} In many
cases, such permanent couplings are easy to build in solid state devices
(in fact a lot of effort usually goes into \emph{suppressing} them).
The qubits can be of the \emph{same type} as the other qubits in the
device, so no interfacing is required.%

\begin{figure}[htbp]
\begin{centering}\includegraphics[width=0.9\columnwidth]{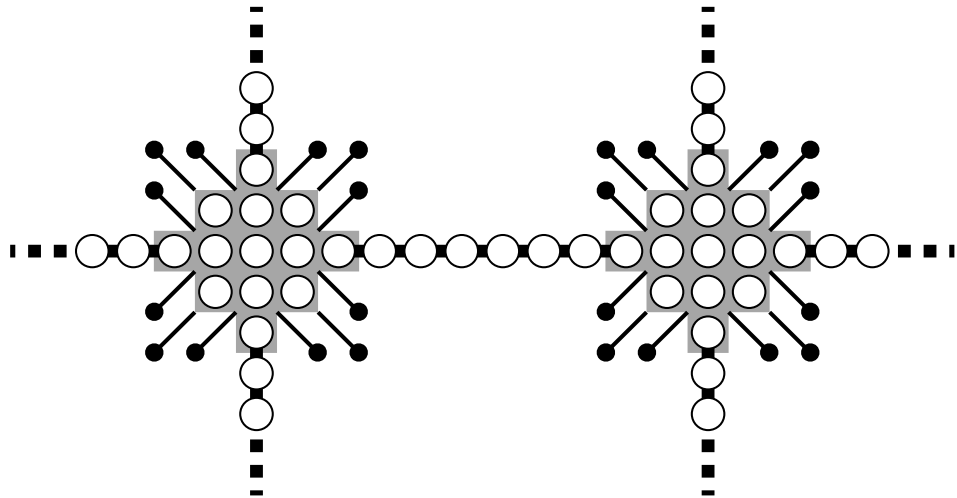}\par\end{centering}

\caption{\label{fig:connect2}Small blocks (grey) of qubits (white circles)
connected by quantum chains. Each block consists of (say) 13 qubits,
4 of which are connected to outgoing quantum chains (the thick black
lines denote their nearest-neighbour couplings). The blocks are connected
to the macroscopic world through classical wires (thin black lines
with black circles at their ends) through which arbitrary unitary
operations can be triggered on the block qubits. The quantum chains
require no external control.}
\end{figure}

Another related motivation to consider quantum chains is that they
can simplify the \emph{layout} of quantum devices on wafers. A typical
chip can contain millions of qubits, and the fabrication of many qubits
is in principle no more difficult than the fabrication of a single
one. In the last couple of years, remarkable progress was made in
experiments with quantum dots~\cite{KOPPENS,HANSON} and super-conducting
qubits~\cite{YAMA,CHIO}. It should however be emphasised that for
initialisation, control and readout, those qubits have to be connected
to the macroscopic world (see Fig.~\ref{fig:quantumcomputer}). For
example, in a typical flux qubit gate, microwave pulses are applied
onto specific qubits of the sample. This requires many (classical)
wires on the chip, which is thus a \emph{compound} of quantum and
classical components. The macroscopic size of the classical control
is likely to be the bottleneck of the scalability as a whole. In this
situation, quantum chains are useful in order to keep some distance
between the controlled quantum parts. A possible layout for such a
quantum computer is shown in Fig.~\ref{fig:connect2}. It is built
out of blocks of qubits, some of which are dedicated to communication
and therefore connected to another block through a quantum chain.
Within each block, arbitrary unitary operations can be performed in
a fast and reliable way (they may be decomposed into single and two-qubit
operations). Such blocks do not currently exist, but they are the
focus of much work in solid state quantum computer architecture. The
distance between the blocks is determined by the length of the quantum
chains between them. It should be large enough to allow for classical
control wiring of each block, but short enough so that the time-scale
of the quantum chain communication is well below the time-scale of
decoherence in the system.%

Finally, an important reason to study quantum state transfer in quantum
chains stems from a more fundamental point of view. Such systems in
principle allow tests of Bell-inequalities and non-locality in solid-state
experiments well before the realisation of a quantum computer. Although
quantum transport is quite an established field, the quantum information
point of view offers many new perspectives. Here, one looks at the
transport of information rather than excitations, and at entanglement~\cite{Lakshminarayan2006,Subrahmanyam2004,Eisert2004,Palma2004}
rather than correlation functions. It has recently been shown that
this sheds new light on well-known physical phenomena such as quantum
phase transitions~\cite{Plenio,Plenio2006,Verstraete,Verstraete2004},
quantum chaos~\cite{Monteiro2006,Twamley2005,Santos2004,Boness2006}
and localisation~\cite{Winter,Apollaro2006}. Furthermore, quantum
information takes on a more \emph{active} attitude. The correlations
of the system are not just calculated, but one also looks at how they
may be \emph{changed}.

\section{Implementations and experiments\index{experiments}\label{sec:Implementations-and-experiments}}

\begin{figure}[tbh]
\begin{centering}\includegraphics[width=0.65\paperwidth]{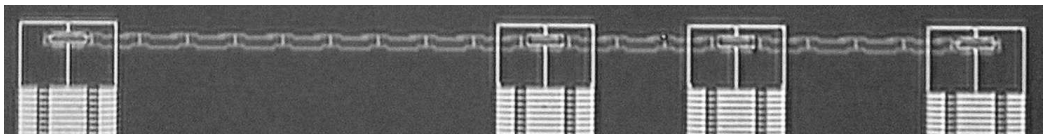}\par\end{centering}

\caption{\label{fig:A-quantum-chain}A quantum chain consisting of $N=20$
flux qubits~\cite{Bruder2005} (picture and experiment by Floor Paauw,
TU Delft). The chain is connected to four larger SQUIDS for readout
and gating.}
\end{figure}

As we have seen above, the main advantage of state transfer with quantum
chains is that the qubits can be of the same type as those used for
the quantum computation. Therefore, most systems that are thought
of as possible realisations of a quantum computer can also be used
to build quantum chains. Of course there has to be some coupling between
the qubits. This is typically easy to achieve in solid state systems,
such as Josephson junctions with charge qubits~\cite{Bruder2005a,Tsomokos2006},
flux qubits~\cite{Bruder,Bruder2005}\index{flux qubits} (see also
Fig.~\ref{fig:A-quantum-chain}) or quantum dots dots using the electrons~\cite{Loss1998,Greentree2004}
or excitons~\cite{DAmico,Lovett}. Other systems where quantum chain
Hamiltonians can at least be \emph{simulated} are NMR qubits~\cite{Suter2006,Jones,Zhang2005}
and optical lattices~\cite{Garcia-Ripoll2003}. Such a simulation
is particularly useful in the latter case, where local control is
extremely difficult. Finally, qubits in cavities~\cite{Falci2005,Paternostro2005}
and coupled arrays of cavities were considered \cite{Bose,Plenioa}.%

For the more fundamental questions, such as studies of entanglement
transfer, non-locality and coherent transport, the quantum chains
could also be realised by systems which are not typically thought
of as qubits, but which are \emph{natural spin chains.} These can
be molecular systems~\cite{EXCITON} or quasi-1D solid state materials~\cite{Motoyama1996,Gambardella2002}.

\section{Basic communication protocol\label{sec:Basic-transport-protocol}}

We now review the most basic transport protocol for quantum state
transfer, initially suggested in~\cite{Bose2003}. For the sake of
simplicity, we concentrate on the linear chain setting, though more
general graphs of qubits can be considered in the same way. The protocol
consists of the following steps:

\begin{enumerate}
\item Initialise the quantum chain in the ground state \begin{equation}
|G\rangle.\end{equation}

\item Put an arbitrary and unknown qubit with (possibly mixed) state $\rho$
at the sending end of the chain \begin{equation}
\rho\otimes\mbox{Tr}_{1}\left\{ |G\rangle\langle G|\right\} .\end{equation}

\item Let the system evolve under its Hamiltonian $H$ for a time $t$\begin{equation}
\exp\left\{ -iHt\right\} \rho\otimes\mbox{Tr}_{1}\left\{ |G\rangle\langle G|\right\} \exp\left\{ iHt\right\} .\end{equation}

\item Pick up the quantum state at the end of the chain\begin{equation}
\sigma\equiv\mbox{Tr}_{1,\ldots,N-1}\left[\exp\left\{ -iHt\right\} \rho\otimes\mbox{Tr}_{1}\left\{ |G\rangle\langle G|\right\} \exp\left\{ iHt\right\} \right].\end{equation}

\end{enumerate}
Some practical aspects how to realise these steps are discussed in
the next section. For the moment, we will concentrate on the \emph{quality}
of quantum state transfer given that the above steps can be performed.
From a quantum information perspective, the above equations describe
a \emph{quantum channel}\index{quantum channel}~\cite{WERNER} $\tau$
that maps input states $\rho$ at one end of the chain to output states
$\tau(\rho)=\sigma$ on the other end. A very simple measure of the
quality of such a quantum channel is the \emph{fidelity}~\cite{Uhlmann1976,Jozsa1994,NIELSEN}\emph{\index{fidelity}}
\begin{equation}
F(\rho,\sigma)\equiv\left(\mbox{Tr}\sqrt{\rho^{1/2}\sigma\rho^{1/2}}\right)^{2}.\end{equation}
More advanced measures of the quality of transfer will be discussed
in Chapter \ref{cha:Multi-rail-and-Capacity}. Note also that some
authors define the fidelity without taking the square of the trace.
It is a real-valued, symmetric function with range between $0$ and
$1,$ assuming unity if and only if $\rho=\sigma.$ Since the transported
state that is an unknown result of some quantum computation, we are
interested in the \emph{minimal fidelity}\index{minimal fidelity}\begin{equation}
F_{0}\equiv\mbox{min}_{\rho}F(\rho,\tau(\rho)).\label{eq:minfidelity}\end{equation}
We remark that some authors also assume an equal distribution of input
states and compute the \emph{average fidelity}~\cite{Bose2003}.
Using the strong concavity of the fidelity~\cite{NIELSEN} and the
linearity of $\tau$ we find that the minimum must be assumed on pure
input states,\begin{equation}
F_{0}=\mbox{min}_{\psi}\langle\psi|\tau(\psi)|\psi\rangle.\label{eq:minfidelity2}\end{equation}
In the present context, $F_{0}=F_{0}(H,t)$ is a function of of the
Hamiltonian $H$ of the quantum chain (through the specific role of
the ground state in the protocol and through the time evolution),
and of the time interval $t$ that the system is evolving in the third
step of the protocol.

\subsection{Initialisation and end-gates\label{sec:Initialisation}}

There are two strong assumptions in the protocol from the last section.
The first one is that the chain can be initialised in the ground state
$|G\rangle.$ How can that be achieved if there is no local control
along the chain? The answer appears to be quite easy: one just applies
a strong global magnetic field and strong cooling (such as laser cooling
or dilution refrigeration) and lets the system reach its ground state
by relaxation. The cooling needs to be done for the remaining parts
of the quantum computer anyway, so no extra devices are required.
However there is a problem with the time-scale of the relaxation.
If the system is brought to the ground state by cooling, it must be
coupled to some environment. But during the quantum computation, one
clearly does not want such an environment. This is usually solved
by having the time-scale of the computation much smaller (say microseconds)
than the time-scale of the cooling (say seconds or minutes). But if
the quantum chain should be used multiple times during one computation,
then how is it reset between each usage? This is important to avoid
memory effects~\cite{Werner2005}, and there are two solutions to
this problem. Either the protocol is such that at the end the chain
is automatically in the ground state. Such a protocol usually corresponds
to \emph{perfect state transfer.} The other way is to use the control
at the ends of the chain to bring it back to the ground state. A simple
\emph{\index{cooling protocol}cooling protocol} is given by the following:
one measures the state of the last qubit of the chain. If it is in
$|0\rangle,$ then one just lets the chain evolve again and repeats.
If however it is found to be in $|1\rangle,$ one applies the Pauli
operator $X$ to flip it before evolving and repeating. It will become
clear later on in the thesis that such a protocol typically converges
exponentially fast to the ground state of the chain.

The second assumption in the last section is that the sender and receiver
are capable of swapping in and out the state much quicker than the
time-scale of the interaction of the chain. Alternatively, it is assumed
that they can switch on and off the interaction between the chain
and their memory in such time-scale. It has recently been shown~\cite{Bruder}
that this is not a fundamental problem, and that finite switching
times can even slightly improve the fidelity if they are carefully
included in the protocol. But this requires to solve the full time-dependent
Schrödinger equation, and introduces further parameters to the model
(i.e. the raise and fall time of the couplings). For the sake of simplicity,
we will therefore assume that the end gates are much faster then the
time evolution of the chain (see also Section~\ref{sec:Practical-Considerations}).

\subsection{Symmetries}

The dimensionality of the Hilbert space $\mathcal{H}$ of a quantum
chain of $N$ qubits is $2^{N}.$ This makes it quite hopeless in
general to determine the minimal fidelity Eq.~(\ref{eq:minfidelity2})
for long quantum chains. Most investigations on quantum state transfer
with quantum chains up to date are therefore concentrating on Hamiltonians
with additional symmetries. With few exceptions \cite{Bruder2005,Plenio,Plenio2006,Kay}
Hamiltonians that conserve the number of excitations are considered.
In this case the Hilbert space is a direct sum of subspaces invariant
under the time evolution,\begin{equation}
\mathcal{H}=\bigoplus_{\ell=0}^{N}\mathcal{H}_{\ell},\end{equation}
with $\mbox{dim}\mathcal{H}_{\ell}=\binom{N}{\ell},$ and where $\ell$
is the number of excitations. These Hamiltonians are \emph{much} easier
to handle both analytically and numerically, and it is also easier
to get an intuition of the dynamics. Furthermore, they occur quite
naturally as a coupling between qubits in the relevant systems. We
stress though that there is \emph{no fundamental} reason to restrict
quantum chain communication to this case.

\subsection{Transfer functions\index{transfer functions}}

The space $\mathcal{H}_{0}$ only contains the state $|\boldsymbol{0}\rangle$
which is thus always an eigenstate of $H.$ We will assume here that
it is also the ground state,\begin{equation}
|G\rangle=|\boldsymbol{0}\rangle.\end{equation}
This can be achieved by applying a strong global magnetic field (or
equivalent) to the system. The space $\mathcal{H}_{1}$ is spanned
by the vectors $\left\{ |\boldsymbol{k}\rangle,k=1,\ldots,N\right\} $
having exactly one excitation. The above protocol becomes:

\begin{enumerate}
\item Initialise the quantum chain in the ground state \begin{equation}
|\boldsymbol{0}\rangle\end{equation}

\item Put an arbitrary and unknown qubit in the pure state $|\psi\rangle=\alpha|0\rangle+\beta|1\rangle$
at the sending end of the chain \begin{equation}
\alpha|\boldsymbol{0}\rangle+\beta|\boldsymbol{1}\rangle\end{equation}

\item Let the system evolve for a time $t$\begin{equation}
\alpha|\boldsymbol{0}\rangle+\beta\exp\left\{ -iHt\right\} |\boldsymbol{1}\rangle\end{equation}

\item Pick up the quantum state at the end of the chain (see~\cite{Bose2003})\begin{equation}
\tau(\psi)=(1-p(t))|0\rangle\langle0|+p(t)|\psi\rangle\langle\psi|,\end{equation}
with the minimal fidelity given by\begin{align}
F_{0} & =\mbox{min}_{\psi}\langle\psi|\tau(\psi)|\psi\rangle\label{eq:minfidelity3}\\
 & =p(t)+(1-p(t))\mbox{min}_{\psi}\left|\langle0|\psi\rangle\right|^{2}=p(t).\end{align}

\end{enumerate}
The function $p(t)$ is the transition probability from the state
$|\boldsymbol{1}\rangle$ to $|\boldsymbol{N}\rangle$ given by\begin{equation}
p(t)=\left|\langle\boldsymbol{N}|\exp\left\{ -iHt\right\} |\boldsymbol{1}\rangle\right|^{2}.\end{equation}
We see that in the context of quantum state transfer, a \emph{single}
parameter suffices to characterise the properties of an excitation
conserving chain. The averaged fidelity~\cite{Bose2003} is also
easily computed as\begin{equation}
\bar{F}=\frac{\sqrt{p(t)}}{3}+\frac{p(t)}{6}+\frac{1}{2}.\label{eq:averagedfid}\end{equation}
Even more complex measures of transfer such as the quantum capacity
only depend on $p(t)$~\cite{Giovannetti2005}. It is also a physically
intuitive quantity, namely a particular matrix element of the time
evolution operator, \begin{align}
f_{n,m}(t) & \equiv\langle\boldsymbol{n}|\exp\left\{ -iHt\right\} |\boldsymbol{m}\rangle\label{eq:spintrans}\\
 & =\sum_{k}e^{-iE_{k}t}\langle\boldsymbol{n}|E_{k}\rangle\langle E_{k}|\boldsymbol{m}\rangle,\end{align}
where $|E_{k}\rangle$ and $E_{k}$ are the eigenstates and energy
levels of the Hamiltonian in $\mathcal{H}_{1}.$

\subsection{Heisenberg Hamiltonian}

The Hamiltonian chosen in~\cite{Bose2003} is a \index{Heisenberg Hamiltonian}Heisenberg
Hamiltonian\begin{equation}
H=-\frac{J}{2}\sum_{n=1}^{N-1}\left(X_{n}X_{n+1}+Y_{n}Y_{n+1}+Z_{n}Z_{n+1}\right)-B\sum_{n=1}^{N}Z_{n}+c,\label{eq:heisenberg}\end{equation}
with a constant term \begin{equation}
c=\frac{J(N-1)}{2}+NB\end{equation}
 added to set the ground state energy to $0.$ For $J>0$ it fulfils
all the assumptions discussed above, namely its ground state is given
by $|\boldsymbol{0}\rangle$ and it conserves the number of excitations
in the chain. The Heisenberg interaction is very common and serves
here as a typical and analytically solvable model for quantum state
transfer. 

In the first excitation subspace $\mathcal{H}_{1}$, the Heisenberg
Hamiltonian Eq.~(\ref{eq:heisenberg}) is expressed in the basis
$\left\{ |\boldsymbol{n}\rangle\right\} $ as \begin{equation}
\left(\begin{array}{cccccc}
1 & -1\\
-1 & 2 & -1\\
 & -1 & 2 & \ddots\\
 &  & \ddots & \ddots & -1\\
 &  &  & -1 & 2 & -1\\
 &  &  &  & -1 & 1\end{array}\right).\label{eq:matrix}\end{equation}
A more general study of such \emph{tridiagonal} matrices can be found
in a series of articles on coherent dynamics~\cite{Eberly1977,Bialynicka-Birula1977,Cook1979,Shore1979}.
Some interesting analytically solvable models have also been identified~\cite{Ekert2004,Bialynicka-Birula1977,Cook1979}
(we shall come back to that point later).

\begin{figure}[htbp]
\begin{centering}\includegraphics[width=0.5\paperwidth]{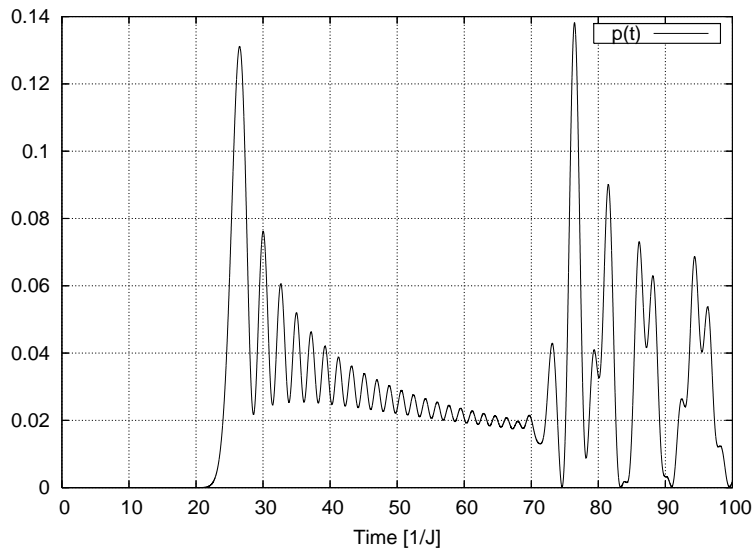}\par\end{centering}

\caption{\label{fig:fidelity}Minimal fidelity $p(t)$ for a Heisenberg chain
of length $N=50.$}
\end{figure}

For the present case, the eigenstates of Eq~(\ref{eq:matrix}) are~\cite{Bose2003}
\begin{equation}
|E_{k}\rangle=\sqrt{\frac{1+\delta_{k0}}{N}}\sum_{n=1}^{N}\cos\left[\frac{\pi k}{2N}(2n-1)\right]|\boldsymbol{n}\rangle\quad(k=0,\ldots,N-1),\end{equation}
with the corresponding energies given by\begin{equation}
E_{k}=2B+2J\left[1-\cos\frac{\pi k}{N}\right].\label{eq:eigenfrequencies}\end{equation}
The parameter $B$ has no relevance for the fidelity but determines
the stability of the ground state (the energy of the first excited
state is given by $2B).$ The minimal fidelity\index{minimal fidelity}
for a Heisenberg chain is given by

\begin{equation}
\boxed{p(t)=N^{-2}\left|1+\sum_{k=1}^{N-1}\exp\left\{ -2iJt(1-\cos\frac{\pi k}{N})\right\} (-1)^{k}\left(1+\cos\frac{\pi k}{N}\right)\right|^{2}.}\label{eq:resultfid}\end{equation}
As an example, Fig~\ref{fig:fidelity} shows $p(t)$ for $N=50$.

\subsection{Dynamic and Dispersion\label{sub:Dynamics-and-Dispersion}}

\begin{figure}[htbp]
\begin{centering}\includegraphics[width=0.55\paperwidth]{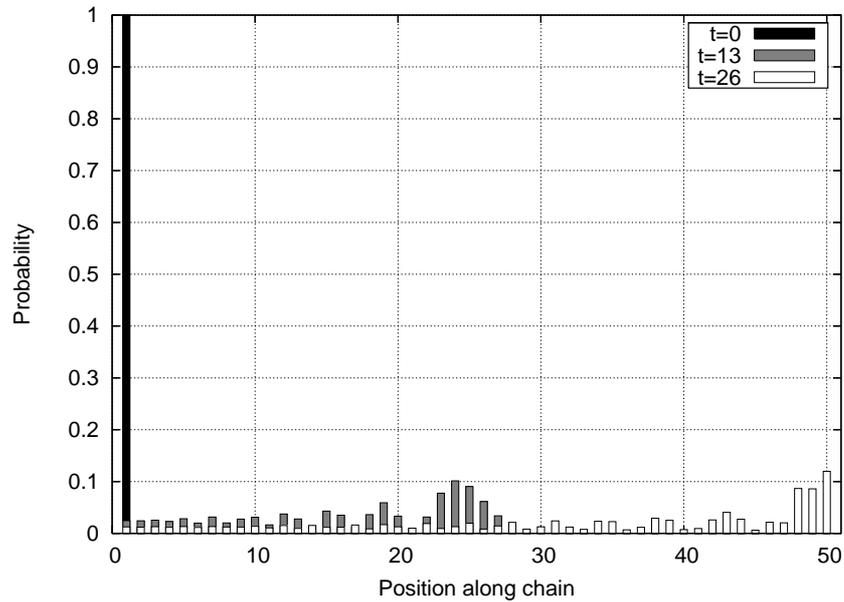}\par\end{centering}

\caption{\label{fig:snapshots}Snapshots of the time evolution of a Heisenberg
chain with $N=50.$ Shown is the distribution $|f_{n,1}(t)|^{2}$
of the wave-function in space at different times if initially localised
at the first qubit.}
\end{figure}

\begin{figure}[htbp]
\begin{centering}\includegraphics[width=0.55\paperwidth]{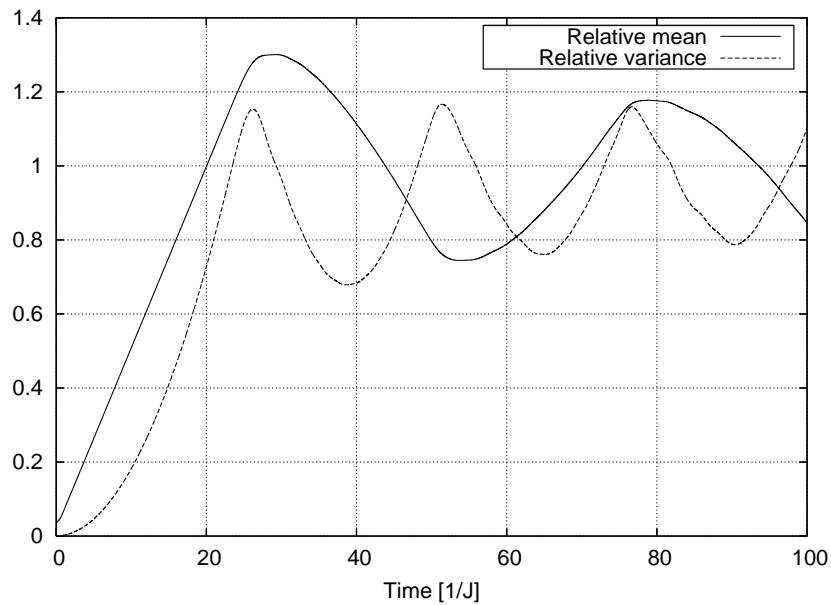}\par\end{centering}

\caption{\label{fig:dispersion}Mean and variance of the state $|\boldsymbol{1}\rangle$
as a function of time. Shown is the case $N=50$ with the y-axis giving
the value \emph{relative} to the mean $N/2+1$ and variance $(N^{2}-1)/12$
of an equal distribution $\frac{1}{\sqrt{N}}\sum|\boldsymbol{n}\rangle.$}
\end{figure}

Already in~\cite{Bose2003} has been realised that the fidelity for
quantum state transfer along spin chains will in general not be perfect.
The reason for the imperfect transfer is the \emph{dispersion}\index{dispersion}~\cite{Linden2004}
of the information along the chain. Initially the quantum information
is localised at the sender, but as it travels through the chain it
also spreads (see Fig.~\ref{fig:snapshots} and Fig.~\ref{fig:dispersion}).
This is not limited to the Heisenberg coupling considered here, but
a very common quantum effect. Due to the dispersion, the probability
amplitude peak that reaches Bob is typically small, and becomes even
smaller as the chains get longer. %

\begin{figure}[htbp]
\begin{centering}\includegraphics[width=0.55\paperwidth]{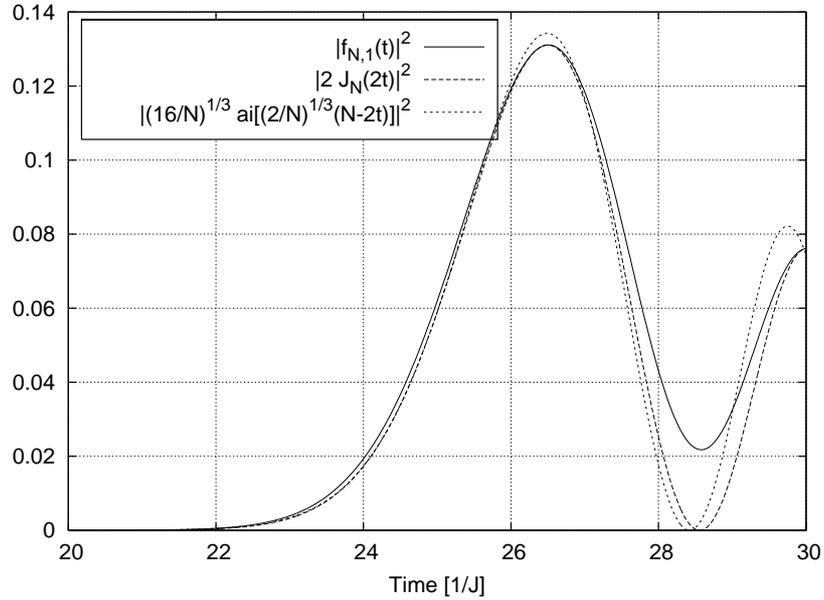}\par\end{centering}

\caption{\label{fig:airyapprox}Approximation of the transfer amplitude for
$N=50$ around the first maximum by Bessel and Airy functions \cite{Bose2003,Abramowitz1972}. }
\end{figure}

\begin{figure}[htbp]
\begin{centering}\includegraphics[width=0.55\paperwidth]{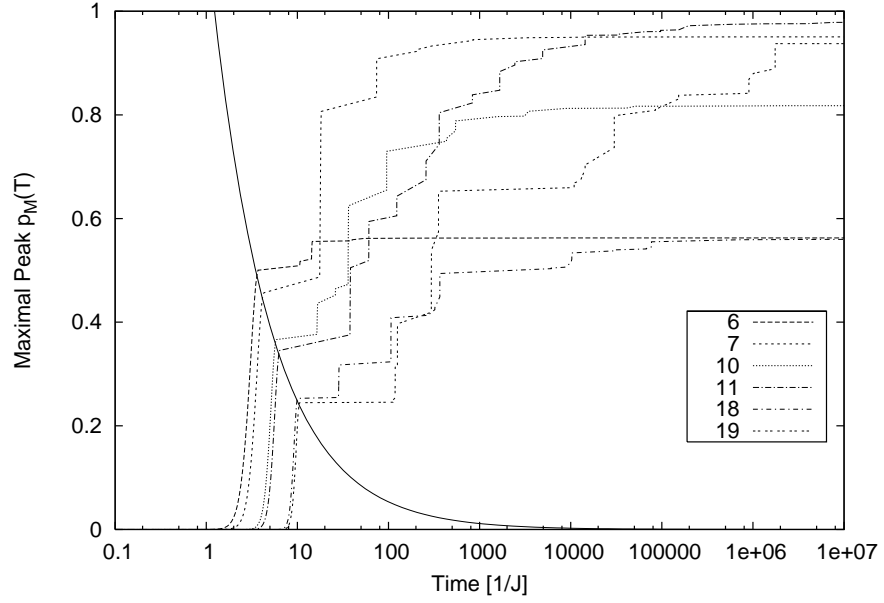}\par\end{centering}

\caption{\label{fig:longtime}$p_{M}(T)$ as a function of $T$ for different
chain lengths. The solid curve is given by $1.82{(2T)}^{-2/3}$ and
corresponds to the first peak of the probability amplitude (Eq.~\ref{eq:estimate})}
\end{figure}

The fidelity given Eq.~(\ref{eq:resultfid}) is shown in Fig.~\ref{fig:fidelity}.
We can see that a wave of quantum information is travelling across
the chain. It reaches the other end at a time of approximately \begin{equation}
t_{\mbox{peak}}\approx\frac{N}{2J}\end{equation}
As a rough estimate of the scaling of the fidelity with respect to
the chain length around this peak we can use~\cite{Bose2003,Abramowitz1972}
(see also Fig.~\ref{fig:airyapprox}) \begin{equation}
|f_{N,1}(t)|^{2}\approx|2J_{N}(\frac{2t}{J})|^{2}\approx|\left(\frac{16}{N}\right)^{1/3}\mbox{ai}[\left(\frac{2}{N}\right)^{1/3}(N-\frac{2t}{J})]|^{2},\label{eq:approxf}\end{equation}
where $J_{N}(x)$ is a Bessel function of first kind and $ai(x)$
is the Airy function. The airy function $ai(x)$ has a maximum of
$0.54$ at $x=-1.02.$ Hence we have \begin{equation}
p(t_{\mbox{peak}})=|f_{N,1}(\frac{N}{2J})|^{2}\approx1.82N^{-2/3}.\label{eq:estimate}\end{equation}
It is however possible to find times where the fidelity of the chain
is much higher. The reason for this is that the wave-packet is reflected
at the ends of the chain and starts interfering with itself (Fig~\ref{fig:fidelity}).
As the time goes on, the probability distribution becomes more and
more random. Sometimes high peaks at the receiving end occur. From
a theoretical point of view, it is interesting to determine the \emph{maximal
peak}\index{maximal peak} occurring, i.e. \begin{equation}
p_{M}(T)\equiv\max_{0<t<T}p(t).\label{eq:maximal}\end{equation}
As we can see in Fig.~\ref{fig:longtime} there is quite a potential
to improve from the estimate Eq.~(\ref{eq:estimate}).%

We will now show a perhaps surprising connection of the function $p_{M}(T)$
to number theory. Some speculations on the dependence of the fidelity
on the chain length being divisible by $3$ were already made in~\cite{Bose2003},
but not rigorously studied. As it turns out, for chains with \emph{prime
number length} the maximum of the fidelity is actually converging
to unity (see Fig.~\ref{fig:longtime}). To show this, we first prove
the following

\begin{lemma}
\label{lem:primindependence}Let $N$ be an odd prime. Then the set\begin{equation}
\left\{ \cos\frac{k\pi}{N}\quad(k=0,1,\ldots,,\frac{N-1}{2})\right\} \end{equation}
is linear independent over the rationals $\mathbb{Q}$.
\end{lemma}
\begin{proof}
Assume that \begin{equation}
\sum_{k=0}^{\frac{N-1}{2}}\lambda_{k}\cos\frac{k\pi}{N}=0\end{equation}
with $\lambda_{k}\in\mathbb{Q}.$ It follows that \begin{equation}
\sum_{k=0}^{\frac{N-1}{2}}\lambda_{k}\left\{ \exp\frac{ik\pi}{N}+\exp\frac{-ik\pi}{N}\right\} =0\end{equation}
and hence\begin{equation}
\sum_{k=0}^{\frac{N-1}{2}}\lambda_{k}\exp\frac{ik\pi}{N}-\sum_{k=0}^{\frac{N-1}{2}}\lambda_{k}\exp\frac{i(N-k)\pi}{N}=0.\end{equation}
Changing indexes on the second sum we get\begin{equation}
\sum_{k=0}^{\frac{N-1}{2}}\lambda_{k}\exp\frac{ik\pi}{N}-\sum_{k=\frac{N+1}{2}}^{N}\lambda_{N-k}\exp\frac{ik\pi}{N}=0.\end{equation}
and finally \begin{equation}
\sum_{k=0}^{N-1}\tilde{\lambda}_{k}\exp\frac{ik\pi}{N}=0,\label{eq:exp}\end{equation}
where\begin{align}
\tilde{\lambda}_{0} & =2\lambda_{0}\\
\tilde{\lambda}_{k} & =\lambda_{k}\quad(k=1,\ldots,\frac{N-1}{2})\\
\tilde{\lambda}_{k} & ={-\lambda}_{N-k}\quad(k=\frac{N+1}{2},\ldots,N-1).\end{align}
Since $N$ is prime, the roots of unity in Eq.~(\ref{eq:exp}) are
all primitive and therefore linearly independent over $\mathbb{Q}$~\cite[Theorem 3.1, p. 313]{Lang1984}.
Hence $\lambda_{k}=0$ for all $k.$
\end{proof}
\begin{framedtheorem}
[Half recurrence]Let $N$ be an odd prime. For a Heisenberg chain
of length $N$ we have\begin{equation}
\lim_{T\rightarrow\infty}p_{M}(T)=\lim_{T\rightarrow\infty}\left[\max_{0<t<T}p(t)\right]=1.\end{equation}

\end{framedtheorem}
\begin{proof}
The eigenfrequencies of the Hamiltonian in the first excitation sector
$\mathcal{H}_{1}$ are given by \begin{equation}
E_{k}=2B+2J\left[1-\cos\frac{\pi k}{N}\right]\quad(k=0,1,\ldots,,N-1).\label{eq:eigenfrequencies2}\end{equation}
Using Kronecker's theorem~\cite{Hemmer1958} and Lemma~\ref{lem:primindependence},
the equalities \begin{equation}
\exp\left\{ itE_{k}\right\} =\left(-1\right)^{k}e^{2(B+J)t}\quad(k=0,1,\ldots,,\frac{N-1}{2})\label{eq:set}\end{equation}
can be fulfilled \emph{arbitrarily well} by choosing an appropriate
$t.$ Since\begin{equation}
\cos\frac{k\pi}{N}=-\cos\frac{(N-k)\pi}{N},\end{equation}
the equalities~(\ref{eq:set}) are then also fulfilled arbitrarily
well for $k=0,\ldots,N-1.$ This is known as as sufficient condition
for perfect state transfer in mirror symmetric chains~\cite{Bose2005},
where the eigenstates can be chosen such that they are alternately
symmetric and antisymmetric. Roughly speaking, Eq.~(\ref{eq:set})
introduces the correct phases (a sign change for the antisymmetric
eigenstates) to move the state $|1\rangle$ to $|N\rangle$ and hence
the theorem.
\end{proof}
\begin{remark}
The time-scale for finding high valued peaks is however \emph{exponential}
in the chain length~\cite{Hemmer1958}. Therefore the above theorem
has little practical use. For non-prime chain lengths, the eigenfrequencies
are not sufficiently independent to guarantee a perfect state transfer,
with the algebraic dimensionality of the roots of unity for non-prime
$N$ given by the Euler totient function $\phi(N)$~\cite[Theorem 3.1, p. 313]{Lang1984}.
We also remark that due to its asymptotic character, the above result
is not contradicting~\cite{Landahl2005}, where it was shown that
chains longer than $N\ge4$ never have perfect fidelity.
\end{remark}
Having proved that there are many chains that can in principle perform
arbitrarily well, it is important to find a cut-off time for the optimisation
Eq.~(\ref{eq:maximal}). Faster transfer than linear in $N$ using
local Hamiltonians is impossible due to the Lieb-Robinson bound~\cite{Bravyi2006,Nachtergaele2006},
which is a ''speed limit'' in non-relativistic quantum mechanics
giving rise to a well defined group velocity. Transport faster than
this group velocity is exponentially suppressed. Going back to the
motivation of quantum state transfer, a natural comparison~\cite{DAmico}
for the time-scale of quantum state transfer is given by the time
it would take to perform a sequence of swap gates\index{swap gates}
(cf. Fig~\ref{fig:swapping0}) that are realised by a pairwise switchable
coupling Hamiltonian \begin{equation}
\frac{J}{2}(X_{n}X_{n+1}+Y_{n}Y_{n+1}).\end{equation}
This time is linear in the chain length: \begin{equation}
t_{\mbox{swap}}=\frac{(N-1)\pi}{2J}.\end{equation}
Ideally one could say that the time for quantum state transfer should
not take much longer than this. However one may argue that there is
a trade-off between quick transfer on one hand, and minimising control
on the other hand. A second cut-off time may be given by the \emph{decoherence
time} of the specific implementation. But short decoherence times
could always be counteracted by increasing the chain coupling $J.$
A more general and implementation independent limit is given by the
requirement that the \emph{peak width\index{peak width}} ${\Delta t}_{\mbox{peak}}$
should not be too small with respect to the total time. Otherwise
it is difficult to pick up the state at the correct time. For the
first peak, we can estimate the width by using the full width at half
height of the airy function. From Eq.~(\ref{eq:approxf}) we get
an absolute peak width of $\Delta t_{\mbox{peak}}\approx0.72N^{1/3}/J$
and a relative width of\begin{equation}
\frac{\Delta t_{\mbox{peak}}}{t_{\mbox{peak}}}\approx1.44N^{-2/3}.\end{equation}
This is already quite demanding from an experimental perspective and
we conclude that the transfer time should not be chosen much longer
than those of the first peak.

\subsection{How high should $p(t)$ be?\label{sub:How-high-should}}

We have not discussed yet what the actual value of $p(t)$ should
be to make such a spin chain useful as a device for quantum state
transfer. $p(t)=0$ corresponds to no state transfer, $p(t)=1$ to
a perfect state transfer. But what are the relevant scales for intermediate
$p(t)$? In practice, the quantum transfer will suffer from additional
external noise (Chapter~\ref{cha:Problems-and-Practical}) and also
the quantum computer itself is likely to be very noisy. From this
point of view, requiring $p(t)=1$ seems a bit too demanding.

From a theoretical perspective, it is interesting that for any $p(t)>0,$
one can already do things which are impossible using classical channels,
namely entanglement transfer\index{entanglement transfer} and distillation\index{distillation}~\cite{NIELSEN}.
The entanglement of formation\index{entanglement of formation} between
the sender (\emph{Alice}) and the receiver (\emph{Bob}) is simply
given by $\sqrt{p(t)}$~\cite{Bose2003}. This entanglement can be
partially distilled~\cite{Horodecki1997} into singlets, which could
be used for state transfer using teleportation~\cite{NIELSEN}. It
is however not known \emph{how much}, i.e. at which rate, entanglement
can be distilled (we will develop lower bounds for the entanglement
of distillation\index{entanglement of distillation} in Section~\ref{sec:con}
and Section~\ref{s:sec3}). Also, entanglement distillation is a
quite complex procedure that requires local unitary operations and
measurements, additional classical communication, and multiple chain
usages; and few explicit protocols are known. This is likely to preponderate
the benefits of using a quantum chain.

When the chain is used without encoding and further operation, the
averaged fidelity Eq.~(\ref{eq:averagedfid}) becomes better than
the classical%
\footnote{By ''classical fidelity'', we mean the fidelity that can be achieved
by optimising the following protocol: Alice performs measurements
on her state and sends Bob the outcome through a classical communication
line. Bob then tries to rebuild the state that Alice had before the
measurement based on the information she sent. For qubits, the classical
fidelity is given by $2/3$\cite{Horodecki1999}.%
} averaged fidelity\index{classical averaged fidelity}~\cite{Bose2003}
when $p(t)>3-2\sqrt{2}.$ Following the conclusion from the last subsection
that the first peak is the most relevant one, this would mean that
only chains with length until $N=33$ perform better than the classical
fidelity.

Finally, the \emph{quantum capacity}\index{quantum capacity}~\cite{Giovannetti2005,SHOR}
of the channel becomes non-zero only when $p(t)>1/2,$ corresponding
to chain lengths up to $N=6.$ Roughly speaking, it is a measure of
the number of perfectly transmitted qubits per chain usage that can
be achieved asymptotically using encoding and decoding operations
on multiple channel usages. The quantum capacity considered here is
not assumed to be assisted by a classical communication, and the threshold
of $p(t)>0.5$ to have a non-zero quantum capacity is a result of
the non-cloning theorem~\cite{NIELSEN}. This is not contradicting
the fact that entanglement distillation is possible for \emph{any}
$p(t)>0,$ as the entanglement distillation protocols require additional
classical communication.

\begin{figure}[htbp]
\begin{centering}\includegraphics[width=0.6\paperwidth]{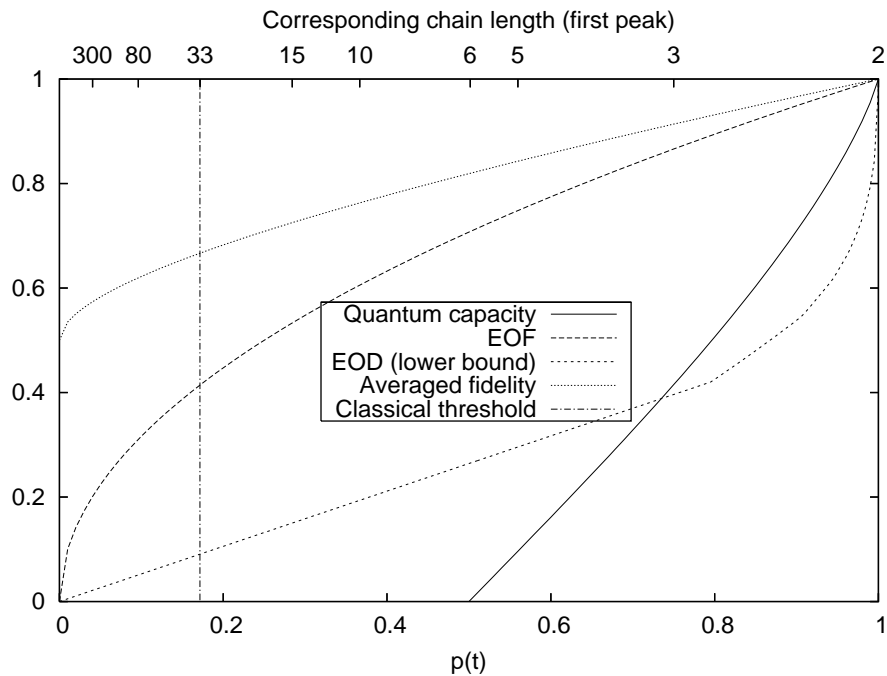}\par\end{centering}

\caption{\label{fig:capacity}Quantum capacity, entanglement of formation
(EOF), a lower bound for the entanglement of distillation (EOD) and
the averaged fidelity as a function of $p(t).$ We also show the corresponding
chain length which reaches this value as a first peak and the classical
threshold $3-2\sqrt{2}$. The explicit expression for the quantum
capacity plotted here is given in~\cite{Giovannetti2005}, and the
lower bound of the entanglement of distillation will be derived in
Section~\ref{s:sec3}.}
\end{figure}

All the above points are summarised in Fig.~\ref{fig:capacity}.
We can see that only very short chains reach reasonable values (say
$>0.6$) for the minimal fidelity. %

\section{Advanced communication protocols\label{sec:Advanced-transfer-protocols}}

We have seen in the last section that without much further effort,
i.e. entanglement distillation, unmodulated Heisenberg chains are
useful only when they are very short. Shortly after the initial proposal~\cite{Bose2003}
it has been shown that there are ways to achieve even \emph{perfect
state transfe}r with arbitrarily long chains. These advanced proposals
can roughly be grouped into four categories, which we will now briefly
describe.

\subsection{Engineered Hamiltonians\label{sub:Engineered-Hamiltonians}}

\begin{figure}[htbp]
\begin{centering}\includegraphics[width=0.6\paperwidth]{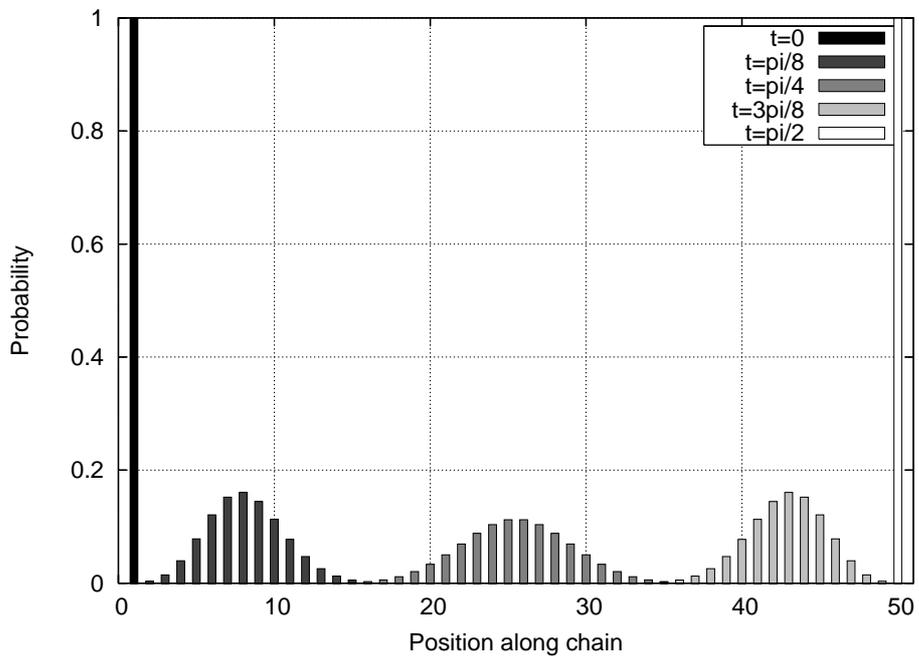}\par\end{centering}

\caption{\label{fig:snapshotseng}Snapshots of the time evolution of a quantum
chain with engineered couplings~(\ref{eq:engineered}) for $N=50.$
Shown is the distribution of the wave-function in space at different
times if initially localised at the first qubit (compare Fig.~\ref{fig:snapshots}).}
\end{figure}

The Heisenberg model chosen by Bose features many typical aspects
of coherent transport, i.e. the wave-like behaviour, the dispersion,
and the almost-periodicity of the fidelity. These features do not
depend so much on the specific choices of the parameters of the chain,
such as the couplings strengths. There are however \emph{specific
couplings} for quantum chains that show a quite different time evolution,
and it was suggested in~\cite{Landahl2004} and independently in~\cite{Lambropoulos2004a}
to use these to achieve a \emph{perfect} state transfer: \begin{equation}
H=-J\sum_{n=1}^{N-1}\sqrt{n(N-n)}\left(X_{n}X_{n+1}+Y_{n}Y_{n+1}\right)\label{eq:engineered}\end{equation}
These values for engineered couplings\index{engineered couplings}
also appear in a different context in~\cite{Cook1979,Peres1985}.
The time evolution under the Hamiltonian~(\ref{eq:engineered}) features
an additional \emph{mirror symmetry:} the wave-packet disperses initially,
but the dispersion is reversed after its centre has passed the middle
of the chain (Fig.~\ref{fig:snapshotseng}).%
 This approach has been extended by various authors~\cite{Bose2005,Yung2006,Sun2006a,Sun2005c,Sun2005a,Lambropoulos2006,Lambropoulos2004,Ericsson2005,Kay,Kay2006,Kay2006a,Stolze2005,Jing-Fu2006,Landahl2005,Ekert2004,Eisert2004},
and many other choices of parameters for perfect or near perfect state
transfer in various settings were found~\cite{Ekert2004,Stolze2005,Kay2006}.

\subsection{Weakly coupled sender and receiver\label{sub:Weakly-coupled-sender}}

A different approach of tuning the Hamiltonian was suggested in~\cite{Semiao2005}.
There, only the first and the last couplings $j$ of the chain are
engineered to be \emph{much weaker} than the remaining couplings $J$
of the chain, which can be quite arbitrary. The fidelity can be made
arbitrarily high by making the edge coupling strengths smaller. It
was shown~\cite{Bednarska,Bednarska2005} that to achieve a fidelity
of $1-\delta$ in a chain of odd length, it takes approximately a
time of

\begin{equation}
2N\pi/\sqrt{\delta}\end{equation}
and the coupling ratio has to be approximately $j/J\approx\sqrt{\delta/N}.$
Some specific types of quantum chains which show high fidelity for
similar reasons were also investigated~\cite{Sun2005,Bose2006,Bose2006a,Pasquale2006}.

\subsection{Encoding\label{sub:Encoding}}

We have seen in Subsec.~\ref{sub:How-high-should} that if $p(t)<1/2,$
the fidelity cannot be improved by using any encoding/decoding strategy
(because the quantum capacity is zero). However it is possible to
\emph{change the protocol} described in Sec.~\ref{sec:Basic-transport-protocol}
slightly such that the fidelity is much higher\emph{.} This can be
thought of as a \emph{''}hardware encoding\emph{''}, and was suggested
first in~\cite{Linden2004}. There, it was assumed that the chain
consists of three sections: one part of length $\approx2N^{1/3}$
controlled by the sending party, one ''free'' part of length $N$
and one part of length $\approx2.8N^{1/3}$ controlled by the receiving
party. The sender encodes the qubit not only in a single qubit of
the chain, but in a \emph{Gaussian-modulated superposition} of his
qubits. These Gaussian packets are known to have minimal dispersion.
Likewise, the receiver performs a decoding operation on all qubits
he controls. Near-perfect fidelity can be reached.

\subsection{Time-dependent control\label{sub:Time-dependent-control}}

Finally, a number of authors found ways of improving the fidelity
by time-dependent control of some parameters of the Hamiltonian. In~\cite{Haselgrove2005}
it is shown that if the end couplings can be controlled as arbitrary
(in general complex valued) smooth functions of time the encoding
scheme~\cite{Linden2004} could be \emph{simulated} without the requirement
of additional operations and qubits. Another possibility to achieve
perfect state transfer is to have an Ising interaction with additionally
pulsed global rotations~\cite{Fitzsimons2006,Jones,Raussendorf2005}.
Further related methods of manipulating the transfer by global fields
were reported in \cite{Sun2006,Maruyama,Monteiro2006,Korepin2005,Yang2006,Boness2006}.

\section{Motivation and outline of this work}

While the advanced transfer protocols have shown that in principle
high fidelity can be achieved with arbitrarily long chains, they have
all come at a cost. Engineering each coupling of the Hamiltonian puts
extra demands on the experimental realisation, which is often already
at its very limits just to ensure the \emph{coherence} of the system.
Furthermore, the more a scheme relies on particular properties of
the Hamiltonian, the more it will be affected by imperfections in
its implementation~\cite{Fazio2005,Jing-Fu2006}. For example, simulating
an engineered chain of length $N=50$ with a (relative) disorder of
$5\%$, we get a fidelity peak of $0.95\pm0.02.$ For a disorder of
$10\%$ we get $0.85\pm0.05.$ The weakly coupled system is very stable
for off-site disorder~\cite{Semiao2005}, but suffers strongly from
on-site disorder (i.e. magnetic fields in $z-$direction) at the ends
of the chain. For example, for a chain of $N=50$ with edge couplings
$j=0.01$ and the remaining couplings being $J=1,$ we find that already
a magnetic field of the order of $0.00001$ lowers the fidelity to
$0.87\pm0.12.$ For fields of the order of $0.00005$ we find $0.45\pm0.32.$
This is because these fluctuations must be small with respect to the
\emph{small} coupling, so there is a double scaling. Also, the time-scale
of the transfer is longer than in other schemes (note though that
this may sometimes even be useful for having enough time to pick up
the received state). On the other hand, encoding and time-dependent
control require additional resources and gating operations. It is
not possible to judge independently of the realisation which of the
above schemes is the ''most practical'' one. We summarise the different
aspects that are important in the following five criteria for quantum
state transfer\index{criteria for quantum state transfer}:

\begin{enumerate}
\item \emph{High efficiency:} How does the fidelity depend on the length
of the chain? Which rate~\cite{Fazio,Kay2006,Yung2006} can be achieved?
\item \emph{Minimal control}: How many operations are required to achieve
a certain fidelity? \emph{Where}%
\footnote{For example, gates at the ends of the chain are always needed for
write-in and read-out, and thus ''cheaper'' than gates along the
chain. Global control along the whole chain is often easier than local
control.%
} is control required?
\item \emph{Minimal resources:} What additional resources are required?
\item \emph{Minimal design}: How general is the coupling type%
\footnote{Often the coupling type is already fixed by the experiment%
}? What values of the coupling strengths are allowed? 
\item \emph{Robustness}: How is the fidelity affected by static disorder,
by time-dependent disorder, by gate and timing errors, and by external
noise such as decoherence and dissipation?
\end{enumerate}
At the start of this research, only the engineering and encoding schemes
were available. The engineering schemes are strong in the points 2
and 3, but quite weak in the points 4 and 5. The encoding scheme on
the other hand has its weakness in points 2 and 3. It was hence desirable
to develop more balanced schemes. Since most experiments in Quantum
Information are extremely sensitive and at the cutting edge of their
parameters (i.e. requiring extremely low temperatures, well tuned
lasers, and so forth, to maintain their quantum behaviour), we particularly
wanted to find schemes which are strong in the points 4 and 5. Also,
from a more fundamental point of view, we were interested in seeing
how much information on the state of a quantum chain could be obtained
by the receiver in principle, and how the receiver might even be able
to \emph{prepare} states on the whole chain. 

The main achievements of this thesis are two schemes for the transfer
of quantum information using measurements (Chapter~\ref{cha:Dual-Rail}
and~\ref{cha:Multi-rail-and-Capacity}) or unitary operations (Chapter~\ref{cha:Full-read-and}
and~\ref{cha:Single-memory}) at the receiving end of the chain.
Since both schemes use convergence properties of quantum operations,
it seemed natural to investigate these properties in a more abstract
way (Chapter~\ref{cha:Ergodicity-and-mixing}). There, we found a
new way of characterising mixing maps, which has applications beyond
quantum state transfer, and may well be relevant for other fields
such as chaos theory or statistical physics. Finally, in Chapter~\ref{cha:Problems-and-Practical}
we discuss problems quantum state transfer in the presence of external
noise. The results in Chapters~\ref{cha:Multi-rail-and-Capacity}-\ref{cha:Single-memory}
were developed in collaboration with Vittorio Giovannetti from Scuola
Normale, Pisa. Much of the material discussed in this thesis has been
published or submitted for publication~\cite{Bose2006b,DUALRAIL,RANDOMRAIL,Giovannetti2006,Giovannetti,Bose2006c,Burgarth,MULTIRAIL,MEMORYSWAP,Bose2005e}.

\chapter{Dual Rail encoding\label{cha:Dual-Rail}}

\section{Introduction}

The role of measurement in quantum information theory has become more
active recently. Measurements are not only useful to obtain information
about some state or for preparation, but also, instead of gates, for
quantum computation \cite{Raussendorf2001}. In the context of quantum
state transfer, it seems first that measurements would spoil the coherence
and destroy the state. The first indication that measurements can
actually be used to transfer quantum information along anti-ferromagnetic
chains was given in~\cite{Verstraete2004}. However there the measurements
had to be performed along the whole chain. This may in some cases
be easier than to perform swap gates, but still requires high local
accessibility. We take a ''hybrid'' approach here: along the chain,
we let the system evolve coherently, but at the receiving end, we
try to \emph{help} the transfer by measuring. The main disadvantage
of the encoding used in the protocols above is that once the information
dispersed, there is no way of finding out where it is without destroying
it. A dual rail encoding~\cite{Chuang1996} as used in quantum optics
on the other hand allows us to perform parity type measurements that
do \emph{not} spoil the coherence of the state that is sent. The outcome
of the measurement tells us if the state has arrived at the end (corresponding
to a perfect state transfer) or not. We call this \emph{conclusively
perfect state transfer}.\index{conclusively perfect state transfer.}
Moreover, by performing repetitive measurements, the probability of
success can be made arbitrarily close to unity. As an example of such
an \emph{\index{amplitude delaying channel}amplitude delaying channel},
we show how two parallel Heisenberg spin chains can be used as quantum
wires. Perfect state transfer with a probability of failure lower
than $P$ in a Heisenberg chain of $N$ qubits can be achieved in
a time-scale of the order of $0.33J^{-1}N^{1.7}|\ln P|$. We demonstrate
that our scheme is more robust to decoherence and non-optimal timing
than any scheme using single spin chains. 

We then generalise the dual rail encoding to disordered quantum chains.
The scheme performs well for both spatially correlated and uncorrelated
fluctuations if they are relatively weak (say 5\%). Furthermore, we
show that given a quite arbitrary pair of quantum chains, one can
check whether it is capable of perfect transfer by only local operations
at the ends of the chains, and the system in the middle being a \emph{black
box}. We argue that unless some specific symmetries are present in
the system, it \emph{will} be capable of perfect transfer when used
with dual rail encoding. Therefore our scheme puts minimal demand
not only on the control of the chains when using them, but also on
the design when building them. 

This Chapter is organised as follows. In Section \ref{sec:con}, we
suggest a scheme for quantum communication using two parallel spin
chains of the most natural type (namely those with constant couplings).
We require modest encodings (or gates) and measurements only at the
ends of the chains. The state transfer is \emph{conclusive}, which
means that it is possible to tell by the outcome of a quantum measurement,
without destroying the state, if the transfer took place or not. If
it did, then the transfer was \emph{perfect}. The transmission time
for conclusive transfer is not longer than for single spin chains.
In Section \ref{sec:Arbit}, we demonstrate that our scheme offers
even more: if the transfer was not successful, then we can wait for
some time and just repeat the measurement, without having to resend
the state. By performing sufficiently many measurements, the probability
for perfect transfer approaches unity. Hence the transfer is \emph{arbitrarily
perfect}. We will show in Section \ref{sec:Estimati} that the time
needed to transfer a state with a given probability scales in a reasonable
way with the length of the chain. In Section \ref{sec:Decohere} we
show that encoding to parallel chains and the conclusiveness also
makes our protocol more robust to decoherence (a hitherto unaddressed
issue in the field of quantum communication through spin chains).
In the last part of this chapter, we show how this scheme can be generalised
to disordered chains (Sections~\ref{sec:Disordered-chains}-\ref{sec:Numerical-Examples})
and even coupled chains (Section~\ref{sec:Coupled-chains}).

\section{Scheme for conclusive transfer\label{sec:con}}

We intend to propose our scheme in a system-independent way with occasional
references to systems where conditions required by our scheme are
achieved. We assume that our system consists of two identical uncoupled
spin-$1/2$-chains $(1)$ and $(2)$ of length $N$, described by
the Hamiltonian\begin{equation}
H=H^{(1)}\otimes I^{(2)}+I^{(1)}\otimes H^{(2)}-E_{g}I^{(1)}\otimes I^{(2)}.\end{equation}
\begin{figure}[htbp]
\begin{centering}\includegraphics[width=0.65\paperwidth]{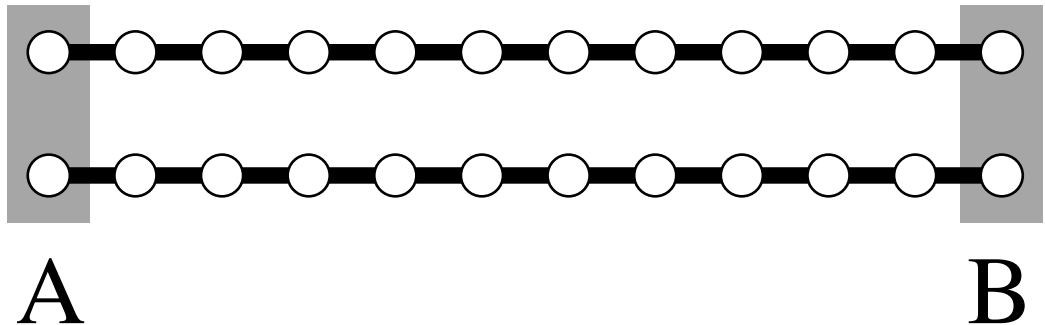}\par\end{centering}

\caption{\label{fig:channelord}Two quantum chains interconnecting $A$ and
$B$. Control of the systems is only possible at the two qubits of
either end.}
\end{figure}
 The term identical states that $H^{(1)}$ and $H^{(2)}$ are the
same apart from the label of the Hilbert space they act on. The requirement
of parallel chains instead of just one is not a real problem, since
in many experimental realisations of spin chains, it is much easier
to produce a whole bunch of parallel uncoupled~\cite{Motoyama1996,Gambardella2002}
chains than just a single one.

\begin{figure*}[tbh]
\begin{centering}\includegraphics[width=0.9\textwidth]{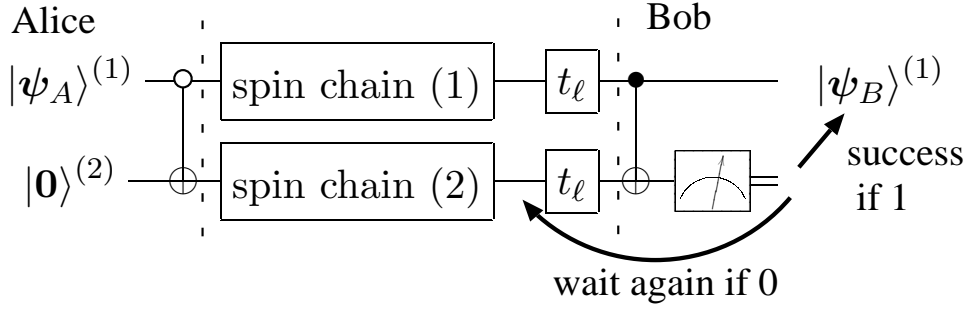}\par\end{centering}

\caption{\label{fig:summary}Quantum circuit representation of conclusive
and arbitrarily perfect state transfer. The first gate at Alice's
qubits represents a NOT gate applied to the second qubit controlled
by the first qubit being zero. The qubit $\left|\boldsymbol{\psi}_{A}\right\rangle _{1}$
on the left hand side represents an arbitrary input state at Alice's
site, and the qubit $\left|\boldsymbol{\psi}_{B}\right\rangle _{1}$
represents the same state, successfully transferred to Bob's site.
The $t_{\ell}$-gate represents the unitary evolution of the spin
chains for a time interval of $t_{\ell}$.}
\end{figure*}

We assume that the ground state of each chain is $\left|\boldsymbol{0}\right\rangle _{i}$,
i.e. a ferromagnetic ground state, with $H^{(i)}\left|\boldsymbol{0}\right\rangle _{i}=E_{g}\left|\boldsymbol{0}\right\rangle _{i},$
and that the subspace consisting of the single spin excitations $\left|\boldsymbol{n}\right\rangle _{i}$
is invariant under $H^{(i)}.$ Let us assume that the state that Alice
wants to send is at the first qubit of the first chain, i.e.\begin{equation}
\left|\boldsymbol{\psi}_{A}\right\rangle _{1}\equiv\alpha\left|\boldsymbol{0}\right\rangle _{1}+\beta\left|\boldsymbol{1}\right\rangle _{1},\end{equation}
and that the second chain is in the ground state $|\boldsymbol{0}\rangle_{2}.$
The aim of our protocol is to transfer quantum information from the
$1$st ({}``Alice'') to the $N$th ({}``Bob'') qubit of the first
chain:\begin{equation}
\left|\boldsymbol{\psi}_{A}\right\rangle _{1}\rightarrow\left|\boldsymbol{\psi}_{B}\right\rangle _{1}\equiv\alpha\left|\boldsymbol{0}\right\rangle _{1}+\beta\left|\boldsymbol{N}\right\rangle _{1}.\end{equation}
 The first step (see also Fig.~\ref{fig:summary}) is to encode the
input qubit in a \emph{dual rail}\index{dual rail}~\cite{Chuang1996}
by applying a NOT gate on the first qubit of system $(2)$ controlled
by the first qubit of system $(1)$ being zero, resulting in a superposition
of excitations in both systems, \begin{equation}
\left|\boldsymbol{s}(0)\right\rangle =\alpha\left|\boldsymbol{0,1}\right\rangle +\beta\left|\boldsymbol{1,0}\right\rangle ,\label{eq:superposition}\end{equation}
where we have introduced the short notation $|\boldsymbol{n,m}\rangle\equiv|\boldsymbol{n}\rangle_{1}\otimes|\boldsymbol{m}\rangle_{2}.$
This is assumed to take place in a much shorter time-scale than the
system dynamics. Even though a 2-qubit gate in solid state systems
is difficult, such a gate for charge qubits has been reported~\cite{YAMA}.
For the same qubits, Josephson arrays have been proposed as single
spin chains for quantum communication~\cite{Bruder2005a}. For this
system, both requisites of our scheme are thus available. In fact,
the demand that Alice and Bob can do measurements and apply gates
to their local qubits (i.e. the ends of the chains) will be naturally
fulfilled in practice since we are suggesting a scheme to transfer
information between quantum computers (as described in Section~\ref{sec:Quantum-state-transfer}).%

Under the system Hamiltonian, the excitation in Eq. (\ref{eq:superposition})
will travel along the two systems. The state after the time $t_{1}$
can be written as \begin{equation}
\left|\boldsymbol{\phi}(t_{1})\right\rangle =\sum_{n=1}^{N}f_{n,1}(t_{1})\left|\boldsymbol{s}(n)\right\rangle ,\end{equation}
 where $\left|\boldsymbol{s}(n)\right\rangle =\alpha\left|\boldsymbol{0,n}\right\rangle +\beta\left|\boldsymbol{n,0}\right\rangle $
and the complex amplitudes $f_{n,1}(t_{1})$ are given by Eq.~(\ref{eq:spintrans}).
We can \emph{decode} the qubit by applying a CNOT gate at Bob's site.
Assuming that this happens on a time-scale much shorter than the evolution
of the chain, the resulting state is given by \begin{equation}
\sum_{n=1}^{N-1}f_{n,1}(t_{1})\left|\boldsymbol{s}(n)\right\rangle +f{}_{N,1}(t_{1})\left|\boldsymbol{\psi}_{B}\right\rangle _{1}\otimes\left|\boldsymbol{N}\right\rangle _{2}.\end{equation}
 Bob can now perform a measurement on his qubit of system $(2).$
If the outcome of this measurement is $1$, he can conclude that the
state $\left|\boldsymbol{\psi}\right\rangle _{1}^{(1)}$ has been
successfully transferred to him. This happens with the probability
$\left|f_{N,1}(t_{1})\right|^{2}.$ If the outcome is $0$, the system
is in the state \begin{equation}
\frac{1}{\sqrt{P(1)}}\sum_{n=1}^{N-1}f_{n,1}(t_{1})\left|\boldsymbol{s}(n)\right\rangle ,\label{eq:firstmeas}\end{equation}
 where $P(1)=1-\left|f_{N,1}(t_{1})\right|^{2}$ is the probability
of \emph{failure} for the first measurement. If the protocol stopped
here, and Bob would just assume his state as the transferred one,
the channel could be described as an \emph{amplitude damping channel}~\cite{Giovannetti2005},
with exactly the same fidelity as the single chain scheme discussed
in~\cite{Bose2003}. Note that here the encoding is symmetric with
respect to $\alpha$ and $\beta,$ so the minimal fidelity is the
same as the averaged one.

But success probability is more valuable than fidelity: Bob has gained
knowledge about his state, and may reject it and ask Alice to retransmit
(this is known as a \emph{\index{quantum erasure channel}quantum
erasure channel}~\cite{Bennett1997}). Of course in general the state
that Alice sends is the unknown result of some quantum computation
and cannot be sent again easily. This can be overcome in the following
way: Alice sends one e-bit on the dual rail first. If Bob measures
a success, he tells Alice, and they both start to teleport the unknown
state. If he measures a failure, they reset the chains and start again.
Since the joint probability of failure converges exponentially fast
to zero this is quite efficient. In fact the conclusive transfer of
entanglement is possible even on a \emph{single chain} by using the
same chain again instead of a second one \cite{Wan}. This can be
seen as a very simple \index{entanglement distillation}entanglement
distillation procedure, achieving a rate of $|f_{N,1}(t)|^{2}/2.$
However the chain needs to be reset between each transmission (see
Section~\ref{sec:Initialisation} for problems related to this),
and Alice and Bob require classical communication. We will show in
the next section, that the reuse of the chain(s) is not necessary,
as arbitrarily perfect state transfer can already achieved in the
first transmission.

\section{Arbitrarily perfect state transfer\label{sec:Arbit}}

\begin{figure}[tbh]
\begin{centering}\includegraphics[width=0.8\columnwidth]{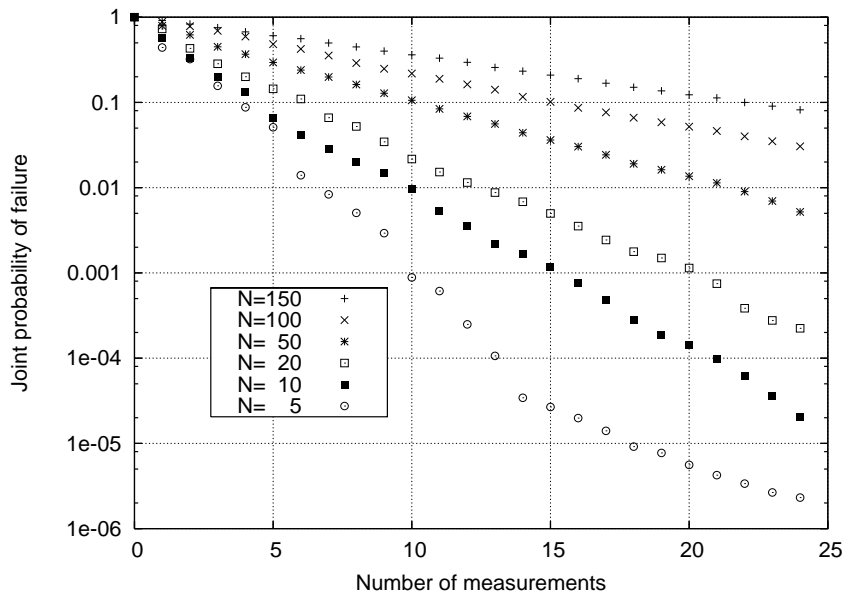}\par\end{centering}

\caption{\label{fig:heisenberg}Semilogarithmic plot of the joint probability
of failure $P(\ell)$ as a function of the number of measurements
$\ell$. Shown are Heisenberg spin-$1/2$-chains with different lengths
$N$. The times between measurements $t_{\ell}$ have been optimised
numerically. }
\end{figure}

Because Bob's measurement has not revealed anything about the input
state (the success probability is independent of the input state),
the information is still residing in the chain. By letting the state
(\ref{eq:firstmeas}) evolve for another time $t_{2}$ and applying
the CNOT gate again, Bob has another chance of receiving the input
state. The state before performing the second measurement is easily
seen to be \begin{equation}
\frac{1}{\sqrt{P(1)}}\sum_{n=1}^{N}\left\{ f_{n,1}(t_{2}+t_{1})-f_{n,N}(t_{2})f_{N,1}(t_{1})\right\} \left|\boldsymbol{s}(n)\right\rangle .\label{eq:second_meas}\end{equation}
 Hence the probability to receive the qubit at Bobs site at the second
measurement is\begin{equation}
\frac{1}{P(1)}\left|f_{N,1}(t_{2}+t_{1})-f_{N,N}(t_{2})f_{N,1}(t_{1})\right|^{2}.\end{equation}
 If the transfer was still unsuccessful, this strategy can be repeated
over and over. Each time Bob has a probability of failed state transfer
that can be obtained from the generalisation of Eq. (\ref{eq:second_meas})
to an arbitrary number of iterations. The joint probability that Bob
fails to receive the state all the time is just the product of these
probabilities. We denote the joint probability of failure for having
done $l$ unsuccessful measurements as $P(\ell)$. This probability
depends on the time intervals $t_{\ell}$ between the $(\ell-1)$th
and $\ell$th measurement, and we are interested in the case where
the $t_{\ell}$ are chosen such that the transfer is fast. It is possible
to write a simple algorithm that computes $P(\ell)$ for any transition
amplitude $f_{r,s}(t).$ Figure \ref{fig:heisenberg} shows some results
for the Heisenberg Hamiltonian given by Eq.~(\ref{eq:heisenberg}).%

An interesting question is whether the joint probability of failure
can be made arbitrarily small with a large number of measurements.
In fact, the times $t_{\ell}$ can be chosen such that the transfer
becomes arbitrarily perfect. We will prove this in the next Chapter,
where a generalisation of the dual rail scheme and a much wider class
of Hamiltonians is considered. In the limit of large number of measurements,
the spin channel will not damp the initial amplitude, but only \emph{delay}
it.

\section{Estimation of the \index{time-scale}time-scale the transfer\label{sec:Estimati}}

The achievable fidelity is an important, but not the only criterion
of a state transfer protocol. In this Section, we give an heuristic
approach to estimate the time that it needs to achieve a certain fidelity
in a Heisenberg spin chain. The comparison with numeric examples is
confirming this approach.

\begin{figure}[tbh]
\includegraphics[width=1\columnwidth]{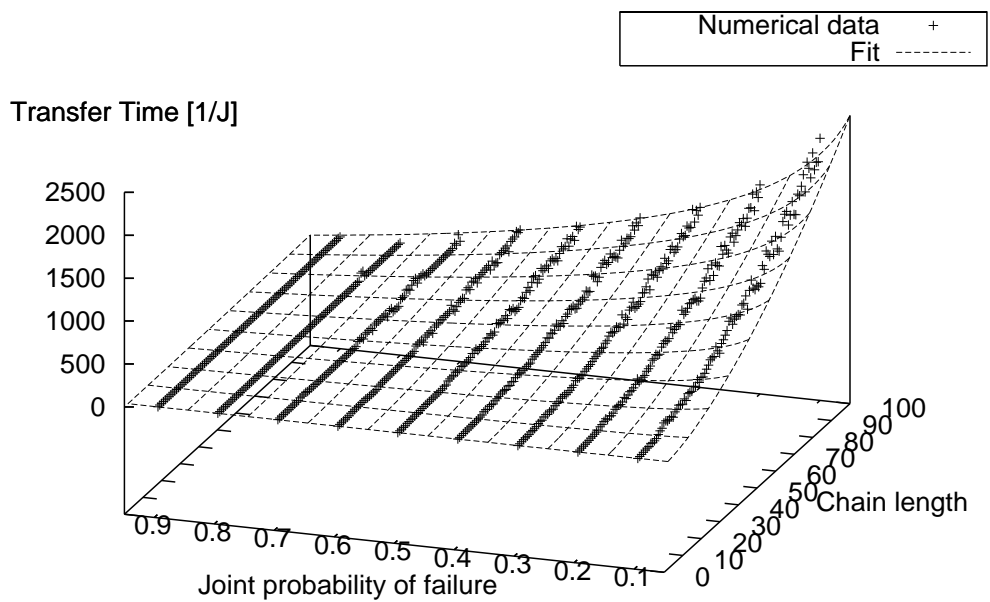}

\caption{\label{cap:Timefit}Time $t$ needed to transfer a state with a given
joint probability of failure $P$ across a chain of length $N$. The
points denote exact numerical data, and the fit is given by Eq. (\ref{eq:fit3}).}
\end{figure}

Let us first describe the dynamic of the chain in a very qualitative
way. Once Alice has initialised the system, an excitation wave packet
will travel along the chain. As shown in Subsection~\ref{sub:Dynamics-and-Dispersion},
it will reach Bob at a time of the order of \begin{equation}
t_{\mbox{peak}}\approx\frac{N}{2J},\end{equation}
 with an amplitude of \begin{equation}
\left|f_{N,1}(t_{\mbox{peak}})\right|^{2}\approx1.82N^{-2/3}.\label{eq:peak}\end{equation}
 It is then reflected and travels back and forth along the chain.
Since the wave packet is also dispersing, it starts interfering with
its tail, and after a couple of reflections the dynamic is becoming
quite randomly. This effect becomes even stronger due to Bobs measurements,
which change the dynamics by projecting away parts of the wave packet.
We now assume that $2t_{\mbox{peak}}$ (the time it takes for a wave
packet to travel twice along the chain) remains a good estimate of
the time-scale in which significant probability amplitude peaks at
Bobs site occur, and that Eq.~(\ref{eq:peak}) remains a good estimate
of the amplitude of these peaks%
\footnote{This is not a strong assumption. If the excitation was fully randomly
distributed, the probability would scale as $N^{-1}.$ By searching
for good arrival times, this can be slightly increased to $N^{-2/3}.$%
}. Therefore, the joint probability of failure is expected to scale
as\begin{equation}
P(\ell)\approx\left(1-1.82N^{-2/3}\right)^{\ell}\label{eq:probscale}\end{equation}
 in a time of the order of \begin{equation}
t(\ell)\approx2t_{max}\ell=J^{-1}N\ell.\label{eq:timescale}\end{equation}
 If we combine Eq. (\ref{eq:probscale}) and (\ref{eq:timescale})
and solve for the time $t(P)$ needed to reach a certain probability
of failure $P$, we get for $N\gg1$ \begin{equation}
t(P)\approx0.55J^{-1}N^{5/3}\left|\ln P\right|.\end{equation}
 We compare this rough estimate with exact numerical results in Fig.
\ref{cap:Timefit}. The best fit for the range shown in the figure
is given by\begin{equation}
\boxed{t(P)=0.33J^{-1}N^{5/3}\left|\ln P\right|.}\label{eq:fit3}\end{equation}
 We can conclude that the transmission time for arbitrarily perfect
transfer is scaling not much worse with the length $N$ of the chains
than the single spin chain schemes. Despite of the logarithmic dependence
on $P,$ the time it takes to achieve high fidelity is still reasonable.
For example, a system with $N=100$ and $J=20K*k_{B}$ will take approximately
$1.3ns$ to achieve a fidelity of 99\%. In many systems, decoherence
is completely negligible within this time-scale. For example, some
Josephson junction systems~\cite{Vion2002} have a decoherence time
of $T_{\phi}\approx500ns$, while trapped ions have even larger decoherence
times.%

\section{Decoherence and imperfections\label{sec:Decohere}}

If the coupling between the spins $J$ is very small, or the chains
are very long, the transmission time may no longer be negligible with
respect to the decoherence time. It is interesting to note that the
dual rail encoding then offers some significant general advantages
over single chain schemes. Since we are suggesting a system-independent
scheme, we will not study the effects of specific environments on
our protocol, but just qualitatively point out its general advantages.

At least theoretically, it is always possible to cool the system down
or to apply a strong magnetic field so that the environment is not
causing further excitations. For example in flux qubit systems, the
system is cooled to $\approx25mK$ to ensure that the energy splitting$\Delta\gg k_{B}T$
\cite{Chiorescu2003}. Then, there are two remaining types of quantum
noise that will occur: phase noise\index{phase noise} and amplitude
damping\index{amplitude damping}. Phase noise is a serious problem
and arises here \emph{only} when an environment can distinguish between
spin flips on the first chain and spin flips on the second chain.
It is therefore important that the environment cannot resolve their
difference. In this case, the environment will only couple with the
total $z$-component\begin{equation}
Z_{n}^{(1)}+Z_{n}^{(2)}\end{equation}
 of the spins of both chains at each position $n$. This has been
discussed for spin-boson models in~\cite{Palma1996,Hwang2000} but
also holds for spin environments as long as the chains are close enough.
The qubit is encoded in a decoherence-free subspace\index{decoherence-free subspace}~\cite{Beige2000}
and the scheme is fully robust to phase noise. Even though this may
not be true for all implementations of dual rail encoding, it is worthwhile
noticing it because such an opportunity does not exist \emph{at all}
for single chain schemes, where the coherence between two states with
different total z-component of the spin has to be preserved. Having
shown one way of avoiding phase noise, at least in some systems, we
now proceed to amplitude damping.

\begin{figure}[tbh]
\begin{centering}\includegraphics[width=0.8\columnwidth]{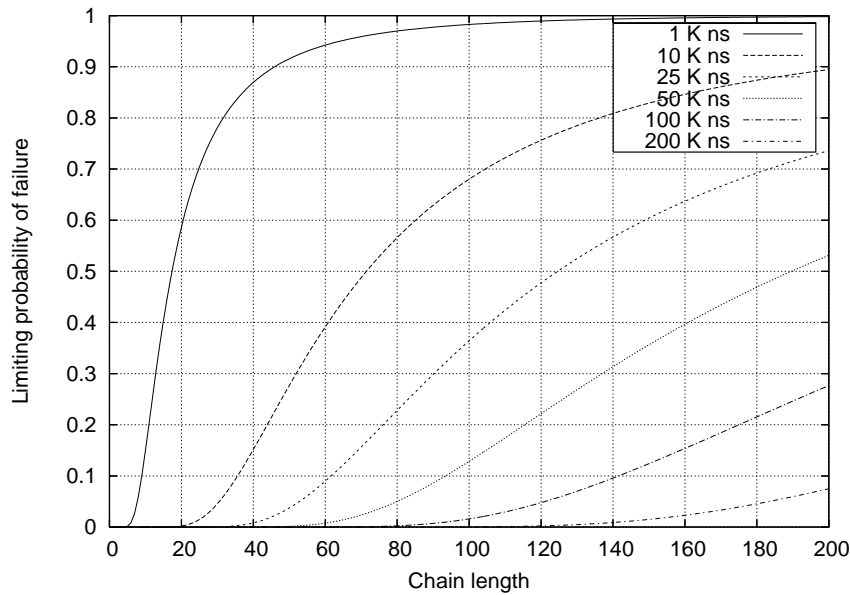}\par\end{centering}

\caption{\label{fig:losses}The minimal joint probability of failure $P(\ell)$
for chains with length $N$ in the presence of amplitude damping.
The parameter $J/\Gamma$ of the curves is the coupling of the chain
(in Kelvin) divided by the decay rate ($ns^{-1}$).}
\end{figure}

The evolution of the system in presence of amplitude damping of a
rate $\Gamma$ can be easily derived using a quantum-jump approach\index{quantum-jump approach}~\cite{Plenio1998}.
This is based on a quantum master equation approach, which is valid
in the Born-Markov approximation~\cite{OPENQUANTUM} (i.e. it holds
for weakly coupled environments without memory effects). Similarly
to phase noise, it is necessary that the environment acts symmetrically
on the chains. The dynamics is then given by an effective non-Hermitian
Hamiltonian\begin{equation}
H_{eff}=H+i\Gamma\sum_{n}\left(Z_{n}^{(1)}+Z_{n}^{(2)}+2\right)/2\end{equation}
 if no jump occurs. If a jump occurs, the system is back in the ground
state $|\boldsymbol{0}\rangle$. The state of the system before the
first measurement conditioned on no jump is given by\begin{equation}
e^{-\Gamma t}\sum_{n=1}^{N}f_{n,1}(t)\left|\boldsymbol{s}(n)\right\rangle ,\label{eq:jump1}\end{equation}
 and this happens with the probability of $e^{-2\Gamma t}$ (the norm
of the above state). If a jump occurs, the system will be in the ground
state \begin{equation}
\sqrt{1-e^{-2\Gamma t}}\left|\boldsymbol{0,0}\right\rangle .\label{eq:jump2}\end{equation}
 The density matrix at the time $t$ is given by a mixture of (\ref{eq:jump1})
and (\ref{eq:jump2}). In case of (\ref{eq:jump2}), the quantum information
is completely lost and Bob will always measure an unsuccessful state
transfer. If Bob however measures a success, it is clear that no jump
has occurred and he has the perfectly transferred state. Therefore
the protocol \emph{remains conclusive}, but the success probability
is lowered by $e^{-2\Gamma t}.$ This result is still valid for multiple
measurements, which leave the state (\ref{eq:jump2}) unaltered. The
probability of a successful transfer at each particular measurement
$\ell$ will decrease by $e^{-2\Gamma t(\ell)}$, where $t(\ell)$
is the time at which the measurement takes place. After a certain
number of measurements, the \emph{joint} probability of failure will
no longer decrease. Thus the transfer will no longer be \emph{arbitrarily}
perfect, but can still reach a very high fidelity. Some numerical
examples of the minimal joint probability of failure that can be achieved,
\begin{eqnarray}
\lim_{l\rightarrow\infty}P(\ell) & \approx & \prod_{\ell=1}^{\infty}\left(1-1.35N^{-2/3}e^{-\frac{2\Gamma N}{J}\ell}\right)\label{eq:plimit}\end{eqnarray}
 are given in Fig. \ref{fig:losses}. For $J/\Gamma=50K\: ns$ nearly
perfect transfer is still possible for chains up to a length of $N\approx40$. 

Even if the amplitude damping is not symmetric, its effect is weaker
than in single spin schemes. This is because it can be split in a
symmetric and asymmetric part. The symmetric part can be overcome
with the above strategies. For example, if the amplitude damping on
the chains is $\Gamma_{1}$ and $\Gamma_{2}$ with $\Gamma_{1}>\Gamma_{2},$
the state (\ref{eq:jump1}) will be\begin{eqnarray}
 &  & \sum_{n=1}^{N}f_{n,1}(t)\left\{ \alpha e^{-\Gamma_{2}t}\left|\boldsymbol{0,n}\right\rangle +\beta e^{-\Gamma_{1}t}\left|\boldsymbol{n,0}\right\rangle \right\} \\
 & \approx & e^{-\Gamma_{2}t}\sum_{n=1}^{N}f_{n,1}(t)\left|\boldsymbol{s}(n)\right\rangle \label{eq:nodeviation}\end{eqnarray}
 provided that $t\ll\left(\Gamma_{1}-\Gamma_{2}\right)^{-1}.$ Using
a chain of length $N=20$ with $J=20K*k_{B}$ and $\Gamma_{1}^{-1}=4ns$,
$\Gamma_{2}^{-1}=4.2ns$ we would have to fulfil $t\ll164ns$. We
could perform approximately $10$ measurements (cf. Eq. (\ref{eq:timescale}))
without deviating too much from the state (\ref{eq:nodeviation}).
In this time, we can use our protocol in the normal way. The resulting
success probability given by the finite version of Eq. (\ref{eq:plimit})
would be $75$\%. A similar reasoning is valid for phase noise, where
the environment can be split into common and separate parts. If the
chains are close, the common part will dominate and the separate parts
can be neglected for short times.%

\section{Disordered chains\label{sec:Disordered-chains}}

The main requirement for perfect transfer with dual rail encoding
in the above is that two \emph{identical} quantum chains have to be
designed. While this is not so much a theoretical problem, for possible
experimental realizations of the scheme~\cite{Bruder2005a} the question
arises naturally how to cope with slight asymmetries of the channels.
We are now going to demonstrate that in many cases, perfect state
transfer with dual rail encoding is possible for quantum chains with
differing Hamiltonians.

By doing so, we also offer a solution to another and perhaps more
\emph{general} problem: if one implements \emph{any} of the schemes
for quantum state transfer, the Hamiltonians will always be different
from the theoretical ones by some random perturbation. This will lead
to a decrease of fidelity in particular where specific energy levels
were assumed (see~\cite{Fazio2005,Jing-Fu2006} for an analysis of
fluctuations affecting the engineered chains described in Subsection~\ref{sub:Engineered-Hamiltonians}).
This problem can be avoided using the scheme described below. In general,
disorder can lead to a Anderson localisation\index{Anderson localisation}~\cite{Anderson1958,Winter,Apollaro2006}
of the eigenstates (and therefore to low fidelity transport of quantum
information). In this section however this is not relevant, as we
consider only short chains $(N<100)$ and small disorder ($\approx10\%$
of the coupling strength), and the localisation length is much longer
then the length of the chain. We will show numerically that the dual
rail scheme can still achieve arbitrarily perfect transfer for a uniformly
coupled Heisenberg Hamiltonian with disordered coupling strengths
(both for the case of spatially correlated and uncorrelated disorder).
Moreover, for any two quantum chains, we show that Bob and Alice can
check whether their system is capable of dual rail transfer without
directly measuring their Hamiltonians or local properties of the system
along the chains but by only measuring \emph{their} part of the system.

\section{Conclusive transfer in the presence of disorder}

\begin{figure}[htbp]
\begin{centering}\includegraphics[width=0.5\paperwidth]{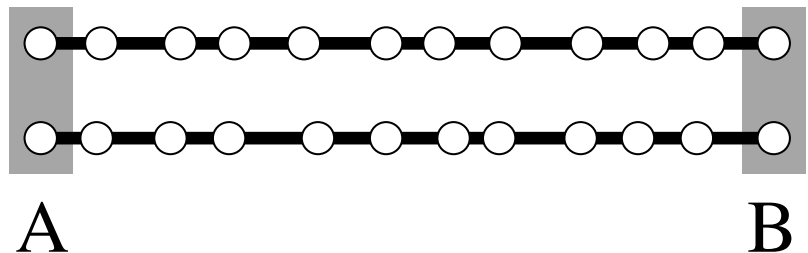}\par\end{centering}

\caption{\label{fig:channel}Two \emph{disordered} quantum chains interconnecting
$A$ and $B$. Control of the systems is only possible at the two
qubits of either end.}
\end{figure}

We consider two uncoupled quantum chains $(1)$ and $(2)$, as shown
in Fig. \ref{fig:channel}. The chains are described by the two Hamiltonians
$H^{(1)}$ and $H^{(2)}$ with total Hamiltonian given by\begin{equation}
H=H^{(1)}\otimes I^{(2)}+I^{(1)}\otimes H^{(2)},\end{equation}
 and the time evolution operator factorising as\begin{eqnarray}
U(t) & = & \exp\left(-iH^{(1)}t\right)\otimes\exp\left(-iH^{(2)}t\right).\end{eqnarray}
 For the moment, we assume that both chains have equal length $N$,
but it will become clear in Section \ref{sec:Tomography} that this
is not a requirement of our scheme. All other assumptions remain as
in the first part of the chapter.%

\begin{figure}[htbp]
\begin{centering}\includegraphics[width=0.8\columnwidth]{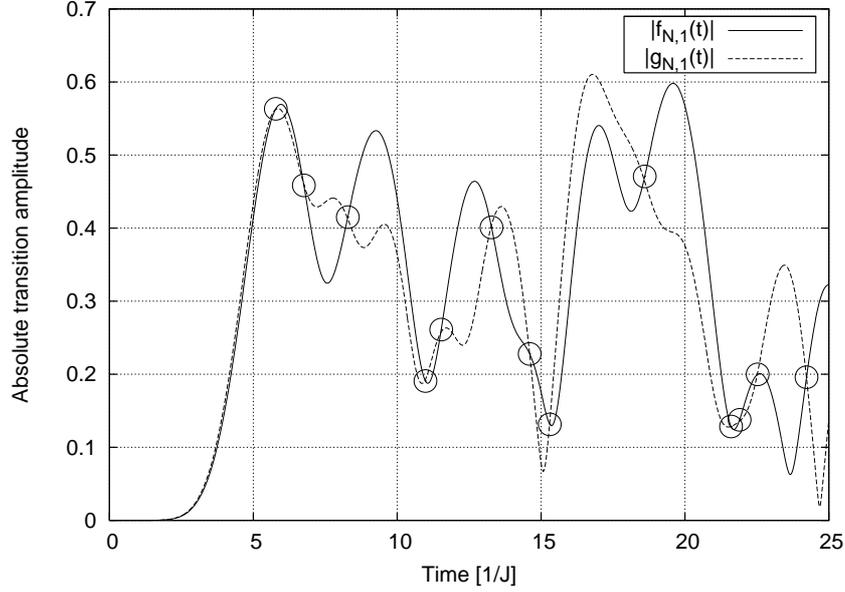}\par\end{centering}

\caption{\label{fig:coincidence}The absolute values of the transition amplitudes
$f_{N,1}(t)$ and $g_{N,1}(t)$ for two Heisenberg chains of length
$N=10$. The couplings strengths of both chains were chosen randomly
from the interval $\left[0.8J,1.2J\right].$ The circles show times
where Bob can perform measurements without gaining information on
$\alpha$ and $\beta.$}
\end{figure}

Initially, Alice encodes the state as\begin{equation}
\alpha\left|\boldsymbol{0,1}\right\rangle +\beta\left|\boldsymbol{1,0}\right\rangle .\end{equation}
 This is a superposition of an excitation in the first qubit of the
first chain and an excitation in the first qubit of the second chain.
The state will evolve into\begin{equation}
\sum_{n=1}^{N}\left\{ \alpha g_{n,1}(t)\left|\boldsymbol{0,n}\right\rangle +\beta f_{n,1}(t)\left|\boldsymbol{n,0}\right\rangle \right\} ,\label{eq:evolved}\end{equation}
 with \begin{eqnarray}
f_{n,1}(t) & \equiv & \left\langle \boldsymbol{n,0}\left|U(t)\right|\boldsymbol{1,0}\right\rangle \\
g_{n,1}(t) & \equiv & \left\langle \boldsymbol{0,n}\left|U(t)\right|\boldsymbol{0,1}\right\rangle .\end{eqnarray}
 In Section~\ref{sec:con}, these functions were identical. For differing
chains this is no longer the case. We may, however, find a time $t_{1}$
such that the modulus of their amplitudes at the last spins are the
same (see Fig. \ref{fig:coincidence}),\begin{equation}
g_{N,1}(t_{1})=e^{i\phi_{1}}f_{N,1}(t_{1}).\label{eq:req1}\end{equation}

At this time, the state (\ref{eq:evolved}) can be written as\begin{eqnarray}
\sum_{n=1}^{N-1}\left\{ \alpha g_{n,1}(t_{1})\left|\boldsymbol{0,n}\right\rangle +\beta f_{n,1}(t_{1})\left|\boldsymbol{n,0}\right\rangle \right\} +\nonumber \\
f_{N,1}(t_{1})\left\{ e^{i\phi_{1}}\alpha\left|\boldsymbol{0,N}\right\rangle +\beta\left|\boldsymbol{N,0}\right\rangle \right\} .\end{eqnarray}
 Bob decodes the state by applying a CNOT gate on his two qubits,
with the first qubit as the control bit. The state thereafter is\begin{eqnarray}
\sum_{n=1}^{N-1}\left\{ \alpha g_{n,1}(t_{1})\left|\boldsymbol{0,n}\right\rangle +\beta f_{n,1}(t_{1})\left|\boldsymbol{n,0}\right\rangle \right\} +\nonumber \\
f_{N,1}(t_{1})\left\{ e^{i\phi_{1}}\alpha\left|\boldsymbol{0}\right\rangle ^{(1)}+\beta\left|\boldsymbol{N}\right\rangle ^{(1)}\right\} \otimes\left|\boldsymbol{N}\right\rangle ^{(2)}.\end{eqnarray}
 Bob then measures his second qubit. Depending on the outcome of this
measurement, the systems will either be in the state\begin{equation}
\frac{1}{\sqrt{p_{1}}}\sum_{n=1}^{N-1}\left\{ \alpha g_{n,1}(t_{1})\left|\boldsymbol{0,n}\right\rangle +\beta f_{n,1}(t_{1})\left|\boldsymbol{n,0}\right\rangle \right\} \label{eq:failure}\end{equation}
 or in\begin{equation}
\left\{ e^{i\phi_{1}}\alpha\left|\boldsymbol{0}\right\rangle ^{(1)}+\beta\left|\boldsymbol{N}\right\rangle ^{(1)}\right\} \otimes\left|\boldsymbol{N}\right\rangle ^{(2)},\label{eq:success}\end{equation}
 where $p_{1}=1-\left|f_{N,1}(t_{1})\right|^{2}=1-\left|g_{N,1}(t_{1})\right|^{2}$
is the probability that Bob has \emph{not} received the state. The
state (\ref{eq:success}) corresponds to the correctly transferred
state with a \emph{known} phase error (which can be corrected by Bob
using a simple phase gate). If Bob finds the system in the state (\ref{eq:failure}),
the transfer has been unsuccessful, but the information is still in
the chain. We thus see that conclusive transfer is still possible
with randomly coupled chains as long as the requirement (\ref{eq:req1})
is met. This requirement will be further discussed and generalised
in the next section.

\section{Arbitrarily perfect transfer in the presence of disorder\label{sec:Arbitrarily-perfect-transferl}}

If the transfer was unsuccessful, the state (\ref{eq:failure}) will
evolve further, offering Bob further opportunities to receive Alice's
message. For identical quantum chains, leads to a success for any
reasonable Hamiltonian (Section~\ref{sec:Quantum-chains-with}).
For differing chains, this is not necessarily the case, because measurements
are only allowed at times where the probability amplitude at the end
of the chains is equal, and there may be systems where this is never
the case. In this section, we will develop a criterion that generalises
Eq. (\ref{eq:req1}) and allows to check numerically whether a given
system is capable of arbitrarily perfect state transfer.

The quantity of interest for conclusive state transfer is the joint
probability $P(\ell)$ that after having checked $l$ times, Bob still
has not received the proper state at his end of the chains. Optimally,
this should approach zero if $\ell$ tends to infinity. In order to
derive an expression for $P(\ell),$ let us assume that the transfer
has been unsuccessful for $\ell-1$ times with time intervals $t_{\ell}$
between the the $\ell$th and the $(\ell-1)$th measurement, and calculate
the probability of failure at the $\ell$th measurement. In a similar
manner, we assume that all the $\ell-1$ measurements have met the
requirement of conclusive transfer (that is, Bob's measurements are
unbiased with respect to $\alpha$ and $\beta$) and derive the requirement
for the $\ell$th measurement.

To calculate the probability of failure for the $\ell$th measurement,
we need to take into account that Bob's measurements disturb the unitary
dynamics of the chain. If the state before a measurement with the
outcome {}``failure'' is $\left|\psi\right\rangle ,$ the state
after the measurement will be\begin{equation}
\frac{1}{\sqrt{p_{\ell}}}Q\left|\psi\right\rangle ,\end{equation}
 where $Q$ is the projector\begin{equation}
Q=I-\left|\boldsymbol{0,N}\right\rangle \left\langle \boldsymbol{0,N}\right|-\left|\boldsymbol{N,0}\right\rangle \left\langle \boldsymbol{N,0}\right|,\end{equation}
 and $p_{\ell}$ is the probability of failure at the $l$th measurement.
The dynamics of the chain is alternating between unitary and projective,
such that the state before the $\ell$th measurement is given by\begin{equation}
\frac{1}{\sqrt{P(\ell-1)}}\prod_{k=1}^{\ell}\left\{ U(t_{k})Q\right\} \left\{ \alpha\left|\boldsymbol{1,0}\right\rangle +\beta\left|\boldsymbol{0,1}\right\rangle \right\} ,\label{eq:product}\end{equation}
 where \begin{equation}
P(\ell-1)=\prod_{\ell=1}^{\ell-1}p_{k}.\label{eq:joint_equal_prod}\end{equation}
 Note that the operators in (\ref{eq:product}) do not commute and
that the time ordering of the product (the index $k$ increases from
right to left) is important. The probability that there is an excitation
at the $N$th site of either chain is given by\begin{equation}
\frac{1}{P(\ell-1)}\left\{ \left|\alpha\right|^{2}\left|F(\ell)\right|^{2}+\left|\beta\right|^{2}\left|G(\ell)\right|^{2}\right\} ,\end{equation}
 with \begin{equation}
F(\ell)\equiv\left\langle \boldsymbol{N,0}\right|\prod_{k=1}^{\ell}\left\{ U(t_{k})Q\right\} \left|\boldsymbol{1,0}\right\rangle ,\label{eq:onlything}\end{equation}
 and \begin{equation}
G(\ell)\equiv\left\langle \boldsymbol{0,N}\right|\prod_{k=1}^{\ell}\left\{ U(t_{k})Q\right\} \left|\boldsymbol{0,1}\right\rangle .\label{eq:onlything2}\end{equation}
 Bob's measurements are therefore unbiased with respect to $\alpha$
and $\beta$ if and only if\begin{equation}
\left|F(\ell)\right|=\left|G(\ell)\right|\quad\forall\ell.\label{eq:condition0}\end{equation}
 In this case, the state can still be transferred conclusively (up
to a known phase). The probability of failure at the $\ell$th measurement
is given by\begin{equation}
p_{\ell}=1-\frac{\left|F(\ell)\right|^{2}}{P(\ell-1)}.\label{eq:prob_fail_single}\end{equation}
It is easy (but not very enlightening) to show~\cite{RANDOMRAIL}
that the condition (\ref{eq:condition0}) is equivalent to\begin{equation}
\left\Vert \prod_{k=1}^{\ell}\left\{ U(t_{k})Q\right\} \left|\boldsymbol{1,0}\right\rangle \right\Vert =\left\Vert \prod_{k=1}^{\ell}\left\{ U(t_{k})Q\right\} \left|\boldsymbol{0,1}\right\rangle \right\Vert \quad\forall\ell,\label{eq:finalcondition}\end{equation}
 and that the joint probability of failure - if at each measurement
the above condition is fulfilled - is simply given by\begin{equation}
P(\ell)=\left\Vert \prod_{k=1}^{\ell+1}\left\{ U(t_{k})Q\right\} \left|\boldsymbol{1,0}\right\rangle \right\Vert ^{2}.\label{eq:finalprob}\end{equation}
It may look as if Eq. (\ref{eq:finalcondition}) was a complicated
multi-time condition for the measuring times $t_{\ell}$, that becomes
increasingly difficult to fulfil with a growing number of measurements.
This is not the case. If proper measuring times have been found for
the first $\ell-1$ measurements, a trivial time $t_{\ell}$ that
fulfils Eq. (\ref{eq:finalcondition}) is $t_{\ell}=0.$ In this case,
Bob measures immediately after the $(\ell-1)$th measurement and the
probability amplitudes on his ends of the chains will be equal - and
zero (a useless measurement). But since the left and right hand side
of Eq. (\ref{eq:finalcondition}) when seen as functions of $t_{\ell}$
are both almost-periodic functions with initial value zero, it is
likely that they intersect many times, unless the system has some
specific symmetry or the systems are completely different. Note that
we do not claim at this point that any pair of chains will be capable
of arbitrary perfect transfer. We will discuss in the next system
how one can check this for a given system by performing some simple
experimental tests.

\section{Tomography\index{tomography}\label{sec:Tomography}}

\begin{figure}[tbh]
\begin{centering}\includegraphics[width=0.8\columnwidth]{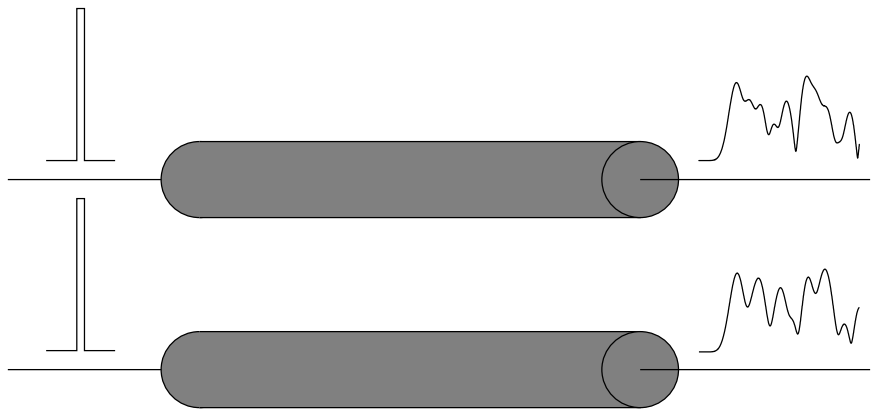}\par\end{centering}

\caption{\label{cap:blackbox}The relevant properties for conclusive transfer
can be determined by measuring the response of the two systems at
their ends only.}
\end{figure}

Suppose someone gives you two different experimentally designed spin
chains. It may seem from the above that knowledge of the full Hamiltonian
of both chains is necessary to check how well the system can be used
for state transfer. This would be a very difficult task, because we
would need access to all the spins along the channel to measure all
the parameters of the Hamiltonian. In fact by expanding the projectors
in Eq. (\ref{eq:finalcondition}) one can easily see that the only
matrix elements of the evolution operator which are relevant for conclusive
transfer are\begin{eqnarray}
f_{N,1}(t) & = & \left\langle \boldsymbol{N,0}\right|U(t)\left|\boldsymbol{1,0}\right\rangle \label{eq:t1}\\
f_{N,N}(t) & = & \left\langle \boldsymbol{N,0}\right|U(t)\left|\boldsymbol{N,0}\right\rangle \label{eq:t2}\\
g_{N,1}(t) & = & \left\langle \boldsymbol{0,N}\right|U(t)\left|\boldsymbol{0,1}\right\rangle \label{eq:t3}\\
g_{N,N}(t) & = & \left\langle \boldsymbol{0,N}\right|U(t)\left|\boldsymbol{0,N}\right\rangle .\label{eq:t4}\end{eqnarray}
 Physically, this means that the only relevant properties of the system
are the transition amplitudes to \emph{arrive} at Bob's ends and to
\emph{stay} there. The modulus of $f_{N,1}(t)$ and $f_{N,N}(t)$
can be measured by initialising the system in the states $\left|\boldsymbol{1,0}\right\rangle $
and $\left|\boldsymbol{N,0}\right\rangle $ and then performing a
reduced density matrix tomography at Bob's site at different times
$t$, and the complex phase of these functions is obtained by initialising
the system in $\left(\left|\boldsymbol{0,0}\right\rangle +\left|\boldsymbol{1,0}\right\rangle \right)/\sqrt{2}$
and $\left(\left|\boldsymbol{0,0}\right\rangle +\left|\boldsymbol{N,0}\right\rangle \right)/\sqrt{2}$
instead. In the same way, $g_{N,1}(t)$ and $g_{N,N}(t)$ are obtained.
All this can be done in the spirit of \emph{minimal control} at the
sending and receiving ends of the chain only, and needs to be done
only once. It is interesting to note that the dynamics in the middle
part of the chain is not relevant at all. It is a \emph{\index{black box}black
box} (see Fig. \ref{cap:blackbox}) that may involve even completely
different interactions, number of spins, etc., as long as the total
number of excitations is conserved. %
Once the transition amplitudes {[}Equations (\ref{eq:t1})-(\ref{eq:t4})]
are known, one can search numerically for optimised measurement times
$t_{\ell}$ using Eq. (\ref{eq:finalprob}) and the condition from
Eq. (\ref{eq:finalcondition}).

One weakness of the scheme described here is that the times at which
Bob measures have to be very precise, because otherwise the measurements
will not be unbiased with respect to $\alpha$ and $\beta.$ This
demand can be relaxed by measuring at times where not only the probability
amplitudes are similar, but also their \emph{slope} (see Fig. \ref{fig:coincidence}).
The computation of these optimal timings for a given system may be
complicated, but they only need to be done once.

\section{Numerical Examples\label{sec:Numerical-Examples}}

In this section, we show some numerical examples for two chains with
Heisenberg couplings $J$ which are fluctuating. The Hamiltonians
of the chains $i=1,2$ are given by\begin{eqnarray}
H^{(i)} & = & \sum_{n=1}^{N-1}J(1+\delta_{n}^{(i)})\left(X_{n}^{(i)}X_{n+1}^{(i)}+Y_{n}^{(i)}Y_{n+1}^{(i)}+Z_{n}^{(i)}Z_{n+1}^{(i)}\right),\end{eqnarray}
 where $\delta_{n}^{(i)}$ are uniformly distributed random numbers
from the interval $\left[-\Delta,\Delta\right].$ We have considered
two different cases: in the first case, the $\delta_{n}^{(i)}$ are
completely uncorrelated (i.e. independent for both chains and all
sites along the chain). In the second case, we have taken into account
a spacial correlation of the signs of the $\delta_{n}^{(i)}$ along
each of the chains, while still keeping the two chains uncorrelated.
For both cases, we find that arbitrarily perfect transfer remains
possible except for some very rare realisations of the $\delta_{n}^{(i)}.$

Because measurements must only be taken at times which fulfil the
condition (\ref{eq:finalcondition}), and these times usually do not
coincide with the optimal probability of finding an excitation at
the ends of the chains, it is clear that the probability of failure
at each measurement will in average be higher than for chains without
fluctuations. Therefore, more measurements have to be performed in
order to achieve the same probability of success. The price for noisy
couplings is thus a longer transmission time and a higher number of
gating operations at the receiving end of the chains. Some averaged
values are given in Table \ref{cap:The-total-time}%
\begin{table}[tbh]
\begin{centering}\begin{tabular}{|c|c|c|c|c|c|}
\hline 
&
$\Delta=0$&
$\Delta=0.01$&
$\Delta=0.03$&
$\Delta=0.05$&
$\Delta=0.1$\tabularnewline
\hline
\hline 
$t$$\left[\frac{1}{J}\right]$&
$377$&
$524\pm27$&
$694\pm32$&
$775\pm40$&
$1106\pm248$\tabularnewline
\hline 
$M$&
$28$&
$43\pm3$&
$58\pm3$&
$65\pm4$&
$110\pm25$\tabularnewline
\hline
\end{tabular}\par\end{centering}

\caption{\label{cap:The-total-time}The total time $t$ and the number of
measurements $M$ needed to achieve a probability of success of $99$\%
for different fluctuation strengths $\Delta$ (uncorrelated case).
Given is the statistical mean and the standard deviation. The length
of the chain is $N=20$ and the number of random samples is $10.$
For strong fluctuations $\Delta=0.1$, we also found particular samples
where the success probability could not be achieved within the time
range searched by the algorithm.}
\end{table}
 for the Heisenberg chain with uncorrelated coupling fluctuations.

For the case where the signs of the $\delta_{n}^{(i)}$ are correlated,
we have used the same model as in~\cite{Fazio2005}, introducing
the parameter $c$ such that\begin{equation}
\delta_{n}^{(i)}\delta_{n-1}^{(i)}>0\qquad\textrm{with propability }c,\label{eq:corr1}\end{equation}
 and \begin{equation}
\delta_{n}^{(i)}\delta_{n-1}^{(i)}<0\qquad\textrm{with propability }1-c.\label{eq:corr2}\end{equation}
 For $c=1$ ($c=0)$ this corresponds to the case where the signs
of the couplings are completely correlated (anti-correlated). For
$c=0.5$ one recovers the case of uncorrelated couplings. We can see
from the numerical results in Table \ref{cap:correlated} that arbitrarily
perfect transfer is possible for the whole range of $c.$ %
\begin{table}[tbh]
\begin{centering}\begin{tabular}{|c|c|c|c|c|c|c|}
\hline 
&
$c=0$&
$c=0.1$&
$c=0.3$&
$c=0.7$&
$c=0.9$&
$c=1$\tabularnewline
\hline
\hline 
$t$$\left[\frac{1}{J}\right]$&
$666\pm20$&
$725\pm32$&
$755\pm41$&
$797\pm35$&
$882\pm83$&
$714\pm41$\tabularnewline
\hline 
$M$&
$256\pm2$&
$62\pm3$&
$65\pm4$&
$67\pm4$&
$77\pm7$&
$60\pm4$\tabularnewline
\hline
\end{tabular}\par\end{centering}

\caption{\label{cap:correlated}The total time $t$ and the number of measurements
$M$ needed to achieve a probability of success of $99$\% for different
correlations $c$ between the couplings {[}see Eq. (\ref{eq:corr1})
and Eq. (\ref{eq:corr2})]. Given is the statistical mean and the
standard deviation for a fluctuation strength of $\Delta=0.05$. The
length of the chain is $N=20$ and the number of random samples is
$20.$ }
\end{table}

\begin{figure}[tbh]
\begin{centering}\includegraphics[width=1\columnwidth]{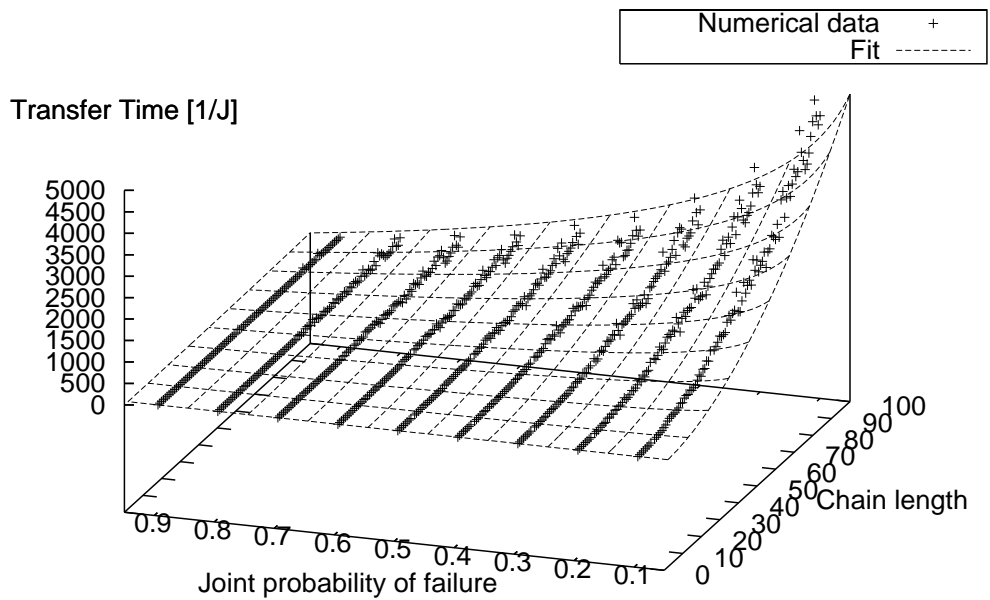}\par\end{centering}

\caption{\label{cap:fitfig}Time $t$ needed to transfer a state with a given
joint probability of failure $P$ across a chain of length $N$ with
uncorrelated fluctuations of $\Delta=0.05.$ The points denote numerical
data averaged over $100$ realisations, and the fit is given by Eq.
(\ref{eq:fit2}). This figure should be compared with Fig. \ref{cap:Timefit}
where $\Delta=0.$ }
\end{figure}

For $\Delta=0$, we know from Section~\ref{sec:Estimati} that the
time to transfer a state with probability of failure $P$ scales as\begin{equation}
t(P)=0.33J^{-1}N^{1.6}\left|\ln P\right|.\label{eq:fit}\end{equation}
 If we want to obtain a similar formula in the presence of noise,
we can perform a fit to the exact numerical data. For uncorrelated
fluctuations of $\Delta=0.05,$ this is shown in Fig. \ref{cap:fitfig}.
The best fit is given by\begin{equation}
t(P)=0.2J^{-1}N^{1.9}\left|\ln P\right|.\label{eq:fit2}\end{equation}
 We conclude that weak fluctuations (say up to $5$\%) in the coupling
strengths do not deteriorate the performance of our scheme much for
the chain lengths considered. Both the transmission time and the number
of measurements raise, but still in a reasonable way {[}cf. Table~\ref{cap:The-total-time}
and Fig.~\ref{cap:fitfig}]. For larger fluctuations, the scheme
is still applicable in principle, but the amount of junk (i.e. chains
not capable of arbitrary perfect transfer) may get too large.%

Note that we have considered the case where the fluctuations $\delta_{n}^{i}$
are constant in time. This is a reasonable assumption if the dynamic
fluctuations (e.g. those arising from thermal noise) can be neglected
with respect to the constant fluctuations (e.g. those arising from
manufacturing errors). If the fluctuations were varying with time,
the tomography measurements in Sec. \ref{sec:Tomography} would involve
a time-average, and Bob would not measure exactly at the correct times.
The transferred state (\ref{eq:success}) would then be affected by
both phase and amplitude noise.

\section{Coupled chains\index{coupled chains}\label{sec:Coupled-chains}}

\begin{figure}[tbh]
\begin{centering}\includegraphics[width=8cm]{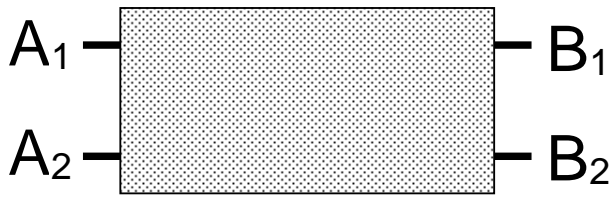}\par\end{centering}

\caption{\label{fig:A-black-box}Most general setting for conclusive transfer:
A \emph{black box} with two inputs and two outputs, acting as an amplitude
damping channel defined by Eqs.~(\ref{eq:ampin}) and (\ref{eq:ampout}) }
\end{figure}

Let us look at the condition for conclusive transfer in the more general
scenario indicated by Fig.~ \ref{fig:A-black-box}: Alice and Bob
have a black box\index{black box} acting as an amplitude damping
channel in the following way. It has two inputs and two outputs. If
Alice puts in state in the dual rail, \begin{equation}
|\psi\rangle=\alpha|01\rangle+\beta|10\rangle,\label{eq:ampin}\end{equation}
where $\alpha$ and $\beta$ are \emph{arbitrary and unknown} normalised
amplitudes, then the output at Bob is given by\begin{equation}
p|\phi\rangle\langle\phi|+\left(1-p\right)|00\rangle\langle00|,\label{eq:ampout}\end{equation}
with a normalised ''success'' state\begin{equation}
|\phi\rangle=\frac{1}{\sqrt{p}}\left[\alpha f|01\rangle+\beta g|10\rangle+\alpha\tilde{f}|10\rangle+\beta\tilde{g}|01\rangle\right].\end{equation}
This black box describes the behaviour of an arbitrarily coupled qubit
system that conserves the number of excitations and that is initialised
in the all zero state, including parallel uncoupled chains, and coupled
chains.%

From the normalisation of $|\phi\rangle$ it follows that\begin{equation}
p=p(\alpha,\beta)=\left|\alpha f+\beta\tilde{g}\right|^{2}+\left|\beta g+\alpha\tilde{f}\right|^{2}.\end{equation}
We are interested in conclusive transfer: by measuring the observable
$|00\rangle\langle00|$ the Bob can project the output onto either
the failure state $|00\rangle$ or $|\phi\rangle.$ This is clearly
possible, but the question is if the output $|\phi\rangle$ and the
input $|\psi\rangle$ are related by a \emph{unitary} operation. 

\begin{figure}[tbh]
\begin{centering}\includegraphics[width=0.6\columnwidth]{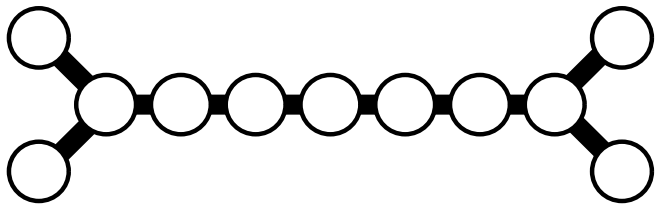}\par\end{centering}

\caption{\label{cap:counter}A simple counterexample for a vertically symmetric
system where dual rail encoding is not possible. The black lines represent
exchange couplings.}
\end{figure}

\begin{figure}[tbh]
\begin{centering}\includegraphics[width=0.6\columnwidth]{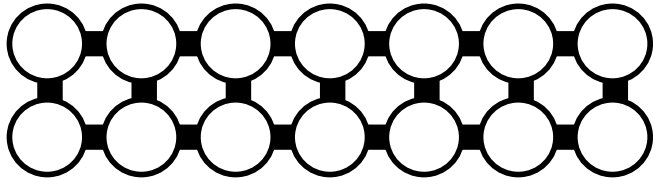}\par\end{centering}

\caption{\label{cap:coupled}An example for a vertically symmetric system
where dual rail encoding is possible. The black lines represent exchange
couplings \emph{of equal strength}.}
\end{figure}

If Bob is able to recover the full information that Alice sent, then
$p(\alpha,\beta)$ must be \emph{independent of $\alpha$ and} $\beta$
(otherwise, some information on these amplitudes could be obtained
by the measurement already, which contradicts the non-cloning theorem~\cite{NIELSEN}).
This implies that $p(1,0)=p(0,1),$ i.e.\begin{equation}
\left|f\right|^{2}+\left|\tilde{f}\right|^{2}=\left|\tilde{g}\right|^{2}+\left|g\right|^{2}.\label{eq:numma1}\end{equation}
Because \begin{eqnarray}
p\left(\frac{1}{\sqrt{2}},\frac{1}{\sqrt{2}}\right) & = & \frac{1}{2}\left|f+\tilde{g}\right|^{2}+\frac{1}{2}\left|g+\tilde{f}\right|^{2}\\
 & = & p(1,0)+\mbox{Re}\left\{ f^{*}\tilde{g}+g\tilde{f}^{*}\right\} \end{eqnarray}
it also implies that \begin{equation}
\mbox{Re}\left\{ f^{*}\tilde{g}+g\tilde{f}^{*}\right\} =0.\end{equation}
Using the same trick for $p\left(\frac{1}{\sqrt{2}},\frac{i}{\sqrt{2}}\right)$
we get that $\mbox{Im}\left\{ f^{*}\tilde{g}+g\tilde{f}^{*}\right\} =0$
and therefore\begin{equation}
f^{*}\tilde{g}+g\tilde{f}^{*}=0.\label{eq:numma2}\end{equation}
If we write $|\psi\rangle=U|\phi\rangle$ we get\begin{equation}
U=\frac{1}{\sqrt{p}}\left(\begin{array}{cc}
f & \tilde{f}\\
\tilde{g} & g\end{array}\right),\end{equation}
which is a unitary operator if Eq.~(\ref{eq:numma1}) and (\ref{eq:numma2})
hold. We thus come to the conclusion that conclusive transfer with
the black box defined above is possible if and only if the probability
$p$ is independent of $\alpha$ and $\beta.$ It is interesting to
note that a vertical mirror symmetry of the system does not guarantee
this. A counterexample is sketched in Fig.~\ref{cap:counter}: clearly
the initial ({}``dark'') state $|01\rangle-|10\rangle$ does not
evolve, whereas $|01\rangle+|10\rangle$ \emph{does.} Hence the probability
must depend on $\alpha$ and $\beta.$ A trivial case where conclusive
transfer works is given by two uncoupled chains, at times where $|f|^{2}=|g|^{2}.$
This was discussed in Sect.~ \ref{sec:Arbitrarily-perfect-transferl}.
A non-trivial example is given by the coupled system sketched in Fig.~\ref{cap:coupled}.
This can be seen by splitting the Hamiltonian in a horizontal and
vertical component,\begin{equation}
H=H_{v}+H_{z}.\end{equation}
 By applying $H_{v}H_{z}$ and $H_{z}H_{v}$ on single-excitation
states it is easily checked that they commute in the first excitation
sector (this is not longer true in higher sectors). Since the probability
is independent of $\alpha$ and $\beta$ in the uncoupled case it
must also be true in the coupled case (a rotation in the subspace
$\left\{ |01\rangle,|10\rangle\right\} $ does not harm).%

A final remark - as Alice and Bob alway only deal with the states
$\left\{ |00\rangle,|10\rangle,|01\rangle\right\} $ it is obvious
that the encoding used in this chapter is really living on \emph{qutrits.}
In some sense it would be more natural to consider permanently coupled
systems of qutrits\index{qutrits}, such as SU(3) chains~\cite{Sutherland1975,DUALRAIL,Bose2005a,Bayat2006}.
The first level of the qutrit $|0\rangle$ is then used as a marker
for ''no information here'', whereas the information is encoded
in the states $|1\rangle$ and $|2\rangle.$ One would have to ensure
that there is no transition between $|0\rangle$ and $|1\rangle,|2\rangle,$
and that the system is initialised in the all zero state.

\section{Conclusion\label{sec:Conclusions}}

In conclusion, we have presented a simple scheme for conclusive and
arbitrarily perfect quantum state transfer. To achieve this, two parallel
spin chains (individually amplitude damping channels) have been used
as one \emph{amplitude delaying channel}. We have shown that our scheme
is more robust to decoherence and imperfect timing than the single
chain schemes.  We have also shown that the scheme is applicable to
disordered and coupled chains. The scheme can be used as a way of
improving any of the other schemes from the introduction. For instance,
one may try to engineer the couplings to have a very high probability
of success already at the first measurement, and use further measurements
to compensate the errors of implementing the correct values for the
couplings. We remark that the dual rail protocol is unrelated to error
filtration~\cite{Gisin2005} where parallel channels are used for
filtering out environmental effects on flying qubits, whereas the
purpose of the dual rail protocol is to ensure the \emph{arrival}
of the qubit. Indeed one could combine both protocols to send a qubit
on say four rails to ensure the arrival \emph{and} filter errors.
Finally, we note that in some recent work~\cite{Ericsson2005} it
was shown that our encoding can be used to perform quantum gates while
the state is transferred, and that it can increase the convergence
speed if one performs measurements at intermediate positions~\cite{Bose2005e,Vaucher2005}.

\chapter{Multi Rail encoding\label{cha:Multi-rail-and-Capacity}}

\section{Introduction}

In quantum information theory the rate $R$ of transferred qubits
per channel is an important efficiency parameter~\cite{SHOR}. Therefore
one question that naturally arises is whether or not there is any
special meaning in the 1/2 value of $R$ achieved in the dual rail
protocol of the last chapter. We will show now that this is not the
case, because there is a way of bringing $R$ arbitrarily close to
$1$ by considering multi rail encodings. Furthermore, in Section~\ref{sec:Arbit}
it was still left open for which Hamiltonians the probability of success
can be made arbitrarily close to $1.$ Here, we give a sufficient
and easily attainable condition for achieving this goal.

This chapter is organised as follows: the model and the notation are
introduced in Sec.~\ref{s:sec1}. The efficiency and the fidelity
of the protocol are discussed in Sec.~\ref{s:sec2} and in Sec.~\ref{s:sec3},
respectively. Finally in Sec.~\ref{s:sec4} we prove a theorem which
provides us with a sufficient condition for achieving efficient and
perfect state transfer in quantum chains.

\begin{figure}[tbh]
\begin{centering}\includegraphics[width=0.9\columnwidth]{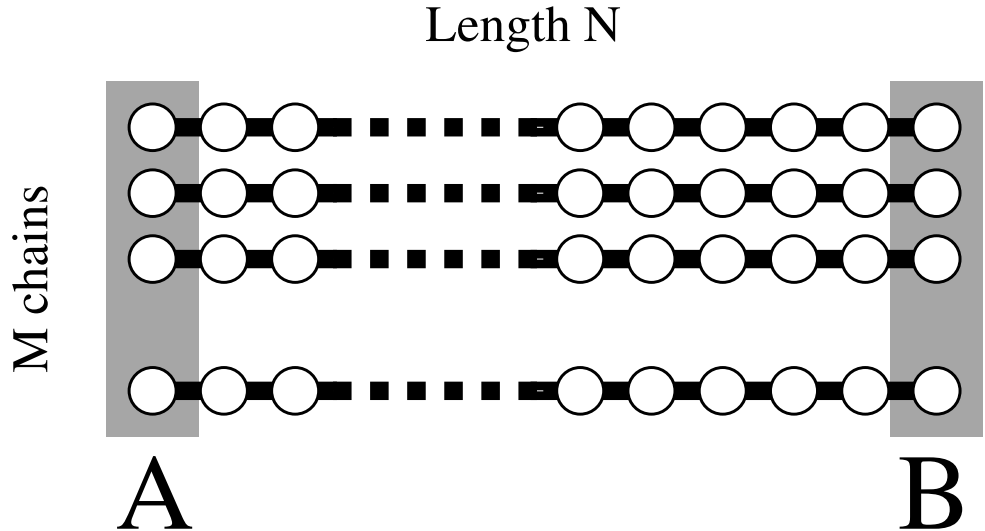}\par\end{centering}

\caption{\label{f:fig1} Schematic of the system: Alice and Bob operate $M$
chains, each containing $N$ spins. The spins belonging to the same
chain interact through the Hamiltonian $H$ which accounts for the
transmission of the signal in the system. Spins of different chains
do not interact. Alice encodes the information in the first spins
of the chains by applying unitary transformations to her qubits. Bob
recovers the message in the last spins of the chains by performing
joint measurements.}
\end{figure}

\section{The model\label{s:sec1}}

Assume that the two communicating parties operate on $M$ independent
(i.e. non interacting) copies of the chain. This is quite a common
attitude in quantum information theory~\cite{SHOR} where successive
uses of a memoryless channel are formally described by introducing
many parallel copies of the channel (see~\cite{Giovannetti2005}
for a discussion on the possibility of applying this formal description
to quantum chain models). Moreover for the case at hand the assumption
of Alice and Bob dealing with {}``real'' parallel chains seems reasonable
also from a practical point of view~\cite{Motoyama1996,Gambardella2002}.
The idea is to use these copies to improve the overall fidelity of
the communication. As usual, we assume Alice and Bob to control respectively
the first and last qubit of each chain (see Fig.~\ref{f:fig1}).
By preparing any superposition of her spins Alice can in principle
transfer up to $M$ logical qubits. However, in order to improve the
communication fidelity the two parties will find it more convenient
to redundantly encode only a small number (say $Q(M)\leqslant M$)
of logical qubits in the $M$ spins. By adopting these strategies
Alice and Bob are effectively sacrificing the efficiency $R(M)=Q(M)/M$
of their communication line in order to increase its fidelity. This
is typical of any communication scheme and it is analogous to what
happens in quantum error correction theory, where a single logical
qubit is stored in many physical qubits. In the last chapter we have
seen that for $M=2$ it is possible to achieve perfect state transfer
of a single logical qubit with an efficiency equal to $1/2$. Here
we will generalise such result by proving that there exist an optimal
encoding-decoding strategy which asymptotically allows to achieve
perfect state transfer \emph{and} optimal efficiency\index{efficiency},
i.e. \begin{eqnarray}
\lim_{M\rightarrow\infty}R(M)=1\;.\label{efficiency}\end{eqnarray}
Our strategy requires Alice to prepare superpositions of the $M$
chains where $\sim M/2$ of them have a single excitation in the first
location while the remaining are in $|{\boldsymbol{0}}\rangle$. Since
in the limit $M>>1$ the number of qubit transmitted is $\log\binom{M}{M/2}\approx M$,
this architecture guarantees optimal efficiency~(\ref{efficiency}).
On the other hand, our protocol requires Bob to perform collective
measurements on his spins to determine if all the $\sim M/2$ excitations
Alice is transmitting arrived at his location. We will prove that
by repeating these detections many times, Bob is able to recover the
messages with asymptotically perfect fidelity.

\begin{figure}[tbh]
\begin{centering}\includegraphics[width=0.6\columnwidth]{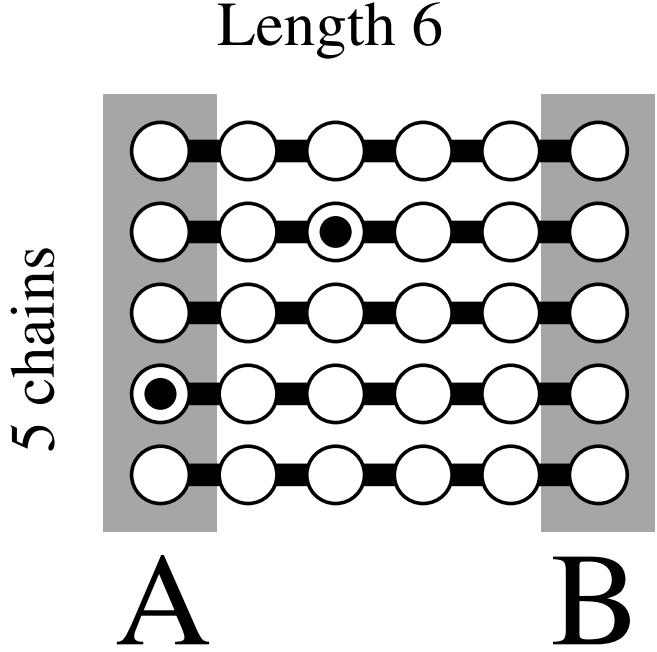}\par\end{centering}

\caption{\label{cap:Example-of-our}Example of our notation for $M=5$ chains
of length $N=6$ with $K=2$ excitations. The state above, given by
$|\boldsymbol{0}\rangle_{1}\otimes|\boldsymbol{3}\rangle_{2}\otimes|\boldsymbol{0}\rangle_{3}\otimes|\boldsymbol{1}\rangle_{4}\otimes|\boldsymbol{0}\rangle_{5},$
has excitations in the chains $m_{1}=2$ and $m_{2}=4$ at the horizontal
position $n_{1}=3$ and $n_{2}=1$. It is in the Hilbert space $\mathcal{H}(S_{6})$
corresponding to the subset $S_{6}=\{2,4\}$ (assuming that the sets
$S_{\ell}$ are ordered in a canonical way, i.e. $S_{1}=\{1,2\},$
$S_{2}=\{1,3\}$ and so on) and will be written as $|(3,1);6\rangle\!\rangle.$
There are $\binom{5}{2}=10$ different sets $S_{\ell}$ and the number
of qubits one can transfer using these states is $\log_{2}10\approx3.$
The efficiency is thus given by $R\approx3/5$ which is already bigger
than in the dual rail scheme.}
\end{figure}

Before beginning the analysis let us introduce some notation. The
following definitions \emph{look} more complicated than they really
\emph{are}; unfortunately we need them to carefully define the states
that Alice uses for encoding the information. In order to distinguish
the $M$ different chains we introduce the label $m=1,\cdots,M$:
in this formalism $|\boldsymbol{n}\rangle_{m}$ represents the state
of $m$-th chain with a single excitation in the $n$-th spin. In
the following we will be interested in those configurations of the
whole system where $K$ chains have a single excitation while the
remaining $M-K$ are in $|\boldsymbol{0}\rangle$, as in the case\begin{equation}
|\boldsymbol{1}\rangle_{1}\otimes|\boldsymbol{1}\rangle_{2}\cdots\otimes|\boldsymbol{1}\rangle_{K}\otimes|\boldsymbol{0}\rangle_{K+1}\cdots\otimes|\boldsymbol{0}\rangle_{M}\label{eq:example}\end{equation}
 where for instance the first $K$ chains have an excitation in the
first chain location. Another more general example is given in Fig.
\ref{cap:Example-of-our}. The complete characterisation of these
vectors is obtained by specifying \emph{i)} \emph{which} chains possess
a single excitation and \emph{ii)} \emph{where} these excitations
are located horizontally along the chains. In answering to the point
\emph{i)} we introduce the $K$-element subsets $S_{\ell}$, composed
by the labels of those chains that contain an excitation. Each of
these subsets $S_{\ell}$ corresponds to a subspace of the Hilbert
space $\mathcal{H}(S_{\ell})$ with a dimension $N^{K}.$ The total
number of such subsets is equal to the binomial coefficient $\binom{M}{K}$,
which counts the number of possibilities in which $K$ objects (excitations)
can be distributed among $M$ parties (parallel chains). In particular
for any $\ell=1,\cdots,\binom{M}{K}$ the $\ell$-th subset $S_{\ell}$
will be specified by assigning its $K$ elements, i.e. $S_{\ell}\equiv\{ m_{1}^{(\ell)},\cdots,m_{K}^{(\ell)}\}$
with $m_{j}^{(\ell)}\in\{1,\cdots,M\}$ and $m_{j}^{(\ell)}<m_{j+1}^{(\ell)}$
for all $j=1,\cdots,K$. To characterise the location of the excitations,
point \emph{ii)}, we will introduce instead the $K$-dimensional vectors
$\vec{n}\equiv(n_{1},\cdots,n_{K})$ where $n_{j}\in\{1,\cdots,N\}$.
We can then define \begin{eqnarray}
|\boldsymbol{\vec{n}};\ell\rangle\!\rangle\equiv\bigotimes_{j=1}^{K}|\boldsymbol{n_{j}}\rangle_{m_{j}^{(\ell)}}\;\bigotimes_{m^{\prime}\in{\overline{S}_{\ell}}}|\boldsymbol{0}\rangle_{m^{\prime}}\;,\label{nvec}\end{eqnarray}
 where $\overline{S}_{\ell}$ is the complementary of $S_{\ell}$
to the whole set of chains.%

The state (\ref{nvec}) represents a configuration where the $j$-th
chain of the subset $S_{\ell}$ is in $|\boldsymbol{n_{j}}\rangle$
while the chains that do not belong to $S_{\ell}$ are in $|\boldsymbol{0}\rangle$
(see Fig. \ref{cap:Example-of-our} for an explicit example). The
kets $|\boldsymbol{\vec{n}};\ell\rangle\!\rangle$ are a natural generalisation
of the states $|\boldsymbol{n}\rangle_{1}\otimes|\boldsymbol{0}\rangle_{2}$
and $|\boldsymbol{0}\rangle_{1}\otimes|\boldsymbol{n}\rangle_{2}$
which were used for the dual rail encoding. They are useful for our
purposes because they are mutually orthogonal, i.e. \begin{eqnarray}
\!\langle\!\langle\boldsymbol{\vec{n}};\ell|\boldsymbol{\vec{n}^{\prime}};\ell^{\prime}\rangle\!\rangle=\delta_{\ell\ell^{\prime}}\;\delta_{\vec{n}\vec{n}^{\prime}}\;,\label{ortho}\end{eqnarray}
 and their time evolution under the Hamiltonian does not depend on
$\ell.$ Among the vectors~(\ref{nvec}) those where all the $K$
excitations are located at the beginning of the $S_{\ell}$ chains
play an important role in our analysis. Here $\vec{n}=\vec{1}\equiv(1,\cdots,1)$
and we can write \begin{eqnarray}
|\boldsymbol{\vec{1}};\ell\rangle\!\rangle\equiv\bigotimes_{m\in S_{\ell}}|\boldsymbol{1}\rangle_{m}\;\bigotimes_{m^{\prime}\in{\overline{S}_{\ell}}}|\boldsymbol{0}\rangle_{m^{\prime}}\;.\label{in}\end{eqnarray}
 According to Eq.~(\ref{ortho}), for $\ell=1,\cdots,\binom{M}{K}$
these states form orthonormal set of $\binom{M}{K}$ elements. Analogously
by choosing $\vec{n}=\vec{N}\equiv(N,\cdots,N)$ we obtain the orthonormal
set of $\binom{M}{K}$ vectors \begin{equation}
|\boldsymbol{\vec{N}};\ell\rangle\!\rangle\equiv\bigotimes_{m\in S_{\ell}}|\boldsymbol{N}\rangle_{m}\;\bigotimes_{m^{\prime}\in{\overline{S}_{\ell}}}|\boldsymbol{0}\rangle_{m^{\prime}},\end{equation}
 where all the $K$ excitations are located at the end of the chains.

\section{Efficient encoding}

\label{s:sec2}

If all the $M$ chains of the system are originally in $|\boldsymbol{0}\rangle$,
the vectors~(\ref{in}) can be prepared by Alice by locally operating
on her spins. Moreover since these vectors span a $\binom{M}{K}$
dimensional subspace, Alice can encode in the chain $Q(M,K)=\log_{2}\binom{M}{K}$
qubits of logical information by preparing the superpositions, \begin{eqnarray}
|\Phi\rangle\!\rangle=\sum_{\ell}A_{\ell}\;|\boldsymbol{\vec{1}};\ell\rangle\!\rangle\;,\label{logicstate}\end{eqnarray}
 with $A_{\ell}$ complex coefficients. The efficiency of such encoding
is hence $R(M,K)=\frac{\log_{2}\binom{M}{K}}{M}$ which maximised
with respect to $K$ gives, \begin{eqnarray}
R(M) & = & \frac{1}{M}\left\{ \begin{array}{ll}
{\log_{2}\binom{M}{M/2}} & \;\mbox{for $M$ even}\\
{\log_{2}\binom{M}{(M-1)/2}} & \;\mbox{for $M$ odd}\;.\end{array}\right.\label{efficiencymax}\end{eqnarray}
 The Stirling approximation can then be used to prove that this encoding
is asymptotically efficient~(\ref{efficiency}) in the limit of large
$M$, e.g. \begin{eqnarray}
\log_{2}\binom{M}{M/2} & \approx & \log_{2}\frac{M^{M}}{(M/2)^{M}}=M.\end{eqnarray}
 Note that already for $M=5$ the encoding is more efficient (cf.
Fig. \ref{cap:Example-of-our}) than in the dual rail encoding. In
the remaining of the chapter we show that the encoding~(\ref{logicstate})
provides perfect state transfer by allowing Bob to perform joint measurements
at his end of the chains.

\section{Perfect transfer\label{s:sec3}}

Since the $M$ chains do not interact with each other and possess
the same free Hamiltonian $H,$ the unitary evolution of the whole
system is described by $U(t)\equiv\otimes_{m}u_{m}(t)$, with $u_{m}(t)$
being the operator acting on the $m$-th chain. The time evolved of
the input $|\boldsymbol{\vec{1}};\ell\rangle\!\rangle$ of Eq.~(\ref{in})
is thus equal to \begin{eqnarray}
 &  & U(t)|\boldsymbol{\vec{1}};\ell\rangle\!\rangle=\sum_{\vec{n}}F[\vec{n},\vec{1};t]\;|\boldsymbol{\vec{n}};\ell\rangle\!\rangle\;,\label{time}\end{eqnarray}
 where the sum is performed for all $n_{j}=1,\cdots,N$ and \begin{eqnarray}
F[{\vec{n},\vec{n^{\prime}}};t]\equiv f_{n_{1},n_{1}^{\prime}}(t)\cdots f_{n_{K},n_{K}^{\prime}}(t)\;,\label{capitolF}\end{eqnarray}
 is a quantity which does \emph{not} depend on $\ell$. In Eq.~(\ref{time})
the term ${\vec{n}}=\vec{N}$ corresponds to having all the $K$ excitations
in the last locations of the chains. We can thus write \begin{eqnarray}
U(t)|\boldsymbol{\vec{1}};\ell\rangle\!\rangle=\gamma_{1}(t)|\boldsymbol{\vec{N}};\ell\rangle\!\rangle+\sqrt{1-|\gamma_{1}(t)|^{2}}\;|\boldsymbol{\xi}(t);\ell\rangle\!\rangle\;,\label{time1}\end{eqnarray}
 where \begin{eqnarray}
\gamma_{1}(t) & \equiv & \langle\!\langle\boldsymbol{\vec{N}};\ell|U(t)|\boldsymbol{\vec{1}};\ell\rangle\!\rangle=F[\vec{N},\vec{1};t]\label{gamma1}\end{eqnarray}
 is the probability amplitude that all the $K$ excitation of $|\boldsymbol{\vec{1}};\ell\rangle\!\rangle$
arrive at the end of the chains, and \begin{eqnarray}
|\boldsymbol{\xi}(t);\ell\rangle\!\rangle\equiv\sum_{\vec{n}\neq\vec{N}}F_{1}[\vec{n},\vec{1};t]\;|\boldsymbol{\vec{n}};\ell\rangle\!\rangle\;,\label{error}\end{eqnarray}
 with \begin{equation}
F_{1}[\vec{n},\vec{1};t]\equiv\frac{F[\vec{n},\vec{1};t]}{\sqrt{1-|\gamma_{1}(t)|^{2}}},\label{eq:13b}\end{equation}
 is a superposition of terms where the number of excitations arrived
to the end of the communication line is strictly less then $K$. It
is worth noticing that Eq.~(\ref{ortho}) yields the following relations,
\begin{eqnarray}
\langle\!\langle\boldsymbol{\vec{N}};\ell|\boldsymbol{\xi}(t);\ell^{\prime}\rangle\!\rangle=0,\quad\!\langle\!\langle\boldsymbol{\xi}(t);\ell|\boldsymbol{\xi}(t);{\ell^{\prime}}\rangle\!\rangle=\delta_{\ell\ell^{\prime}}\;,\label{ortho1}\end{eqnarray}
 which shows that $\left\{ ||\boldsymbol{\xi}(t);\ell\rangle\!\rangle\right\} $
is an orthonormal set of vectors which spans a subspace orthogonal
to the states $|\boldsymbol{\vec{N}};\ell\rangle\!\rangle.$ The time
evolution of the input state~(\ref{logicstate}) follows by linearity
from Eq.~(\ref{time1}), i.e. \begin{eqnarray}
|\Phi(t)\rangle\!\rangle=\gamma_{1}(t)\;|\Psi\rangle\!\rangle+\sqrt{1-|\gamma_{1}(t)|^{2}}\;|\overline{\Psi}(t)\rangle\!\rangle\;,\label{logicstateout}\end{eqnarray}
 with \begin{eqnarray}
|\overline{\Psi}(t)\rangle\!\rangle & \equiv & \sum_{\ell}A_{\ell}\;|\boldsymbol{\xi}(t);\ell\rangle\!\rangle\;,\nonumber \\
|\Psi\rangle\!\rangle & \equiv & \sum_{\ell}A_{\ell}\;|\boldsymbol{\vec{N}};\ell\rangle\!\rangle\;.\label{ok}\end{eqnarray}
 The vectors $|\Psi\rangle\!\rangle$ and $|\overline{\Psi}(t)\rangle\!\rangle$
are unitary transformations of the input message~(\ref{logicstate})
where the orthonormal set $\{|\boldsymbol{\vec{1}};\ell\rangle\!\rangle\}$
has been rotated into $\{|\boldsymbol{\vec{N}};\ell\rangle\!\rangle\}$
and $\{|\boldsymbol{\xi}(t);\ell\rangle\!\rangle\}$ respectively.
Moreover $|\Psi\rangle\!\rangle$ is the configuration we need to
have for perfect state transfer at the end of the chain. In fact it
is obtained from the input message~(\ref{logicstate}) by replacing
the components $|\boldsymbol{1}\rangle$ (excitation in the first
spin) with $|\boldsymbol{N}\rangle$ (excitation in the last spin).
From Eq.~(\ref{ortho1}) we know that $|\Psi\rangle\!\rangle$ and
$|\overline{\Psi}(t)\rangle\!\rangle$ are orthogonal. This property
helps Bob to recover the message $|\Psi\rangle\!\rangle$ from $|\Phi(t)\rangle\!\rangle$:
he only needs to perform a collective measurement on the $M$ spins
he is controlling to establish if there are $K$ or less excitations
in those locations. The above is clearly a projective measurement
that can be performed without destroying the quantum coherence associated
with the coefficients $A_{\ell}$. Formally this can described by
introducing the observable \begin{eqnarray}
\Theta\equiv1-\sum_{\ell}|\boldsymbol{\vec{N}};\ell\rangle\!\rangle\langle\!\langle\boldsymbol{\vec{N}};\ell|\;.\label{observable}\end{eqnarray}
 A single measurement of $\Theta$ on $|\Phi(t_{1})\rangle\!\rangle$
yields the outcome $0$ with probability $p_{1}\equiv|\gamma_{1}(t_{1})|^{2}$,
and the outcome $+1$ with probability $1-p_{1}$. In the first case
the system will be projected in $|\Psi\rangle\!\rangle$ and Bob will
get the message. In the second case instead the state of the system
will become $|\overline{\Psi}(t_{1})\rangle\!\rangle$. Already at
this stage the two communicating parties have a success probability
equal to $p_{1}$. Moreover, as in the dual rail protocol, the channels
have been transformed into a quantum erasure channel~\cite{Bennett1997}
where the receiver knows if the transfer was successful. Just like
the dual rail encoding, this encoding can be used as a simple entanglement
purification method in quantum chain transfer (see end of Section~\ref{sec:con}).
The rate of entanglement that can be distilled is given by\begin{equation}
R(M)\left|F[\vec{N},\vec{1};t]\right|^{2}=R(M)p(t)^{\left\lfloor M/2\right\rfloor },\end{equation}
where we used Eq.~(\ref{capitolF}) and $p(t)\equiv\left|f_{N,1}(t)\right|^{2}.$
As we can see, increasing $M$ on one hand increases $R(M),$ but
on the other hand decreases the factor $p(t)^{\left\lfloor M/2\right\rfloor }.$
Its maximum with respect to $M$ gives us a lower bound of the entanglement
of distillation\index{entanglement of distillation} for a single
spin chain, as shown in Fig.~\ref{fig:capacity}. We can also see
that it becomes worth encoding on more than \emph{three} chains for
conclusive transfer only when $p(t)>0.8.$

Consider now what happens when Bob fails to get the right answer from
the measurement. The state on which the chains is projected is explicitly
given by \begin{eqnarray}
|\overline{\Psi}(t_{1})\rangle\!\rangle & = & \sum_{\vec{n}\neq\vec{N}}F_{1}[\vec{n},\vec{1};t_{1}]\sum_{\ell}A_{\ell}|\boldsymbol{\vec{n}};\ell\rangle\!\rangle\;.\label{explicit}\end{eqnarray}
 Let us now consider the evolution of this state for another time
interval $t_{2}$. By repeating the same analysis given above we obtain
an expression similar to (\ref{logicstateout}), i.e. \begin{eqnarray}
|\Phi(t_{2},t_{1})\rangle\!\rangle & = & \gamma_{2}\;|\Psi\rangle\!\rangle+\sqrt{1-|\gamma_{2}|^{2}}\;|\overline{\Psi}(t_{2},t_{1})\rangle\!\rangle\;,\label{logicstateout2}\end{eqnarray}
 where now the probability amplitude of getting all excitation in
the $N$-th locations is described by \begin{equation}
\gamma_{2}\equiv\sum_{\vec{n}\neq\vec{N}}F[\vec{N},\vec{n};t_{2}]\; F_{1}[\vec{n},\vec{1};t_{1}].\end{equation}
 In this case $|\overline{\Psi}(t)\rangle\!\rangle$ is replaced by
\begin{eqnarray}
|\overline{\Psi}(t_{2},t_{1})\rangle\!\rangle & = & \sum_{\ell}A_{\ell}\;|\boldsymbol{\xi}(t_{2},t_{1});\ell\rangle\!\rangle\;,\label{ko2}\end{eqnarray}
 with \begin{equation}
|\boldsymbol{\xi}(t_{2},t_{1});\ell\rangle\!\rangle=\sum_{\vec{n}\neq\vec{N}}F_{2}[\vec{n},\vec{1};t_{2},t_{1}]|\boldsymbol{\vec{n}};\ell\rangle\!\rangle,\end{equation}
 and $F_{2}$ defined as in Eq.~(\ref{effeq}) (see below). In other
words, the state $|\Phi(t_{2},t_{1})\rangle\!\rangle$ can be obtained
from Eq.~(\ref{logicstateout}) by replacing $\gamma_{1}$ and $F_{1}$
with $\gamma_{2}$ and $F_{2}$. Bob can hence try to use the same
strategy he used at time $t_{1}$: i.e. he will check whether or not
his $M$ qubits contain $K$ excitations. With (conditional) probability
$p_{2}\equiv|\gamma_{2}|^{2}$ he will get a positive answer and his
quantum register will be projected in the state $|\Psi\rangle\!\rangle$
of Eq.~(\ref{ok}). Otherwise he will let the system evolve for another
time interval $t_{3}$ and repeat the protocol. By reiterating the
above analysis it is possible to give a recursive expression for the
conditional probability of success $p_{q}\equiv|\gamma_{q}|^{2}$
after $q-1$ successive unsuccessful steps. The quantity $\gamma_{q}$
is the analogue of $\gamma_{2}$ and $\gamma_{1}$ of Eqs.~(\ref{gamma1})
and (\ref{logicstateout2}). It is given by \begin{eqnarray}
\gamma_{q}\equiv\sum_{\vec{n}\neq\vec{N}}F[\vec{N},\vec{n};t_{q}]\; F_{q-1}[\vec{n},\vec{1},t_{q-1},\cdots,t_{1}]\;,\label{gammaq}\end{eqnarray}
 where \begin{eqnarray}
\lefteqn{F_{q-1}[\vec{n},\vec{1};t_{q-1},\cdots,t_{1}]}\label{effeq}\\
 & \equiv & \sum_{\vec{n}^{\prime}\neq\vec{N}}\frac{F[\vec{N},\vec{n}^{\prime};t_{q-1}]}{\sqrt{1-|\gamma_{q-1}|^{2}}}F_{q-2}[\vec{n}^{\prime},\vec{1};t_{q-2},\cdots,t_{1}]\nonumber \end{eqnarray}
 and $F_{1}[\vec{n},\vec{1},t]$ is given by Eq. (\ref{eq:13b}).
In these equations $t_{q},\cdots,t_{1}$ are the \emph{time-intervals}
that occurred between the various protocol steps. Analogously the
conditional probability of failure at the step $q$ is equal to $1-p_{q}$.
The probability of having $j-1$ failures and a success at the step
$j$-th can thus be expressed as \begin{eqnarray}
\pi(j) & = & p_{j}(1-p_{j-1})(1-p_{j-2})\cdots(1-p_{1})\;,\label{proba}\end{eqnarray}
 while the total probability of success after $q$ steps is obtained
by the sum of $\pi(j)$ for all $j=1,\cdots,q$, i.e. \begin{eqnarray}
P_{q} & = & \sum_{j=1}^{q}\pi(j)\;.\label{probtot}\end{eqnarray}
 Since $p_{j}\geqslant0$, Eq.~(\ref{probtot}) is a monotonic function
of $q$. As a matter of fact in the next section we prove that under
a very general hypothesis on the system Hamiltonian, the probability
of success $P_{q}$ converges to $1$ in the limit of $q\rightarrow\infty$.
This means that by repeating many times the collective measure described
by $\Theta$ Bob is guaranteed to get, sooner or later, the answer
$0$ and hence the message Alice sent to him. In other words our protocol
allows perfect state transfer in the limit of repetitive collective
measures. Notice that the above analysis applies for all classes of
subsets $S_{\ell}$. The only difference between different choices
of $K$ is in the velocity of the convergence of $P_{q}\rightarrow1$.
In any case, by choosing $K\sim M/2$ Alice and Bob can achieve perfect
fidelity \emph{and} optimal efficiency.

\section{Convergence theorem\label{s:sec4}}

\begin{framedtheorem}
[Arbitrarly perfect transfer]\label{fra:If-there-exists}If there
is no eigenvector $|e_{m}\rangle$ of the quantum chain Hamiltonian
$H$ which is orthogonal to $|\boldsymbol{N}\rangle$, then there
is a choice of the times intervals $t_{q},t_{q-1},\cdots,t_{1}$ such
that the fidelity converges to $1$ as $q\rightarrow\infty.$
\end{framedtheorem}
Before proving this Theorem, let us give an intuitive reasoning for
the convergence. The unitary evolution can be thought of of a \emph{rotation}
in some abstract space, while the measurement corresponds to a \emph{projection.}
The dynamics of the system is then represented by alternating rotations
and projections. In general this will decrease the norm of each vector
to null, unless the rotation axis is \emph{the same} as the projection
axis.

\begin{proof}
The state of the system at a time interval of $t_{q}$ after the $(q-1)$-th
failure can be expressed in compact form as follows \begin{eqnarray}
|\Phi(t_{q},\cdots,t_{1})\rangle\!\rangle & = & \frac{U(t_{q})\Theta U(t_{q-1})\Theta\cdots U(t_{1})\Theta|\Phi\rangle\!\rangle}{\sqrt{(1-p_{q-1})\cdots(1-p_{1})}}\end{eqnarray}
 with $U(t)$ the unitary time evolution generated by the system Hamiltonian,
and with $\Theta$ the projection defined in Eq.~(\ref{observable}).
One can verify for instance that for $q=2$, the above equation coincides
with Eq.~(\ref{logicstateout2}). {[}For $q=1$ this is just (\ref{logicstateout})
evaluated at time $t_{1}$]. By definition the conditional probability
of success at step $q$-th is equal to \begin{equation}
p_{q}\equiv|\langle\!\langle\Psi|\Phi(t_{q},\cdots,t_{1})\rangle\!\rangle|^{2}.\end{equation}
 Therefore, Eq.~(\ref{proba}) yields \begin{eqnarray}
\pi(q) & = & |\langle\!\langle\Psi|U(t_{q})\Theta U(t_{q-1})\Theta\cdots U(t_{1})\Theta|\Phi\rangle\!\rangle|^{2}\label{proba0}\\
 & = & |\langle\!\langle\boldsymbol{\vec{N}};\ell|U(t_{q})\Theta U(t_{q-1})\Theta\cdots U(t_{1})\Theta|\boldsymbol{\vec{1}};\ell\rangle\!\rangle|^{2}\;,\nonumber \end{eqnarray}
 where the second identity stems from the fact that, according to
Eq.~(\ref{ortho}), $U(t)\Theta$ preserves the orthogonality relation
among states $|\boldsymbol{\vec{n}};\ell\rangle\!\rangle$ with distinct
values of $\ell.$ In analogy to the cases of Eqs.~(\ref{capitolF})
and (\ref{gamma1}), the second identity of~(\ref{proba0}) establishes
that $\pi(q)$ can be computed by considering the transfer of the
input $|\boldsymbol{\vec{1}};\ell\rangle\!\rangle$ for \emph{arbitrary}
$\ell$. The expression (\ref{proba0}) can be further simplified
by noticing that for a given $\ell$ the chains of the subset $\overline{S}_{\ell}$
contribute with a unitary factor to $\pi(q)$ and can be thus neglected
(according to~(\ref{in}) they are prepared in $|\boldsymbol{0}\rangle$
and do not evolve under $U(t)\Theta$). Identify $|\boldsymbol{\vec{1}}\rangle\!\rangle_{\ell}$
and $|\boldsymbol{\vec{N}}\rangle\!\rangle_{\ell}$ with the components
of $|\boldsymbol{\vec{1}};\ell\rangle\!\rangle$ and $|\boldsymbol{\vec{N}};\ell\rangle\!\rangle$
relative to the chains belonging to the subset $S_{\ell}$. In this
notation we can rewrite Eq.~(\ref{proba0}) as \begin{eqnarray}
\pi(q) & = & |_{\ell}\!\langle\!\langle\boldsymbol{\vec{N}}|U_{\ell}(t_{q})\Theta_{\ell}\;\cdots U_{\ell}(t_{1})\Theta_{\ell}|\boldsymbol{\vec{1}}\rangle\!\rangle_{\ell}|^{2}\;,\label{proba1}\end{eqnarray}
 where $\Theta_{\ell}=1-|\boldsymbol{\vec{N}}\rangle\!\rangle_{\ell}\langle\!\langle\boldsymbol{\vec{N}}|$
and $U_{\ell}(t)$ is the unitary operator $\otimes_{m\in S_{\ell}}u_{m}(t)$
which describes the time evolution of the chains of $S_{\ell}$. To
prove that there exist suitable choices of $t_{\ell}$ such that the
series~(\ref{probtot}) converges to $1$ it is sufficient to consider
the case $t_{\ell}=t>0$ for all $j=1,\cdots,q$: this is equivalent
to selecting decoding protocols with constant measuring intervals.
By introducing the operator $T_{\ell}\equiv U_{\ell}(t)\Theta_{\ell}$,
Eq.~(\ref{proba1}) becomes thus\begin{eqnarray}
 &  & \pi(q)=|_{\ell}\!\langle\!\langle\boldsymbol{\vec{N}}|\;(T_{\ell})^{q}|\boldsymbol{\vec{1}}\rangle\!\rangle_{\ell}|^{2}\label{proba2}\\
 &  & =_{\ell}\!\!\langle\!\langle\boldsymbol{\vec{1}}|(T_{\ell}^{\dag})^{q}|\boldsymbol{\vec{N}}\rangle\!\rangle_{\!\ell}\!\langle\!\langle\boldsymbol{\vec{N}}|\;(T_{\ell})^{q}|\boldsymbol{\vec{1}}\rangle\!\rangle_{\ell}=w(q)-w(q+1)\;,\nonumber \end{eqnarray}
 where \begin{equation}
w(j)\equiv_{\ell}\!\langle\!\langle\boldsymbol{\vec{1}}|(T_{\ell}^{\dag})^{j}\;(T_{\ell})^{j}|\boldsymbol{\vec{1}}\rangle\!\rangle_{\ell}=\Vert(T_{\ell})^{j}|\boldsymbol{\vec{1}}\rangle\!\rangle_{\ell}\Vert^{2}\;,\end{equation}
 is the norm of the vector $(T_{\ell})^{j}|\boldsymbol{\vec{1}}\rangle\!\rangle_{\ell}$.
Substituting Eq.~(\ref{proba2}) in Eq.~(\ref{probtot}) yields
\begin{eqnarray}
P_{q} & = & \sum_{j=1}^{q}\left[w(j)-w(j+1)\right]=1-w(q+1)\label{probtot1}\end{eqnarray}
 where the property $w(1)={_{\ell}\langle\!\langle}\boldsymbol{\vec{1}}|\Theta_{\ell}|\boldsymbol{\vec{1}}\rangle\!\rangle_{\ell}=1$
was employed. Proving the thesis is hence equivalent to prove that
for $q\rightarrow\infty$ the succession $w(q)$ nullifies. This last
relation can be studied using properties of power bounded matrices~\cite{Schott1996}.
In fact, by introducing the norm of the operator $(T_{\ell})^{q}$
we have, \begin{eqnarray}
w(q)=\Vert(T_{\ell})^{q}|\boldsymbol{\vec{1}}\rangle\!\rangle_{\ell}\Vert^{2}\leqslant\Vert(T_{\ell})^{q}\Vert^{2}\leqslant c\left(\frac{1+\rho(T_{\ell})}{2}\right)^{2q}\label{thesis}\end{eqnarray}
 where $c$ is a positive constant which does not depend on $q$ (if
$S$ is the similarity transformation that puts $T_{\ell}$ into the
Jordan canonical form, i.e. $J=S^{-1}T_{\ell}S,$ then $c$ is given
explicitly by $c=\| S\|\:\| S^{-1}\|$) and where $\rho(T_{\ell})$
is the spectral radius\index{spectral radius} of $T_{\ell}$, i.e.
the eigenvalue of $T_{\ell}$ with maximum absolute value (N.B. even
when $T_{\ell}$ is not diagonalisable this is a well defined quantity).
Equation~(\ref{thesis}) shows that $\rho(T_{\ell})<1$ is a sufficient
condition for $w(q)\rightarrow0$. In our case we note that, given
any normalised eigenvector $|\lambda\rangle\!\rangle_{\ell}$ of $T_{\ell}$
with eigenvalue $\lambda$ we have \begin{eqnarray}
|\lambda|=\Vert T_{\ell}|\lambda\rangle\!\rangle_{\ell}\Vert=\Vert\Theta_{\ell}|\lambda\rangle\!\rangle_{\ell}\Vert\leqslant1\;,\label{thesis1}\end{eqnarray}
 where the inequality follows from the fact that $\Theta_{\ell}$
is a projector. Notice that in Eq.~(\ref{thesis1}) the identity
holds only if $|\lambda\rangle\!\rangle$ is also an eigenvector of
$\Theta_{\ell}$ with eigenvalue $+1$, i.e. only if $|\lambda\rangle\!\rangle_{\ell}$
is orthogonal to $|\boldsymbol{\vec{N}}\rangle\!\rangle_{\ell}$.
By definition $|\lambda\rangle\!\rangle_{\ell}$ is eigenvector $T_{\ell}=U_{\ell}(t)\Theta_{\ell}$:
therefore the only possibility to have the equality in Eq.~(\ref{thesis1})
is that \emph{i)} $|\lambda\rangle\!\rangle_{\ell}$ is an eigenvector
of $U_{\ell}(t)$ (i.e. an eigenvector of the Hamiltonian%
\footnote{Notice that strictly speaking the eigenvectors of the Hamiltonian
are not the same as those of the time evolution operators. The latter
still can have evolution times at which additional degeneracy can
increase the set of eigenstates. A trivial example is given for $t=0$
where \emph{all} states become eigenstates. But it is always possible
to find times $t$ at which the eigenstates of $U(t)$ coincide with
those of $H$. %
} $H_{\ell}^{\mbox{\small{tot}}}$ of the chain subset $S_{\ell}$)
and \emph{ii)} it is orthogonal to $|\boldsymbol{\vec{N}}\rangle\!\rangle_{\ell}$.
By negating the above statement we get a sufficient condition for
the thesis. Namely, if all the eigenvectors $|\vec{E}\rangle\!\rangle_{\ell}$
of $H_{\ell}^{\mbox{\small{tot}}}$ are not orthogonal to $|\boldsymbol{\vec{N}}\rangle\!\rangle_{\ell}$
than the absolute values of the eigenvalues $\lambda$ of $T_{\ell}$
are strictly smaller than $1$ which implies $\rho(T_{\ell})<1$ and
hence the thesis. Since the $S_{\ell}$ channels are identical and
do not interact, the eigenvectors $|\vec{E}\rangle\!\rangle_{\ell}\equiv\bigotimes_{m\in S_{\ell}}|e_{m}\rangle_{m}$
are tensor product of eigenvectors $|e_{m}\rangle$ of the single
chain Hamiltonians $H$. Therefore the sufficient condition becomes
\begin{eqnarray}
_{\ell}\langle\!\langle\vec{E}|\boldsymbol{\vec{N}}\rangle\!\rangle_{\ell}=\prod_{m\in S_{\ell}}{_{m}\!\langle\boldsymbol{N}}|e_{m}\rangle_{m}\neq0\;,\label{last}\end{eqnarray}
 which can be satisfied only if ${\langle\boldsymbol{N}}|e_{m}\rangle\neq0$
for all eigenvectors $|e_{m}\rangle$ of the single chain Hamiltonian
$H$.
\end{proof}
\begin{remark}
While we have proven here that for equal time intervals the probability
of success is converging to unity, in practice one may use \emph{optimal}
measuring time intervals $t_{i}$ for a faster transfer (see also
Section~\ref{sec:Estimati}). We also point out that timing errors
may delay the transfer, but will not decrease its fidelity.
\end{remark}

\section{Quantum chains with nearest-neighbour interactions\label{sec:Quantum-chains-with}}

It is worth noticing that Eq. (\ref{last}) is a very weak condition,
because eigenstates of Hamiltonians are typically entangled. For instance,
it holds for open chains with nearest neighbour-interactions:

\begin{framedtheorem}
[Multi rail protocol]\label{thm:Let2}Let $H$ be the Hamiltonian
of an open nearest-neighbour quantum chain that conserves the number
of excitations. If there is a time $t$ such that $f_{1,N}(t)\neq0$
(i.e. the Hamiltonian is capable of transport between Alice and Bob)
then the state transfer can be made arbitrarily perfect by using the
\index{multi rail}multi rail protocol.
\end{framedtheorem}
\begin{proof}
We show by contradiction that the criterion of Theorem~\ref{fra:If-there-exists}
is fulfilled. Assume there exists a normalised eigenvector $\left|e\right\rangle $
of the single chain Hamiltonian $H$ such that \begin{equation}
\langle\boldsymbol{N}|e\rangle=0.\end{equation}
 Because $\left|e\right\rangle $ is an eigenstate, we can conclude
that also \begin{equation}
\left\langle e\left|H\right|\boldsymbol{N}\right\rangle =0.\label{eq:h_null}\end{equation}
 If we act with the Hamiltonian on the ket in Eq. (\ref{eq:h_null})
we may get some term proportional to $\langle e|\boldsymbol{N}\rangle$
(corresponding to an Ising-like interaction) and some part proportional
to $\langle e|\boldsymbol{N-1}\rangle$ (corresponding to a hopping
term; if this term did not exist, then clearly $f_{1,N}(t)=0$ for
all times). We can thus conclude that \begin{equation}
\langle e|\boldsymbol{N-1}\rangle=0.\label{eq:steptwo}\end{equation}
 Note that for a closed chain, e.g. a ring, this need not be the case,
because then also a term proportional to $\langle e|\boldsymbol{N+1}\rangle=\langle e|\boldsymbol{1}\rangle$
would occur. If we insert the Hamiltonian into Eq. (\ref{eq:steptwo})
again, we can use the same reasoning to see that \begin{equation}
\langle e|\boldsymbol{N-2}\rangle=\cdots=\langle e|\boldsymbol{1}\rangle=0\end{equation}
 and hence $\left|e\right\rangle =0,$ which is a contradiction to
$\left|e\right\rangle $ being normalised. 
\end{proof}

\section{Comparison with Dual Rail}

As we have seen above, the Multi Rail protocol allows us in principle
to reach in principle a rate arbitrarily close to one. However for
a fair comparison with the Dual Rail protocol, we should also take
into account the time-scale of the transfer. For the conclusive transfer
of entanglement, we have seen in Section~\ref{s:sec3} that only
for chains which have a success probability higher than $p(t)=0.8$
it is worth encoding on more than three rails. The reason is that
if the probability of success for a single excitation is $p,$ then
the probability of success for $\left\lfloor M/2\right\rfloor $ excitations
on on $M$ parallel chains is lowered to $p^{\left\lfloor M/2\right\rfloor }.$
The protocol for three rails is always more efficient than on two,
as still only one excitation is being used, but three complex amplitudes
can be transferred per usage. 

\begin{figure}[tbh]
\begin{centering}\includegraphics[width=0.9\columnwidth]{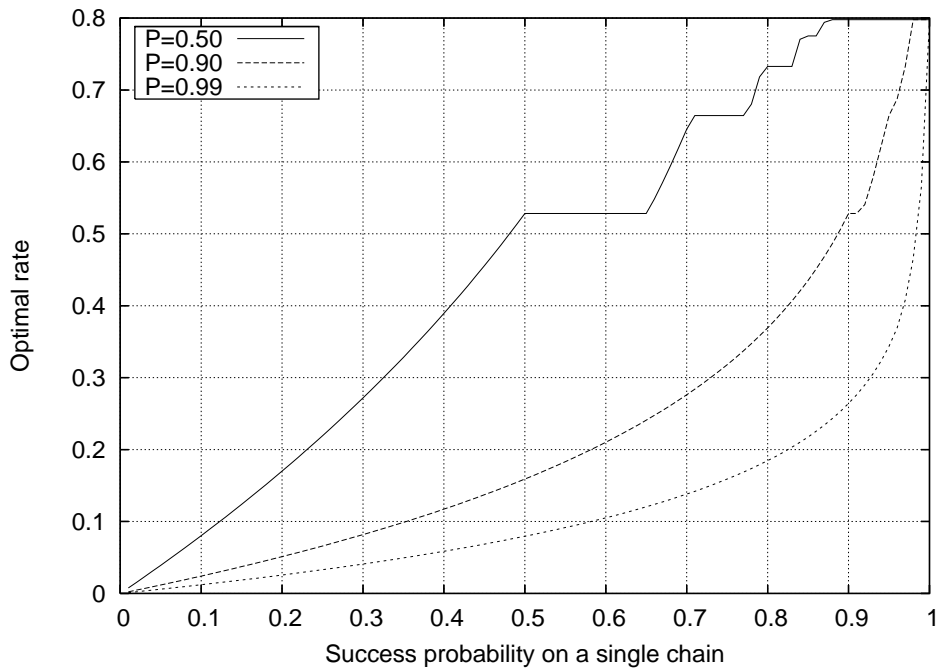}\par\end{centering}

\caption{\label{f:rates} Optimal rates (maximisation of Eq.~(\ref{eq:rate}
with respect to $M$) for the Multi Rail protocol. Shown are three
curves corresponding to different values of the joint probability
of failure $P$ one plans to achieve.}
\end{figure}

For arbitrarily perfect transfer, the situation is slightly more complicated
as the optimal choice of $M$ also depends on the joint probability
of failure that one plans to achieve. Let us assume that at each step
of the protocol, the success probability on a single chain is $p.$
Then the number of steps to achieve a given probability of failure
$P$ using $M$ chains is given by\begin{equation}
\ell(P,M)=\max\left\{ \frac{\ln P}{\ln(1-p^{\left\lfloor M/2\right\rfloor })},1\right\} .\end{equation}
If we assume that the total time-scale of the transfer is proportional
to the number of steps, then the number of qubits that can be transferred
per time interval is given by\begin{equation}
v(P,M)\propto R(M)/\ell(P,M).\label{eq:rate}\end{equation}
Optimising this rate with respect to $M$ we find three different
regimes of the joint probability of failure (see Fig.~\ref{f:rates}).
If one is happy with a large $P,$ then the Multi Rail protocol becomes
superior to the Dual Rail for medium $p.$ For intermediate $P,$
the threshold is comparable to the threshold of $p=0.8$ for conclusive
transfer of entanglement. Finally for very low $P$ the Multi Rail
only becomes useful for $p$ very close to one. In all three cases
the threshold is higher than the $p(t)$ that can usually achieved
with unmodulated Heisenberg chains. We can thus conclude that the
Multi Rail protocol only becomes useful for chains which already have
a very good performance.%

\section{Conclusion}

We thus conclude that any nearest-neighbour Hamiltonian that can transfer
quantum information with nonzero fidelity (including the Heisenberg
chains analysed above) is capable of efficient \emph{and} perfect
transfer when used in the context of parallel chains. Hamiltonians
with non-nearest neighbour interactions~\cite{Bose2006,Kay2006}
can also be used as long as the criterion of Theorem~\ref{fra:If-there-exists}
is fulfilled.

\chapter{Ergodicity and mixing\label{cha:Ergodicity-and-mixing}}

\section{Introduction}

We have seen above that by applying measurements at the end of parallel
chains, the state of the chain is converging to the ground state,
and the quantum information is transferred to the receiver. Indeed,
repetitive application of the same transformation is the key ingredient
of many controls techniques. Beside quantum state transfer, they have
been exploited to inhibit the decoherence of a system by frequently
perturbing its dynamical evolution~\cite{Viola1998,VIOLA2,VIOLA3,VITALI,SIMON}
(\emph{Bang-Bang control}) or to improve the fidelity of quantum gates~\cite{FRANSON}
by means of frequent measurements (\emph{quantum Zeno-effect}~\cite{PERES}).
Recently analogous strategies have also been proposed in the context
of state preparation~\cite{KUMMERER,WELLENS,HOMOGENIZATION1,HOMOGENIZATION2,TERHAL,YUASA1,YUASA2}.
In Refs.~\cite{HOMOGENIZATION1,HOMOGENIZATION2} for instance, a
\emph{homogenisation} protocol was presented which allows one to transform
any input state of a qubit into a some pre-fixed target state by repetitively
coupling it with an external bath. A similar \emph{thermalisation}
protocol was discussed in Ref.~\cite{TERHAL} to study the efficiency
of simulating classical equilibration processes on a quantum computer.
In Refs.~\cite{YUASA1,YUASA2} repetitive interactions with an externally
monitored environment were exploited instead to implement \emph{purification}
schemes which would allow one to extract pure state components from
arbitrary mixed inputs. %
\begin{figure}[tbh]
\begin{centering}\includegraphics[width=0.75\textwidth]{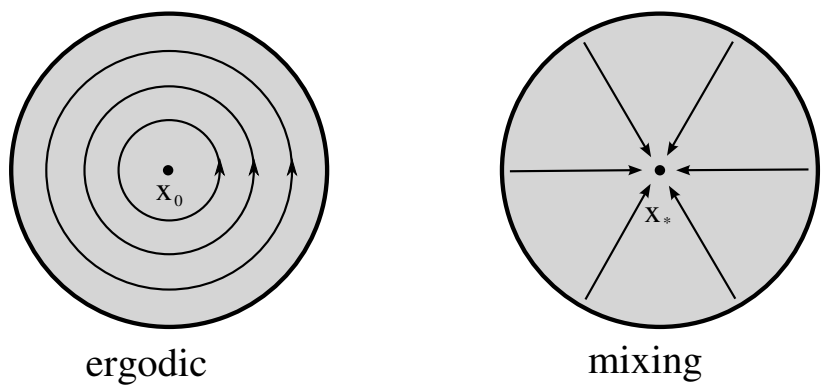}\par\end{centering}

\caption{\label{F1}Schematic examples of the orbits of a ergodic and a mixing
map.}
\end{figure}

The common trait of the proposals~\cite{KUMMERER,WELLENS,HOMOGENIZATION1,HOMOGENIZATION2,TERHAL,YUASA1,YUASA2}
and the dual  and multi rail protocols is the requirement that repeated
applications of a properly chosen quantum operation $\tau$ converges
to a fixed density matrix $x_{*}$ independently from the input state
$x$ of the system, i.e. \begin{eqnarray}
\tau^{n}(x)\equiv\underbrace{\tau\circ\tau\circ\cdots\circ\tau}_{n}\;(x)\Big|_{n\rightarrow\infty}\longrightarrow\;\; x_{*}\;,\label{mixing0}\end{eqnarray}
 with {}``$\circ$'' representing the composition of maps. Following
the notation of Refs.~\cite{RAGINSKY,RICHTER} we call Eq.~(\ref{mixing0})
the \emph{mixing} property of $\tau$. It is related with another
important property of maps, namely \emph{ergodicity} (see Fig. \ref{F1}).
The latter requires the existence of a unique input state $x_{0}$
which is left invariant under a single application of the map%
\footnote{Definition~(\ref{ergodic0}) may sound unusual for readers who are
familiar with a definition of ergodicity from statistical mechanics,
where a map is ergodic if its invariant sets have measure $0$ or
$1.$ The notion of ergodicity used here is completely different,
and was introduced in \cite{RAGINSKY,Raginsky2002,STRICTCONTRATIONS}.
The set $\mathcal{X}$ one should have in mind here is not a measurable
space, but the compact convex set of quantum states. A perhaps more
intuitive definition of ergodicity based on the time average of observables
is given by Lemma~\ref{thm:ergodic}).%
}, i.e. \begin{eqnarray}
\tau(x)=x\qquad\Longleftrightarrow\qquad x=x_{0}\;.\label{ergodic0}\end{eqnarray}
Ergodicity and the mixing property are of high interest not only in
the context of the above quantum information schemes. They also occur
on a more fundamental level in statistical mechanics~\cite{STREATER}
and open quantum systems~\cite{OPENQUANTUM,ALICKI}, where one would
like to study irreversibility and relaxation to thermal equilibrium.

In the case of quantum transformations one can show that mixing maps
with convergence point $x_{*}$ are also ergodic with fixed point
$x_{0}=x_{*}$. The opposite implication however is not generally
true since there are examples of ergodic quantum maps which are not
mixing (see the following). Sufficient conditions for mixing have
been discussed both in the specific case of quantum channel~\cite{TERHAL,RAGINSKY,STRICTCONTRATIONS}
and in the more abstract case of maps operating on topological spaces~\cite{STREATER}.
In particular the Lyapunov direct method~\cite{STREATER} allows
one to prove that an ergodic map $\tau$ is mixing if there exists
a continuous functional $S$ which, for all points but the fixed one,
is strictly increasing under $\tau$. Here we strengthen this criterion
by weakening the requirement on $S$: our \emph{generalised} Lyapunov
functions are requested only to have limiting values $S(\tau^{n}(x))|_{n\rightarrow\infty}$
which differ from $S(x)$ for all $x\neq x_{0}$. It turns out that
the existence of such $S$ is not just a \emph{sufficient} condition
but also a \emph{necessary} condition for mixing. Exploiting this
fact one can easily generalise a previous result on \emph{strictly
contractive} maps~\cite{RAGINSKY} by showing that maps which are
\emph{asymptotic deformations} (see Definition \ref{asymptdef}) are
mixing. This has, unlike contractivity, the advantage of being a property
independent of the choice of metric (see however~\cite{RICHTER}
for methods of finding {}``tight'' norms). In some cases, the generalised
Lyapunov method permits also to derive an optimal mixing condition
for quantum channels based on the quantum relative entropy. Finally
a slightly modified version of our approach which employs \emph{multi-central}
Lyapunov functions yields a characterisation of (not necessarily mixing)
maps which in the limit of infinitely many applications move all points
toward a proper \emph{subset} (rather than a single point) of the
input space.

The introduction of a generalised Lyapunov method seems to be sound
not only from a mathematical point of view, but also from a physical
point of view. In effect, it often happens that the informations available
on the dynamics of a system are only those related on its asymptotic
behaviour (e.g. its thermalisation process), its finite time evolution
being instead difficult to characterise. Since our method is explicitly
constructed to exploit asymptotic features of the mapping, it provides
a more effective way to probe the mixing property of the process.

Presenting our results we will not restrict ourself to the case of
quantum operations. Instead, following~\cite{STREATER} we will derive
them in the more general context of continuous maps operating on topological
spaces~\cite{TOPOLOGYBOOK}. This approach makes our results stronger
by allowing us to invoke only those hypotheses which, to our knowledge,
are strictly necessary for the derivation. It is important to stress
however that, as a particular instance, all the Theorems and Lemmas
presented in this chapter hold for any linear, completely positive,
trace preserving map (i.e. quantum channels) operating on a compact
subset of normed vectors (i.e. the space of the density matrices of
a finite dimensional quantum system). Therefore readers who are not
familiar with topological spaces can simply interpret our derivations
as if they were just obtained for quantum channels acting on a finite
dimensional quantum system.

This chapter is organised as follows. In Sec.~\ref{GLT} the generalised
Lyapunov method along with some minor results is presented in the
context of topological spaces. Then quantum channels are analysed
in Sec.~\ref{QC} providing a comprehensive summary of the necessary
and sufficient conditions for the mixing property of these maps. Conclusions
and remarks form the end of the chapter in Sec.~\ref{CONCLUSION}.

\section{Topological background}

Let us first introduce some basic topological background required
for this chapter. A more detailed introduction is given in~\cite{TOPOLOGYBOOK}.
Topological spaces are a very elegant way of defining compactness,
convergence and continuity without requiring more than the following
structure:

\begin{definition}
A \emph{topological space\index{topological space}} is a pair $(\mathcal{X},\mathcal{O})$
of a set $\mathcal{X}$ and a set $\mathcal{O}$ of subsets of $\mathcal{X}$
(called \emph{open} sets) such that
\begin{enumerate}
\item $\mathcal{X}$ and $\emptyset$ are open
\item Arbitrary unions of open sets are open
\item Intersections of two open sets are open
\end{enumerate}
\end{definition}

\begin{example}
If $\mathcal{X}$ is an arbitrary set, and $\mathcal{O}=\{\mathcal{X},\emptyset\}$, then $(\mathcal{X},\mathcal{O})$ is a topological space. $\mathcal{O}$ is called the \emph{trivial topology}.
\end{example}

\begin{figure}[htbp]
\begin{centering}\includegraphics[width=0.5\textwidth]{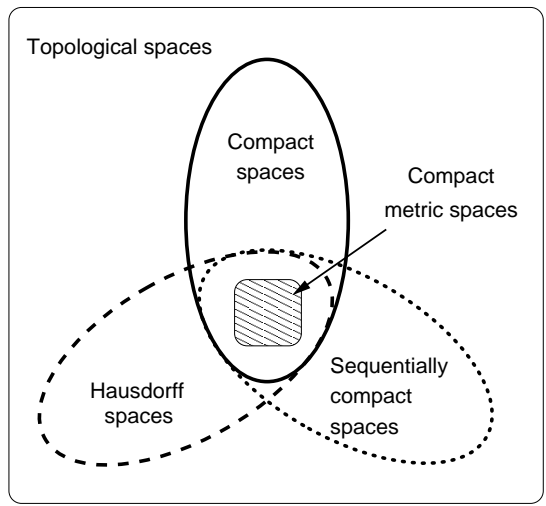}\par\end{centering}

\caption{\label{fig:topologic} Relations between topological spaces~\cite{TOPOLOGYBOOK}.
The space of density matrices on which quantum channels are defined,
is a compact and convex subset of a normed vectors space (the space
of linear operators of the system) which, in the above graphical representation
fits within the set of compact metric spaces.}
\end{figure}

\begin{definition}
A topological space $\mathcal{X}$ is \emph{compact} if any open cover (i.e. a set of open sets such that $\mathcal{X}$ is contained in their union) contains a finite sub-cover.
\end{definition}
\begin{definition}
A sequence $x_n \in \mathcal{X}$ is \emph{convergent} with limit $x_*$ if each open neighbourhood $O(x_*)$ (i.e. a set such that $x_*  \in O(x_*) \in \mathcal{O}$ contains all but finitely many points of the sequence.
\end{definition}
\begin{definition}
A map on a topological space is \emph{continous} if the preimage of any open set is open.
\end{definition}
This is already all we require to make useful statements about ergodicity and mixing. However, there are some subtleties which we need to take care of:
\begin{definition}
A topological space is \emph{sequentially compact} if every sequence has a convergent subsequence.
\end{definition}
Sequentially compactness is in general not related to compactness! Another subtlety is that with the above definition, a sequence can converge to many different points. For example, in the trivial topology, \emph{any} sequence converges to \emph{any} point. This motivates
\begin{definition}
A topological space is \emph{Hausdorff} if any two distinct points can by separated by open neighbourhoods.
\end{definition}
A limit of a sequence in a Hausdorff space is unique. All these problems disappear in metrical spaces:
\begin{definition}
A \emph{metric space} is a pair $(\mathcal{X},d)$ of a set $\mathcal{X}$ and a function $d:\mathcal{X}\times\mathcal{X}\rightarrow\mathbb{R}$ such that
\begin{enumerate}
\item $d(x,y)\ge 0$ and $d(x,y)=0 \Leftrightarrow x=y$
\item $d(x,y)=d(y,x)$
\item $d(x,z)\le d(x,y)+d(y,z)$
\end{enumerate}
\end{definition}
A metric space becomes a topological space with the canonical topology
\begin{definition}
A subset $O$ of a metric space $\mathcal{X}$ is \emph{open} if $\forall x\in O$  there is an $\epsilon >0$ such that $\{y \in  \mathcal{X} | d(x,y) \le \epsilon\} \subset O.$
\end{definition}In a metric space with the canonical topology, compactness and sequentially
compactness become equivalent. Furthermore, it is automatically Hausdorff
(see Fig.~\ref{fig:topologic}).%

\section{Generalised Lyapunov Theorem\label{GLT}}

\subsection{Topological spaces}

In this section we introduce the notation and derive our main result
(the Generalised Lyapunov Theorem). 

\begin{definition}
Let $\mathcal{X}$ be a topological space and let $\tau:\mathcal{X}\rightarrow\mathcal{X}$
be a map. The sequence $x_{n}\equiv\tau^{n}(x)$, where $\tau^{n}$
is a short-hand notation for the $n-$fold composition of $\tau,$
is called the \emph{orbit} of $x.$ An element $x_{*}\in\mathcal{X}$
is called a \emph{fixed point} of $\tau$ if and only if \begin{eqnarray}
\tau(x_{*})=x_{*}\;.\label{defmixing}\end{eqnarray}
 $\tau$ is called \emph{ergodic}\index{ergodic} if and only if it
has exactly one \index{fix-point}fixed point. $\tau$ is called \emph{mixing}\index{mixing}
if and only if there exists a \emph{convergence} point $x_{*}\in\mathcal{X}$
such that any orbit converges to it, i.e. \begin{eqnarray}
\lim_{n\rightarrow\infty}x_{n}=x_{*}\quad\forall x\in\mathcal{X}\;.\label{defergo10}\end{eqnarray}

\end{definition}
A direct connection between ergodicity and mixing can be established
as follows.

\begin{lemma}
\label{lem:hausdorff} Let $\tau:\mathcal{X}\rightarrow\mathcal{X}$
be a continuous mixing map on a topological Hausdorff space $\mathcal{X}.$
Then $\tau$ is ergodic. 
\end{lemma}
\begin{proof}
Let $x_{*}$ be the convergence point of $\tau$ and let $x\in\mathcal{X}$
arbitrary. Since $\tau$ is continuous we can perform the limit in
the argument of $\tau,$ i.e.\begin{equation}
\tau(x_{*})=\tau\left(\lim_{n\rightarrow\infty}\tau^{n}(x)\right)=\lim_{n\rightarrow\infty}\tau^{n+1}(x)=x_{*},\end{equation}
 which shows that $x_{*}$ is a fixed point of $\tau$. To prove that
it is unique assume by contradiction that $\tau$ possesses a second
fixed point $y_{*}\neq x_{*}$. Then $\lim_{n\rightarrow\infty}\tau^{n}(y_{*})=y_{*}\neq x_{*}$,
so $\tau$ could not be mixing (since the limit is unique in a Hausdorff
space -- see Fig.~\ref{fig:topologic}). Hence $\tau$ is ergodic. 
\end{proof}
\begin{remarknn}
The converse is not true in general, i.e. not every ergodic map is
mixing (not even in Hausdorff topological spaces). A simple counterexample
is given by $\tau:[-1,1]\rightarrow[-1,1]$ with $\tau(x)\equiv-x$
and the usual topology of $\mathbb{R}$, which is ergodic with fixed
point $0,$ but not mixing since for $x\neq0$, $\tau^{n}(x)=(-1)^{n}x$
is alternating between two points. A similar counterexample will be
discussed in the quantum channel section (see Example~\ref{exm:ergodic}). 
\end{remarknn}
A well known criterion for mixing is the existence of a \emph{Lyapunov
function}~\cite{STREATER}.

\begin{definition}
Let $\tau:\mathcal{X}\rightarrow\mathcal{X}$ be a map on a topological
space $\mathcal{X}.$ A continuous map $S:\mathcal{X}\rightarrow\mathbb{R}$
is called a \emph{(strict) Lyapunov function for $\tau$ around $x_{*}\in\mathcal{X}$}
if and only if\begin{equation}
S\left(\tau(x)\right)>S(x)\quad\forall x\neq x_{*}.\end{equation}

\end{definition}
\begin{remarknn}
At this point is is neither assumed that $x_{*}$ \emph{is} a fixed
point, nor that $\tau$ is ergodic. Both follows from the theorem
below. 
\end{remarknn}
\begin{theorem}
[Lyapunov function]Let $\tau:\mathcal{X}\rightarrow\mathcal{X}$
be a continuous map on a sequentially compact topological space $\mathcal{X}$.
Let $S:\mathcal{X}\rightarrow\mathbb{R}$ be a Lyapunov function for
$\tau$ around $x_{*}.$ Then $\tau$ is mixing with the fixed point
$x_{*}$. 
\end{theorem}
The proof of this theorem is given in~\cite{STREATER}. We will not
reproduce it here, because we will provide a general theorem that
includes this as a special case. In fact, we will show that the requirement
of the strict monotonicity can be \emph{much} weakened, which motivates
the following definition.

\begin{definition}
Let $\tau:\mathcal{X}\rightarrow\mathcal{X}$ be a map on a topological
space $\mathcal{X}.$ A continuous map $S:\mathcal{X}\rightarrow\mathbb{R}$
is called a \emph{generalised Lyapunov function\index{generalised Lyapunov function}
for $\tau$ around $x_{*}\in\mathcal{X}$} if and only if the sequence
$S\left(\tau^{n}(x)\right)$ is point-wise convergent%
\footnote{Point-wise convergence in this context means that for any fixed $x$
the sequence $S_{n}\equiv S\left(\tau^{n}(x)\right)$ is convergent.%
} for any $x\in\mathcal{X}$ and $S$ fulfils \begin{equation}
S_{*}(x)\equiv\lim_{n\rightarrow\infty}S\left(\tau^{n}(x)\right)\neq S(x)\quad\forall x\neq x_{*}.\label{eq:weakinequality}\end{equation}

\end{definition}
In general it may be difficult to prove the point-wise convergence.
However if $S$ is monotonic under the action of $\tau$ and the space
is compact, the situation becomes considerably simpler. This is summarised
in the following Lemma.

\begin{lemma}
\label{lem:monotonlyapov}Let $\tau:\mathcal{X}\rightarrow\mathcal{X}$
be map on a compact topological space. A continuous map $S:\mathcal{X}\rightarrow\mathbb{R}$
which fulfils\begin{equation}
S\left(\tau(x)\right)\geqslant S(x)\quad\forall x\in\mathcal{X},\label{eq:monotonic}\end{equation}
 and\begin{equation}
S_{*}(x)\equiv\lim_{n\rightarrow\infty}S\left(\tau^{n}(x)\right)>S(x)\quad\forall x\neq x_{*}.\end{equation}
 for some fixed $x_{*}\in\mathcal{X}$ is a generalised Lyapunov function
for $\tau$ around $x_{*}$. 
\end{lemma}
\begin{proof}
It only remains to show the (point-wise) convergence of $S\left(\tau^{n}(x)\right)$.
Since $S$ is a continuous function on a compact space, it is bounded.
By Eq. (\ref{eq:monotonic}) the sequence is monotonic. Any bounded
monotonic sequence converges. 
\end{proof}
\begin{corollary}
Let $\tau:\mathcal{X}\rightarrow\mathcal{X}$ be a map on a compact
topological space. A continuous map $S:\mathcal{X}\rightarrow\mathbb{R}$
which fulfils\begin{equation}
S\left(\tau(x)\right)\geqslant S(x)\quad\forall x\in\mathcal{X},\end{equation}
 and\begin{equation}
S\left(\tau^{N}(x)\right)>S(x)\quad\forall x\neq x_{*},\end{equation}
 for some fixed $N\in\mathbb{N}$ and for some $x_{*}\in\mathcal{X}$
is a generalised Lyapunov function for $\tau$ around $x_{*}$. 
\end{corollary}
\begin{remarknn}
This implies that a strict Lyapunov function is a generalised Lyapunov
function (with $N=1$). 
\end{remarknn}
We can now state the main result of this section:

\begin{framedtheorem}
[Generalized Lyapunov function]\label{thm:weaklyapov} Let $\tau:\mathcal{X}\rightarrow\mathcal{X}$
be a continuous map on a sequentially compact topological space $\mathcal{X}.$
Let $S:\mathcal{X}\rightarrow\mathbb{R}$ be a generalised Lyapunov
function for $\tau$ around $x_{*}.$ Then $\tau$ is mixing with
fixed point $x_{*}$. 
\end{framedtheorem}
\begin{proof}
Consider the orbit $x_{n}\equiv\tau^{n}(x)$ of a given $x\in\mathcal{X}.$
Because $\mathcal{X}$ is sequentially compact, the sequence $x_{n}$
has a convergent subsequence (see Fig.~\ref{fig:topologic}), i.e.
$\lim_{k\rightarrow\infty}x_{n_{k}}\equiv\tilde{x}$. Let us assume
that $\tilde{x}\neq x_{*}$ and show that this leads to a contradiction.
By Eq. (\ref{eq:weakinequality}) we know that there exists a finite
$N\in\mathbb{N}$ such that \begin{equation}
S\left(\tau^{N}(\tilde{x})\right)\neq S(\tilde{x}).\label{eq:contra}\end{equation}
 Since $\tau^{N}$ is continuous we can perform the limit in the argument,
i.e. \begin{equation}
\lim_{k\rightarrow\infty}\tau^{N}\left(x_{n_{k}}\right)=\tau^{N}(\tilde{x}).\end{equation}
Likewise, by continuity of $S$ we have \begin{equation}
\lim_{k\rightarrow\infty}S\left(x_{n_{k}}\right)=S(\tilde{x}),\label{eq:ss1}\end{equation}
 and on the other hand\begin{equation}
\lim_{k\rightarrow\infty}S\left(x_{N+n_{k}}\right)=\lim_{k\rightarrow\infty}S\left(\tau^{N}\left(x_{n_{k}}\right)\right)=S(\tau^{N}\tilde{x}),\label{eq:ss2}\end{equation}
 where the second equality stems from the continuity of the map $S$
and $\tau^{N}$. Because $S$ is a generalised Lyapunov function,
the sequence $S\left(x_{n}\right)$ is convergent. Therefore the subsequences
(\ref{eq:ss1}) and (\ref{eq:ss2}) must have the same limit. We conclude
that $S(\tau^{N}\tilde{x})=S(\tilde{x})$ which contradicts Eq. (\ref{eq:contra}).
Hence $\tilde{x}=x_{*}.$ Since we have shown that any convergent
subsequence of $\tau^{n}(x)$ converges to the same limit $x_{*}$,
it follows by Lemma~\ref{lem:subsequences} that $\tau^{n}(x)$ is
converging to $x_{*}.$ Since that holds for arbitrary $x$, it follows
that $\tau$ is mixing. 
\end{proof}
\begin{lemma}
\label{lem:subsequences} Let $x_{n}$ be a sequence in a sequentially
compact topological space $\mathcal{X}$ such that any convergent
subsequence converges to $x_{*}.$ Then the sequence converges to
$x_{*}.$ 
\end{lemma}
\begin{proof}
We prove by contradiction: assume that the sequence does not converge
to $x_{*}.$ Then there exists an open neighbourhood $O(x_{*})$ of
$x_{*}$ such that for all $k\in\mathbb{N},$ there is a $n_{k}$
such that $x_{n_{k}}\notin O(x_{*}).$ Thus the subsequence $x_{n_{k}}$
is in the closed space $\mathcal{X}\backslash O(x_{*}),$ which is
again sequentially compact. $x_{n_{k}}$ has a convergent subsequence
with a limit in $\mathcal{X}\backslash O(x_{*}),$ in particular this
limit is not equal to $x_{*}.$ 
\end{proof}
There is an even more general way of defining Lyapunov functions which
we state here for completeness. It requires the concept of the quotient
topology~\cite{TOPOLOGYBOOK}.

\begin{definition}
Let $\tau:\mathcal{X}\rightarrow\mathcal{X}$ be a map on a topological
space $\mathcal{X}.$ A continuous map $S:\mathcal{X}\rightarrow\mathbb{R}$
is called a \emph{multi-central Lyapunov function for $\tau$ around
$\mathcal{F}\subseteq\mathcal{X}$} if and only if the sequence $S\left(\tau^{n}(x)\right)$
is point-wise convergent for any $x\in\mathcal{X}$ and if $S$ and
$\tau$ fulfil the following three conditions: $S$ is constant on
$\mathcal{F}$, $\tau(\mathcal{F})\subseteq\mathcal{F}$, and \begin{equation}
S_{*}(x)\equiv\lim_{n\rightarrow\infty}S\left(\tau^{n}(x)\right)\neq S(x)\quad\forall x\notin\mathcal{F}.\end{equation}

\end{definition}
For these functions we cannot hope that the orbit is mixing. We can
however show that the orbit is {}``converging'' to the set $\mathcal{F}$
in the following sense:

\begin{theorem}
[Multi-central Lyapunov function]Let $\tau:\mathcal{X}\rightarrow\mathcal{X}$
be a continuous map on a sequentially compact topological space $\mathcal{X}.$
Let $S:\mathcal{X}\rightarrow\mathbb{R}$ be a multi-central Lyapunov
function for $\tau$ around $\mathcal{F}.$ Let $\varphi:\mathcal{X}\rightarrow\mathcal{X}/\mathcal{F}$
be the continuous mapping into the quotient space (i.e. $\varphi(x)=[x]$
for $x\in\mathcal{X}\backslash\mathcal{F}$ and $\varphi(x)=[\mathcal{F}]$
for $x\in\mathcal{F})$. Then $\tilde{\tau}:\mathcal{X}/\mathcal{F}\rightarrow\mathcal{X}/\mathcal{F}$
given by $\tilde{\tau}([x])=\varphi\left(\tau\left(\varphi^{-1}([x])\right)\right)$
is mixing with fixed point $[\mathcal{F}]$. 
\end{theorem}
\begin{proof}
First note that $\tilde{\tau}$ is well defined because $\varphi$
is invertible on $\mathcal{X}/\mathcal{F}\backslash[\mathcal{F}]$
and $\tau(\mathcal{F})\subseteq\mathcal{F},$ so that $\tilde{\tau}([\mathcal{F}])=[\mathcal{F}]$.
Since $\mathcal{X}$ is sequentially compact, the quotient space $\mathcal{X}/\mathcal{F}$
is also sequentially compact. Note that for $O$ open, $\tilde{\tau}^{-1}(O)=\varphi\left(\tau^{-1}\left(\varphi^{-1}\left(O\right)\right)\right)$
is the image of $\varphi$ of an open set in $\mathcal{X}$ and therefore
(by definition of the quotient topology) open in $\mathcal{X}/\mathcal{F}.$
Hence $\tilde{\tau}$ is continuous. The function $\tilde{S}([x]):\mathcal{X}/\mathcal{F}\rightarrow\mathcal{X}/\mathcal{F}$
given by $\tilde{S}([x])=S(\varphi^{-1}([x]))$ is continuous and
easily seen to be a generalised Lyapunov function around $[\mathcal{F}].$
By Theorem \ref{thm:weaklyapov} it follows that $\tilde{\tau}$ is
mixing. 
\end{proof}

\subsection{Metric spaces}

We now show that for the particular class of compact topological sets
which posses a metric, the existence of a generalised Lyapunov function
is also a necessary condition for mixing. 

\begin{theorem}
[Lyapunov  criterion]\label{thm:glyap} Let $\tau:\mathcal{X}\rightarrow\mathcal{X}$
be a continuous map on a compact metric space $\mathcal{X}.$ Then
$\tau$ is mixing with fixed point $x_{*}$ if and only if a generalised
Lyapunov function around $x_{*}$ exists. 
\end{theorem}
\begin{proof}
Firstly, in metric spaces compactness and sequential compactness are
equivalent, so the requirements of Theorem \ref{thm:weaklyapov} are
met. Secondly, for any mixing map $\tau$ with fixed point $x_{*},$
a generalised Lyapunov function around $x_{*}$ is given by $S(x)\equiv d(x_{*},x)$.
In fact, it is continuous because of the continuity of the metric
and satisfies \begin{equation}
\lim_{n\rightarrow\infty}S\left(\tau^{n}(x)\right)=d(x_{*},x_{*})=0\leqslant d(x_{*},x)=S(x),\end{equation}
 where the equality holds if and only $x=x_{*}.$ We call $d(x_{*},x)$
the \emph{trivial generalised Lyapunov function}. 
\end{proof}
\begin{remark}
In the above Theorem we have not used all the properties of the metric.
In fact a continuous \emph{semi-metric} (i.e. without the triangle
inequality) would suffice.
\end{remark}
The trivial Lyapunov function requires knowledge of the fixed point
of the map. There is another way of characterising mixing maps as
those which bring elements closer to \emph{each other} (rather than
closer to the fixed point). 

\begin{definition}
A map $\tau:\mathcal{X}\rightarrow\mathcal{X}$ is on a metric space
is called a \emph{non-expansive map}\index{non-expansive map} if
and only if \begin{equation}
d(\tau(x),\tau(y))\leqslant d(x,y)\quad\forall x,y\in\mathcal{X},\end{equation}
 a \emph{weak contraction}\index{weak contraction} if and only if
\begin{equation}
d(\tau(x),\tau(y))<d(x,y)\quad\forall x,y\in\mathcal{X},\, x\neq y,\end{equation}
 and a \emph{strict contraction}\index{strict contraction} if and
only if there exists a $k<1$ such that\begin{equation}
d(\tau(x),\tau(y))\leqslant k\, d(x,y)\quad\forall x,y\in\mathcal{X}\,.\end{equation}
 
\end{definition}
\begin{remark}
The notation adopted here is slightly different from the definitions
used by other Authors~\cite{RAGINSKY,RUSKAI,WERNER} who use contraction
to indicate our non-expansive maps. Our choice is motivated by the
need to clearly distinguish between non-expansive transformation and
weak contractions. 
\end{remark}
We can generalise the above definition in the following way:

\begin{definition}
\label{asymptdef}A map $\tau:\mathcal{X}\rightarrow\mathcal{X}$
on a metric space is called an \emph{asymptotic deformation}\index{asymptotic deformation}
if and only if the sequence $d(\tau^{n}(x),\tau^{n}(y))$ converges
point-wise for all $x,y\in\mathcal{X}$ and\begin{equation}
\lim_{n\rightarrow\infty}d(\tau^{n}(x),\tau^{n}(y))\neq d(x,y)\quad\forall x,y\in\mathcal{X},\, x\neq y.\end{equation}

\end{definition}
\begin{lemma}
\label{asymptlem}Let $\tau:\mathcal{X}\rightarrow\mathcal{X}$ be
a non-expansive map on a metric space $\mathcal{X},$ and let \begin{equation}
d(\tau^{N}(x),\tau^{N}(y))<d(x,y)\quad\forall x,y\in\mathcal{X},\, x\neq y\end{equation}
 for some fixed $N\in\mathbb{N}.$ Then $\tau$ is an asymptotic deformation.
Then $\tau$ is an asymptotic deformation. 
\end{lemma}
\begin{proof}
The existence of the limit $\lim_{n\rightarrow\infty}d(\tau^{n}(x),\tau^{n}(y))$
follows from the monotonicity and the fact the any metric is lower
bounded. 
\end{proof}
\begin{remarknn}
Any weak contraction is an asymptotic deformation (with $N=1$). 
\end{remarknn}
\begin{framedtheorem}
[Asymptotic deformations]\label{thm:weakBanach} Let $\tau:\mathcal{X}\rightarrow\mathcal{X}$
be a continuous map on a compact metric space $\mathcal{X}$ with
at least one fixed point. Then $\tau$ is mixing if and only if $\tau$
is an asymptotic deformation. 
\end{framedtheorem}
\begin{proof}
Firstly assume that $\tau$ is an asymptotic deformation. Let $x_{*}$
be a fixed point and define $S(x)=d(x_{*},x).$\begin{eqnarray}
\lim_{n\rightarrow\infty}S(\tau^{n}(x)) & = & \lim_{n\rightarrow\infty}d(x_{*},\tau^{n}(x))\nonumber \\
 & = & \lim_{n\rightarrow\infty}d(\tau^{n}(x_{*}),\tau^{n}(x))\neq d(x_{*},x)=S(x)\quad\forall x\neq x_{*},\end{eqnarray}
 hence $S(x)$ is a generalised Lyapunov function. By Theorem \ref{thm:weaklyapov}
it follows that $\tau$ is mixing. Secondly, if $\tau$ is mixing,
then \begin{equation}
\lim_{n\rightarrow\infty}d(\tau^{n}(x),\tau^{n}(y))=d(x_{*},x_{*})=0\neq d(x,y)\quad\forall x,y\in\mathcal{X},\, x\neq y,\end{equation}
 so $\tau$ is an asymptotic deformation.
\end{proof}
\begin{remarknn}
Note that the existence of a fixed point is assured if $\tau$ is
a weak contraction on a compact space~\cite{STAKGOLD}, or if the
metric space is convex compact \cite{DUGUNDJI}. 
\end{remarknn}
As a special case, we get the following result:

\begin{corollary}
\label{cor:contraction1} Any weak contraction $\tau$ on a compact
metric space is mixing. 
\end{corollary}
\begin{proof}
Since the space is compact $\tau$ has at least one fixed point. Moreover
from Lemma~\ref{asymptlem} we know that $\tau$ is an asymptotic
deformation. Then Theorem~\ref{thm:weakBanach} applies. 
\end{proof}
\begin{remarknn}
This result can be seen as an instance of Banach contraction principle
on compact spaces. In the second part of the chapter we will present
a counterexample which shows that weak contractivity is only a sufficient
criterion for mixing (see Example~\ref{exm:mixing}). In the context
of quantum channels an analogous criterion was suggested in~\cite{STRICTCONTRATIONS,RAGINSKY}
which applied to strict contractions. We also note that for weak and
strict contractions, the trivial generalised Lyapunov function (Theorem
\ref{thm:glyap}) is a strict Lyapunov function. 
\end{remarknn}
Lemma~\ref{thm:ergodic} states the ergodic theorem by Birkhoff~\cite{BIRKHOFF}
which, in the context of normed vector spaces, shows the equivalence
between the definition of ergodicity of Eq.~(\ref{defergo10}) and
the standard time average definition. 

\begin{lemma}
\label{thm:ergodic}Let $\mathcal{X}$ be a convex and compact subset
of a normed vector space, and let $\tau:\mathcal{X}\rightarrow\mathcal{X}$
be a continuous map. If $\tau$ is ergodic with fixed point $x_{*},$
then \begin{eqnarray}
\lim_{n\rightarrow\infty}\frac{1}{n+1}\sum_{\ell=0}^{n}\tau^{\ell}(x)=x_{*}\;.\label{average}\end{eqnarray}

\end{lemma}
\begin{proof}
Define the sequence $A_{n}\equiv\frac{1}{n+1}\sum_{\ell=0}^{n}\tau^{\ell}(x)$.
Let then $M$ be the upper bound for the norm of vectors in $\mathcal{X}$,
i.e. $M\equiv\sup_{x\in\mathcal{X}}\| x\|<\infty$. which exists because
$\mathcal{X}$ is compact. The sequence $A_{n}$ has a convergent
subsequence $A_{n_{k}}$ with limit $\tilde{A}.$ Since $\tau$ is
continuous one has $\lim_{k\rightarrow\infty}\tau(A_{n_{k}})=\tau(\tilde{A})$.
On the other hand, we have\begin{equation}
\|\tau(A_{n_{k}})-A_{n_{k}}\|=\frac{1}{n_{k}+1}\|\tau^{n_{k}+1}(x)-x\|\leqslant\frac{\|\tau^{n_{k}+1}(x)\|+\| x\|}{n_{k}+1}\leqslant\frac{2M}{n_{k}+1},\end{equation}
 so the two sequences must have the same limit, i.e. $\tau(\tilde{A})=\tilde{A}$.
Since $\tau$ is ergodic, we have $\tilde{A}=x_{*}$ and $\lim_{n\rightarrow\infty}A_{n}=x_{*}$
by Lemma \ref{lem:subsequences}. 
\end{proof}
\begin{remarknn}
Note that if $\tau$ has a second fixed point $y_{*}\neq x_{*}$,
then for all $n$ one has $\frac{1}{n+1}\sum_{\ell=0}^{n}\tau^{\ell}(y_{*})=y_{*}$,
so Eq.~(\ref{average}) would not apply.
\end{remarknn}

\section{Quantum Channels}

\label{QC}

\begin{figure}[t]
\begin{centering}\includegraphics[width=1\textwidth]{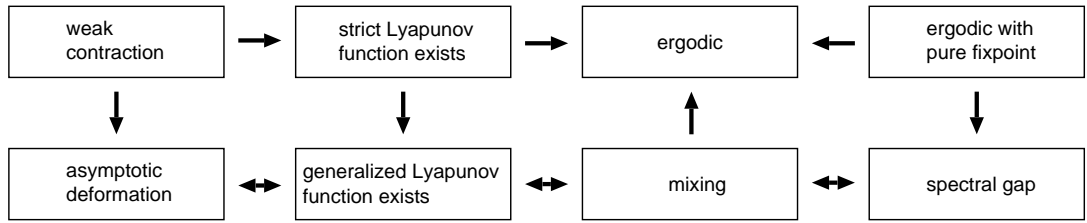}\par\end{centering}

\caption{\label{fig:relations}Relations between the different properties
of a quantum channel.}
\end{figure}

In this Section we discuss the mixing properties of quantum channels\index{quantum channel}~\cite{NIELSEN}
which account for the most general evolution a quantum system can
undergo including measurements and coupling with external environments.
In this context solving the mixing problem~(\ref{mixing0}) is equivalent
to determine if repetitive application of a certain physical transformation
will drive any input state of the system (i.e. its density matrices)
into a unique output configuration. The relationship between the different
mixing criteria one can obtain in this case is summarised in Fig.~\ref{fig:relations}.

At a mathematical level quantum channels correspond to linear maps
acting on the density operators $\rho$ of the system and satisfying
the requirement of being completely positive and trace preserving
(\index{CPT}CPT). For a formal definition of these properties we
refer the reader to~\cite{KRAUS,WERNER,KEYL}: here we note only
that a necessary and sufficient condition to being CPT is to allow
Kraus decomposition~\cite{KRAUS} or, equivalently, Stinespring dilation~\cite{STINE}.
Our results are applicable if the underlying Hilbert space is finite-dimensional.
In such regime there is no ambiguity in defining the convergence of
a sequence since all operator norms are equivalent (i.e. given two
norms one can construct an upper and a lower bound for the first one
by properly scaling the second one). Also the set of bounded operators
and the set of operators of Hilbert-Schmidt class coincide. For the
sake of definiteness, however, we will adopt the trace-norm which,
given the linear operator $\Theta:\mathcal{H}\rightarrow\mathcal{H}$,
is defined as $\|\Theta\|_{1}=\mbox{Tr}[\sqrt{\Theta^{\dag}\Theta}]$
with $\mbox{Tr}[\cdots]$ being the trace over $\mathcal{H}$ and
$\Theta^{\dag}$ being the adjoint of $\Theta$. This choice is in
part motivated by the fact~\cite{RUSKAI} that any quantum channel
is non-expansive with respect to the metric induced%
\footnote{This is just the trace distance $d(\rho,\sigma)=\|\rho-\sigma\|_{1}$.%
} by $\|\cdot\|_{1}$ (the same property does not necessarily apply
to other operator norms, e.g. the Hilbert-Schmidt norm, also when
these are equivalent to $\|\cdot\|_{1}$).

We start by showing that the mixing criteria discussed in the first
half of the chapter do apply to the case of quantum channel. Then
we will analyse these maps by studying their linear extensions in
the whole vector space formed by the linear operators of $\mathcal{H}$.%

\subsection{Mixing criteria for Quantum Channels}

\label{sec:mixing}

Let $\mathcal{H}$ be a finite dimensional Hilbert space and let $\mathcal{S}(\mathcal{H})$
be the set of its density matrices $\rho$. The latter is a convex
and compact subset of the larger normed vector space $\mathcal{L}(\mathcal{H})$
composed by the linear operators $\Theta:\mathcal{H}\rightarrow\mathcal{H}$
of $\mathcal{H}$. From this and from the fact that CPT maps are continuous
(indeed they are linear) it follows that for a quantum channel there
always exists at least one density operator which is a fixed point~\cite{TERHAL}.
It also follows that all the results of the previous section apply
to quantum channels. In particular Lemma~\ref{lem:hausdorff} holds,
implying that any mixing quantum channel must be ergodic. The following
example shows, however, that it is possible to have ergodic quantum
channels which are not mixing.

\begin{example}
\label{exm:ergodic} Consider the qubit quantum channel $\tau$ obtained
by cascading a completely decoherent channel with a NOT gate. Explicitly
$\tau$ is defined by the transformations $\tau(|0\rangle\langle0|)=|1\rangle\langle1|$,
$\tau(|1\rangle\langle1|)=|0\rangle\langle0|$, and $\tau(|0\rangle\langle1|)=\tau(|1\rangle\langle0|)=0$
with $|0\rangle,|1\rangle$ being the computational basis of the qubit.
This map is ergodic with fixed point given by the completely mixed
state $(|0\rangle\langle0|+|1\rangle\langle1|)/2$. However it is
trivially not mixing since, for instance, repetitive application of
$\tau$ on $|0\rangle\langle0|$ will oscillate between $|0\rangle\langle0|$
and $|1\rangle\langle1|$. 
\end{example}
Theorems~\ref{thm:weakBanach} implies that a quantum channel $\tau:\mathcal{S}(\mathcal{H})\rightarrow\mathcal{S}(\mathcal{H})$
is mixing if and only if it is an asymptotic deformation. As already
pointed out in the introduction, this property is \emph{metric independent}
(as opposed to contractivity). Alternatively, if the fixed point of
a quantum channel is known, then one may use the trivial generalised
Lyapunov function (Theorem~\ref{thm:glyap}) to check if it is mixing.
However both criteria depend on the metric distance, which usually
has no easy physical interpretation. A more useful choice is the quantum
relative entropy\index{quantum relative entropy}, which is defined
as \begin{eqnarray}
H(\rho,\sigma)\equiv\textrm{Tr}\rho(\log\rho-\log\sigma).\end{eqnarray}
 The quantum relative entropy is continuous in finite dimension \cite{JENS}
and can be used as a measure of \emph{distance} (though it is not
a metric). It is finite if the support of $\rho$ is contained in
the support of $\sigma.$ To ensure that it is a continuous function
on a compact space, we choose $\sigma$ to be faithful:

\begin{theorem}
[Relative entropy criterion]A quantum channel with faithful fixed
point $\rho_{*}$ is mixing if and only if the quantum relative entropy
with respect to $\rho_{*}$ is a generalised Lyapunov function.
\end{theorem}
\begin{proof}
Because of Theorem~\ref{thm:weaklyapov} we only need to prove the
second part of the thesis, i.e. that mixing channels admit the quantum
relative entropy with respect to the fixed point, $S(\rho)\equiv H(\rho,\rho_{*})$,
as a generalised Lyapunov function. Firstly notice that the quantum
relative entropy is monotonic under quantum channels~\cite{RUSKAI2,RUSKAI2b}.
Therefore the limit $S_{*}(\rho)\equiv\lim_{n\rightarrow\infty}S\left(\tau^{n}(\rho)\right)$
does exist and satisfies the condition $S_{*}(\rho)\geqslant S(\rho)$.
Suppose now there exists a $\rho$ such that $S_{*}(\rho)=S(\rho)$.
Because $\tau$ is mixing and $S$ is continuous we have \begin{equation}
S(\rho)=S_{*}(\rho)=\lim_{n\rightarrow\infty}S\left(\tau^{n}(\rho)\right)=S(\rho_{*})=0,\end{equation}
 and hence $H(\rho,\rho_{*})=0$. Since $H(\rho,\sigma)=0$ if and
only if $\rho=\sigma$ it follows that $S$ is a Lyapunov function
around $\rho_{*}$. 
\end{proof}
Another important investigation tool is Corollary~\ref{cor:contraction1}:
weak contractivity of a quantum channel is a sufficient condition
for mixing. As already mentioned in the previous section, unfortunately
this not a necessary condition. Here we present an explicit counterexample
based on a quantum channel introduced in Ref.~\cite{TERHAL}.

\begin{example}
\label{exm:mixing} Consider a three-level quantum system characterised
by the orthogonal vectors $|0\rangle,|1\rangle,|2\rangle$ and the
quantum channel $\tau$ defined by the transformations $\tau(|2\rangle\langle2|)=|1\rangle\langle1|$,
$\tau(|1\rangle\langle1|)=\tau(|0\rangle\langle0|)=|0\rangle\langle0|$,
and $\tau(|i\rangle\langle j|)=0$ for all $i\neq j$. Its easy to
verify that after just two iterations any input state $\rho$ will
be transformed into the vector $|0\rangle\langle0|$. Therefore the
map is mixing. On the other hand it is explicitly not a weak contraction
with respect to the trace norm since, for instance, one has \begin{equation}
\|\;\tau(|2\rangle\langle2|)-\tau(|0\rangle\langle0|)\;\|_{1}=\|\;|1\rangle\langle1|-|0\rangle\langle0|\;\|_{1}=\|\;|2\rangle\langle2|-|0\rangle\langle0|\;\|_{1}\;,\end{equation}
 where in the last identity we used the invariance of $\|\cdot\|_{1}$
with respect to unitary transformations. 
\end{example}

\subsection{Beyond the density matrix operator space: spectral properties}

\label{sec:newmixing}

Exploiting linearity quantum channels can be extended beyond the space
$\mathcal{S}(\mathcal{H})$ of density operators to become maps defined
on the full vector space $\mathcal{L}(\mathcal{H})$ of the linear
operators of the system, in which basic linear algebra results hold.
This allows one to simplify the analysis even though the mixing property~(\ref{mixing0})
is still defined with respect to the density operators of the system.

Mixing conditions for quantum channels can be obtained by considering
the structure of their eigenvectors in the extended space $\mathcal{L}(\mathcal{H})$.
For example, it is easily shown that the spectral radius~\cite{HORNJOHNSON}
of any quantum channel is equal to unity~\cite{TERHAL}, so its eigenvalues
are contained in the unit circle. The eigenvalues $\lambda$ on the
unit circle (i.e. $|\lambda|=1$) are referred to as \emph{peripheral
eigenvalues\index{peripheral eigenvalues}.} Also, as already mentioned,
since $\mathcal{S}(\mathcal{H})$ is compact and convex, CPT maps
have always at least one fixed point which is a density matrix~\cite{TERHAL}.

\begin{theorem}
[Spectral gap criterion]\label{thm:peri}\label{thm:necessary} Let
$\tau$ be a quantum channel. $\tau$ is mixing if and only if its
only peripheral eigenvalue is $1$ and this eigenvalue is simple.
\end{theorem}
\begin{proof}
The ''if'' direction of the proof is a well known result from linear
algebra (see for example~\cite[Lemma 8.2.7]{HORNJOHNSON}). Now let
us assume $\tau$ is mixing towards $\rho_{*}.$ Let $\Theta$ be
a generic operator in $\mathcal{L}(\mathcal{H})$. Then $\Theta$
can be decomposed in a finite set of non-orthogonal density operators%
\footnote{To show that this is possible, consider an arbitrary operator basis
of $\mathcal{L}(\mathcal{H})$. If $N$ is the finite dimension of
$\mathcal{H}$ the basis will contain $N^{2}$ elements. Each element
of the basis can then be decomposed into two Hermitian operators,
which themselves can be written as linear combinations of at most
$N$ projectors. Therefore there exists a generating set of at most
$2N^{3}$ positive operators, which can be normalised such that they
are quantum states. There even exists a basis (i.e. a minimal generating
set) consisting of density operators, but in general it cannot be
orthogonalised.%
}, i.e. $\Theta=\sum_{\ell}c_{\ell}\rho_{\ell}$, with $\rho_{\ell}\in\mathcal{S}(\mathcal{H})$
and $c_{\ell}$ complex. Since $\textrm{Tr}\left[\rho_{\ell}\right]=1$,
we have have $\textrm{Tr}\left[\Theta\right]=\sum_{\ell}c_{\ell}$.
Moreover since $\tau$ is mixing we have $\lim_{n\rightarrow\infty}\tau^{n}\left(\rho_{\ell}\right)=\rho_{*}$
for all $\ell$, with convergence with respect to the trace-norm.
Because of linearity this implies \begin{eqnarray}
\lim_{n\rightarrow\infty}\tau^{n}\left(\Theta\right)=\sum_{\ell}c_{\ell}\;\rho_{*}=\textrm{Tr}\left[\Theta\right]\;\rho_{*}\;.\label{limit}\end{eqnarray}
 If there existed any other eigenvector $\Theta_{*}$ of $\tau$ with
eigenvalue on the unit circle, then $\lim_{n\rightarrow\infty}\tau^{n}(\Theta_{*})$
would not satisfy Eq.~(\ref{limit}). 
\end{proof}
The speed of convergence can also be estimated by~\cite{TERHAL}
\begin{eqnarray}
\|\tau^{n}\left(\rho\right)-\rho_{*}\|_{1}\;\leqslant C_{N}\; n^{N}\;\kappa^{n}\;,\label{speed}\end{eqnarray}
 where $N$ is the dimensionality of the underlying Hilbert space,
$\kappa$ is the second largest eigenvalue of $\tau$, and $C_{N}$
is some constant depending only on $N$ and on the chosen norm. Hence,
for $n\gg N$ the convergence becomes exponentially fast. As mentioned
in~\cite{RAGINSKY}, the criterion of Theorem~\ref{thm:peri} is
in general difficult to check. This is because one has to find all
eigenvalues of the quantum channel, which is hard especially in the
high dimensional case. Also, if one only wants to check if a particular
channel is mixing or not, then the amount of information obtained
is much higher than the required amount.

\begin{example}
As an application consider the non mixing CPT map of Example~\ref{exm:ergodic}.
One can verify that apart from the eigenvalue $1$ associated with
its fixed point (i.e. the completely mixed state), it possess another
peripheral eigenvalue. This is $\lambda=-1$ which is associated with
the Pauli operator $|0\rangle\langle0|-|1\rangle\langle1|$. 
\end{example}
\begin{corollary}
\label{cor:speed} The convergence speed of any mixing quantum channel
is exponentially fast for sufficiently high values of $n$. 
\end{corollary}
\begin{proof}
From Theorem~\ref{thm:necessary} mixing channels have exactly one
peripheral eigenvalue, which is also simple. Therefore the derivation
of Ref.~\cite{TERHAL} applies and Eq.~(\ref{speed}) holds. 
\end{proof}
This result should be compared with the case of strictly contractive
quantum channels whose convergence was shown to be exponentially fast
along to whole trajectory~\cite{RAGINSKY,STRICTCONTRATIONS}.

\subsection{Ergodic channels with pure fixed points\index{pure fix-points}\label{sub:Ergodic-channels-with}}

\label{sec:pure}

An interesting class of ergodic quantum channel is formed by those
CPT maps whose fixed point is a \emph{pure} density matrix. Among
them we find for instance the maps employed in the communication protocols
discussed in this thesis or those of the purification schemes of Refs.~\cite{YUASA2,YUASA1}.
We will now show that within this particular class, ergodicity and
mixing are indeed equivalent properties.

We first need the following Lemma, which discusses a useful property
of quantum channels (see also~\cite{SCHRADER}).

\begin{lemma}
\label{lem:OBS5} Let $\tau$ be a quantum channel and $\Theta$ be
an eigenvector of $\tau$ with peripheral eigenvalue $\lambda=e^{i\varphi}$.
Then, given $g=\mbox{\emph{Tr}}\left[\sqrt{\Theta^{\dag}\Theta}\right]>0$,
the density matrices $\rho=\sqrt{\Theta\Theta^{\dag}}/g$ and $\sigma=\sqrt{\Theta^{\dag}\Theta}/g$
are fixed points of $\tau$. 
\end{lemma}
\begin{proof}
Use the left polar decomposition to write $\Theta=g\;\rho U$ where
$U$ is a unitary operator. The operator $\rho U$ is clearly an eigenvector
of $\tau$ with eigenvalue $e^{i\varphi}$, i.e. \begin{eqnarray}
\tau(\rho U)=\lambda\;\rho U\;.\label{uuu}\end{eqnarray}
 Hence introducing a Kraus set $\{ K_{n}\}_{n}$ of $\tau$~\cite{KRAUS}
and the spectral decomposition of the density matrix $\rho=\sum_{j}p_{j}|\psi_{j}\rangle\langle\psi_{j}|$
with $p_{j}>0$ being its positive eigenvalues, one gets \begin{eqnarray}
\lambda=\mbox{Tr}[\tau(\rho U)U^{\dag}]=\sum_{j,\ell,n}p_{j}\langle\phi_{\ell}|K_{n}|\psi_{j}\rangle\langle\psi_{j}|UK_{n}^{\dag}U^{\dag}|\phi_{\ell}\rangle\;,\end{eqnarray}
 where the trace has been performed with respect to an orthonormal
basis $\{|\phi_{\ell}\rangle\}_{\ell}$ of $\mathcal{H}$. Taking
the absolute values of both terms gives \begin{eqnarray}
|\lambda| & = & |\sum_{j,\ell,n}p_{j}\langle\phi_{\ell}|K_{n}|\psi_{j}\rangle\langle\psi_{j}|UK_{n}^{\dag}U^{\dag}|\phi_{\ell}\rangle|\nonumber \\
 & \leqslant & \sqrt{\sum_{j,\ell,n}p_{j}\langle\phi_{\ell}|K_{n}|\psi_{j}\rangle\langle\psi_{j}|K_{n}^{\dag}|\phi_{\ell}\rangle}\sqrt{\sum_{j,\ell,n}p_{j}\langle\phi_{\ell}|UK_{n}U^{\dag}|\psi_{j}\rangle\langle\psi_{j}|UK_{n}^{\dag}U^{\dag}|\phi_{\ell}\rangle}\nonumber \\
 & = & \sqrt{\mbox{Tr}[\tau(\rho)}]\sqrt{\mbox{Tr}[\tilde{\tau}(\rho)]}=1,\end{eqnarray}
 where the inequality follows from the Cauchy-Schwartz inequality.
The last identity instead is a consequence of the fact that the transformation
$\tilde{\tau}(\rho)=U\tau(U^{\dag}\rho U)U^{\dag}$ is CPT and thus
trace preserving. Since $|\lambda|=1$ it follows that the inequality
must be replaced by an identity. This happens if and only if there
exist $e^{i\vartheta}$ such that \begin{eqnarray}
\sqrt{p_{j}}\{\langle\phi_{\ell}|K_{n}|\psi_{j}\rangle\}^{*}=\sqrt{p_{j}}\langle\psi_{j}|K_{n}^{\dag}|\phi_{\ell}\rangle=e^{i\vartheta}\sqrt{p_{j}}\langle\psi_{j}|UK_{n}^{\dag}U^{\dag}|\phi_{\ell}\rangle\;,\end{eqnarray}
 for all $j,\ell$ and $n$. Since the $|\phi_{\ell}\rangle$ form
a basis of $\mathcal{H}$, and $p_{j}>0$ this implies \begin{eqnarray}
\langle\psi_{j}|K_{n}^{\dag}=e^{i\vartheta}\;\langle\psi_{j}|UK_{n}^{\dag}U^{\dag}\quad\Rightarrow\quad\langle\psi_{j}|UK_{n}^{\dag}=e^{-i\vartheta}\;\langle\psi_{j}|K_{n}^{\dag}U\;,\end{eqnarray}
 for all $n$ and for all the not null eigenvectors $|\psi_{j}\rangle$
of $\rho$. This yields \begin{eqnarray}
\tau(\rho U) & = & \sum_{j}p_{j}\sum_{n}K_{n}|\psi_{j}\rangle\langle\psi_{j}|UK_{n}^{\dag}=e^{-i\vartheta}\;\sum_{j}p_{j}\sum_{n}K_{n}|\psi_{j}\rangle\langle\psi_{j}|K_{n}^{\dag}U\nonumber \\
 & = & e^{-i\vartheta}\;\tau(\rho)U\end{eqnarray}
 which, replaced in (\ref{uuu}) gives $e^{-i\vartheta}\;\tau(\rho)=e^{i\varphi}\;\rho$,
whose only solution is $e^{-i\vartheta}=e^{i\varphi}$. Therefore
$\tau(\rho)=\rho$ and $\rho$ is a fixed point of $\tau$. The proof
for $\sigma$ goes along similar lines: simply consider the right
polar decomposition of $\Theta$ instead of the left polar decomposition. 
\end{proof}
\begin{corollary}
Let $\tau$ be an ergodic quantum channel. It follows that its eigenvectors
associated with peripheral eigenvalues are normal operators. 
\end{corollary}
\begin{proof}
Let $\Theta$ be an eigenoperator with peripheral eigenvalue $e^{i\varphi}$
such that $\tau\left(\Theta\right)=e^{i\varphi}\;\Theta$. By Lemma
\ref{lem:OBS5} we know that, given $g=\mbox{Tr}\left[\sqrt{\Theta^{\dag}\Theta}\right]$
the density matrices $\rho=\sqrt{\Theta\Theta^{\dag}}/g$ and $\sigma=\sqrt{\Theta^{\dag}\Theta}/g$
must be fixed points of $\tau$. Since the map is ergodic we must
have $\rho=\sigma$, i.e. $\Theta\Theta^{\dag}=\Theta^{\dag}\Theta$.
\end{proof}
\begin{framedtheorem}
[Purely ergodic maps]\label{thm:ergodicmixing}Let $|\psi_{1}\rangle\langle\psi_{1}|$
be the pure fixed point of an ergodic quantum channel $\tau$. It
follows that $\tau$ is mixing. 
\end{framedtheorem}
\begin{proof}
We will use the spectral gap criterion showing that $|\psi_{1}\rangle\langle\psi_{1}|$
is the only peripheral eigenvector of $\tau$. Assume in fact that
$\Theta\in L(\mathcal{H})$ is a eigenvector of $\tau$ with peripheral
eigenvalue, i.e. \begin{eqnarray}
\tau\left(\Theta\right)=e^{i\varphi}\Theta\;.\label{identity1}\end{eqnarray}
 From Lemma~\ref{lem:OBS5} we know that the density matrix \begin{equation}
\rho=\sqrt{\Theta\Theta^{\dag}}/g,\end{equation}
with $g=\mbox{Tr}\left[\sqrt{\Theta^{\dag}\Theta}\right]>0$, must
be a fixed point of $\tau$. Since this is an ergodic map we must
have $\rho=|\psi_{1}\rangle\langle\psi_{1}|$. This implies $\Theta=g|\psi_{1}\rangle\langle\psi_{2}|$,
with $|\psi_{2}\rangle$ some normalised vector of $\mathcal{H}$.
Replacing it into Eq.~(\ref{identity1}) and dividing both terms
by $g$ yields $\tau\left(|\psi_{1}\rangle\langle\psi_{2}|\right)=e^{i\varphi}|\psi_{1}\rangle\langle\psi_{2}|$
and \begin{eqnarray}
|\langle\psi_{1}|\tau(|\psi_{1}\rangle\langle\psi_{2}|)|\psi_{2}\rangle|=1\;.\end{eqnarray}
 Introducing a Kraus set $\{ K_{n}\}_{n}$ of $\tau$ and employing
Cauchy-Schwartz inequality one can then write \begin{eqnarray}
1 & = & |\langle\psi_{1}|\tau(|\psi_{1}\rangle\langle\psi_{2}|)|\psi_{2}\rangle|=|\sum_{n}\langle\psi_{1}|K_{n}|\psi_{1}\rangle\langle\psi_{2}|K_{n}^{\dag}|\psi_{2}\rangle|\\
 & \leqslant & \sqrt{\sum_{n}\langle\psi_{1}|K_{n}|\psi_{1}\rangle\langle\psi_{1}|K_{n}^{\dag}|\psi_{1}\rangle}\sqrt{\sum_{n}\langle\psi_{2}|K_{n}|\psi_{2}\rangle\langle\psi_{2}|K_{n}^{\dag}|\psi_{2}\rangle}\nonumber \\
 & = & \sqrt{\langle\psi_{1}|\tau(|\psi_{1}\rangle\langle\psi_{1}|)|\psi_{1}\rangle}\sqrt{\langle\psi_{2}|\tau(|\psi_{2}\rangle\langle\psi_{2}|)|\psi_{2}\rangle}=\sqrt{\langle\psi_{2}|\tau(|\psi_{2}\rangle\langle\psi_{2}|)|\psi_{2}\rangle}\;,\nonumber \end{eqnarray}
 where we used the fact that $|\psi_{1}\rangle$ is the fixed point
of $\tau$. Since $\tau$ is CPT the quantity $\langle\psi_{2}|\tau(|\psi_{2}\rangle\langle\psi_{2}|)|\psi_{2}\rangle$
is upper bounded by $1$. Therefore in the above expression the inequality
must be replaced by an identity, i.e. \begin{eqnarray}
\langle\psi_{2}|\tau(|\psi_{2}\rangle\langle\psi_{2}|)|\psi_{2}\rangle=1\qquad\Longleftrightarrow\qquad\tau(|\psi_{2}\rangle\langle\psi_{2}|)=|\psi_{2}\rangle\langle\psi_{2}|\;.\end{eqnarray}
 Since $\tau$ is ergodic, we must have $|\psi_{2}\rangle\langle\psi_{2}|=|\psi_{1}\rangle\langle\psi_{1}|$.
Therefore $\Theta\propto|\psi_{1}\rangle\langle\psi_{1}|$ which shows
that $|\psi_{1}\rangle\langle\psi_{1}|$ is the only eigenvector of
$\tau$ with peripheral eigenvalue of. 
\end{proof}
An application of the previous Theorem is obtained as follows.

\begin{lemma}
\label{lem:add} Let $M_{AB}=M_{A}\otimes1_{B}+1_{A}\otimes M_{B}$
be an observable of the composite system $\mathcal{H}_{A}\otimes\mathcal{H}_{B}$
and $\tau$ the CPT linear map on $\mathcal{H}_{A}$ of Stinespring
form~\emph{\cite{STINE}} \begin{equation}
\tau(\rho)=\emph{\mbox{Tr}}_{B}\left[U\left(\rho\otimes|\phi\rangle_{B}\langle\phi|\right)U^{\dag}\right]\;,\label{eq:rep}\end{equation}
 (here $\emph{\mbox{Tr}}_{X}\left[\cdots\right]$ is the partial trace
over the system $X$, and $U$ is a unitary operator of $\mathcal{H}_{A}\otimes\mathcal{H}_{B}$).
Assume that $\left[M_{AB},U\right]=0$ and that $|\phi\rangle_{B}$
is the eigenvector corresponding to a non-degenerate maximal or minimal
eigenvalue of $M_{B}.$ Then $\tau$ is mixing if and only if $U$
has one and only one eigenstate that factorises as $|\nu\rangle_{A}\otimes|\phi\rangle_{B}.$
\end{lemma}
\begin{proof}
Let $\rho$ be an arbitrary fixed point of $\tau$ (since $\tau$
is CPT it has always at least one), i.e. $\textrm{Tr}_{B}\left[U\left(\rho\otimes|\phi\rangle_{B}\langle\phi|\right)U^{\dag}\right]=\rho$.
Since $M_{AB}$ is conserved and $\textrm{Tr}_{A}\left[M_{A}\rho\right]=\textrm{Tr}_{A}\left[M_{A}\tau(\rho)\right]$,
the system $B$ must remain in the maximal state, which we have assumed
to be unique and pure, i.e. \begin{equation}
U\left(\rho\otimes|\phi\rangle_{B}\langle\phi|\right)U^{\dag}=\rho\otimes|\phi\rangle_{B}\langle\phi|\qquad\Longrightarrow\qquad\left[U,\rho\otimes|\phi\rangle_{B}\langle\phi|\right]=0\;.\end{equation}
 Thus there exists a orthonormal basis $\left\{ |u_{k}\rangle\right\} _{k}$
of $\mathcal{H}_{A}\otimes\mathcal{H}_{B}$ diagonalising simultaneously
both $U$ and $\rho\otimes|\phi\rangle_{B}\langle\phi|$. We express
the latter in this basis, i.e. $\rho\otimes|\phi\rangle_{B}\langle\phi|=\sum_{k}p_{k}|u_{k}\rangle\langle u_{k}|$
with $p_{k}>0$, and compute the von Neumann entropy of subsystem
$B$. This yields \begin{eqnarray}
0 & = & H(|\phi\rangle_{B}\langle\phi|)=H\left(\textrm{Tr}_{A}\left[\sum_{k}p_{k}|u_{k}\rangle\langle u_{k}|\right]\right)\geqslant\sum_{k}p_{k}\; H\left(\textrm{Tr}_{A}\left[|u_{k}\rangle\langle u_{k}|\right]\right)\;.\end{eqnarray}
 From the convexity of the von Neumann entropy the above inequality
leads to a contradiction unless $\textrm{Tr}_{A}\left[|u_{k}\rangle\langle u_{k}|\right]=|\phi\rangle_{B}\langle\phi|$
for all $k$. The $|u_{k}\rangle$ must therefore be factorising,
\begin{equation}
|u_{k}\rangle=|\nu_{k}\rangle_{A}\otimes|\phi\rangle_{B}.\label{eq:fact}\end{equation}
 If the factorising eigenstate of $U$ is unique, it must follow that
$\rho=|\nu\rangle\langle\nu|$ for some $|\nu\rangle$ and that $\tau$
is ergodic. By Theorem \ref{thm:ergodicmixing} it then follows that
$\tau$ is also mixing. If on the other hand there exists more than
one factorising eigenstate, than all states of the form of Eq. (\ref{eq:fact})
correspond to a fixed point $\rho_{k}=|\nu_{k}\rangle\langle\nu_{k}|$
and $\tau$ is neither ergodic nor mixing. 
\end{proof}
\begin{remark}
An application of this Lemma is the protocol for read and write access
by local control discussed in the next chapter.
\end{remark}

\section{Conclusion}

\label{CONCLUSION} In reviewing some known results on the mixing
property of continuous maps, we obtained a stronger version of the
direct Lyapunov method. For compact metric spaces (including quantum
channels operating over density matrices) it provides a necessary
and sufficient condition for mixing. Moreover it allows us to prove
that asymptotic deformations with at least one fixed point must be
mixing.

In the specific context of quantum channels we employed the generalised
Lyapunov method to analyse the mixing properties. Here we also analysed
different mixing criteria. In particular we have shown that an ergodic
quantum channel with a pure fixed point is also mixing.

\chapter{Read and write access by local control\label{cha:Full-read-and}}

\section{Introduction}

The unitarity of Quantum Mechanics implies that information is conserved.
Whatever happens to a quantum system - as long as it is unitary, the
original state can in principle be recovered by applying the inverse
unitary transformation. However it is well known that in open quantum
systems~\cite{OPENQUANTUM} the reduced dynamics is no longer unitary.
The reduced dynamics is described by a completely positive, trace
preserving maps, and we have seen in the last chapter that there are
extreme examples, namely \emph{mixing} maps, where all information
about the initial state is eventually lost. Where has it gone? If
the whole system evolves unitary, then this information must have
been transferred in the \emph{correlations} between reduced system
and environment~\cite{Hayden2004}, and/or in the environment. We
can see that this may be useful for quantum state transfer, in particular
the case where all information is transferred into the ''environment'',
which could be another quantum system (the receiver). A particularly
useful case is given by mixing maps with pure convergence points,
because a pure state cannot be correlated, and because we have a simple
convergence criterion in this case (Subsection~\ref{sub:Ergodic-channels-with}).
This is an example of \emph{homogenisation\index{homogenisation}}~\cite{HOMOGENIZATION1,HOMOGENIZATION2}\emph{.}
Furthermore, if the mixing property arises from some operations, we
can expect that by applying the inverse operations, information can
also be transferred back to the system. This property was used in~\cite{WELLENS,Wellens2002}
to generate arbitrary states of a cavity field by sending atoms through
the cavity. The crucial difference is that in our system control is
only assumed to be available on a subsystem (such as, for example,
the ends of a quantum chain). Hence we will show in this chapter how
arbitrary quantum states can be written to (i.e. prepared on) a large
system, and read from it, by \emph{local} control only. This is similar
in spirit to universal quantum interfaces~\cite{Lloyd2004}, but
our different approach allows us to specify explicit protocols and
to give lower bounds for fidelities. We also demonstrate how this
can be used to significantly improve the quantum communication between
two parties if the receiver is allowed to store the received signals
in a quantum memory before decoding them. In the limit of an infinite
memory, the transfer is perfect. We prove that this scheme allows
the transfer of arbitrary multi-partite states along Heisenberg chains
of spin-$1/2$ particles with random coupling strengths.

Even though the convergence of a mixing map is essentially exponentially
fast (Corollary~\ref{cor:speed}), we still have to deal with infinite
limits. Looking at the environment this in turn would require to study
states on an infinite dimensional Hilbert space, and unfortunately
this can introduce many mathematical difficulties. We are mainly interested
in bounds for the finite case: if the protocol stops after finitely
many steps, what is the fidelity of the reading/writing? Which encoding
and decoding operations must be applied? By stressing on these questions,
we can actually avoid the infinite dimensional case, but the price
we have to pay is that our considerations become a bit technically
involved.

\section{Protocol\label{sec:Protocol}}

\begin{figure}[t]
\begin{centering}\includegraphics[width=0.8\columnwidth]{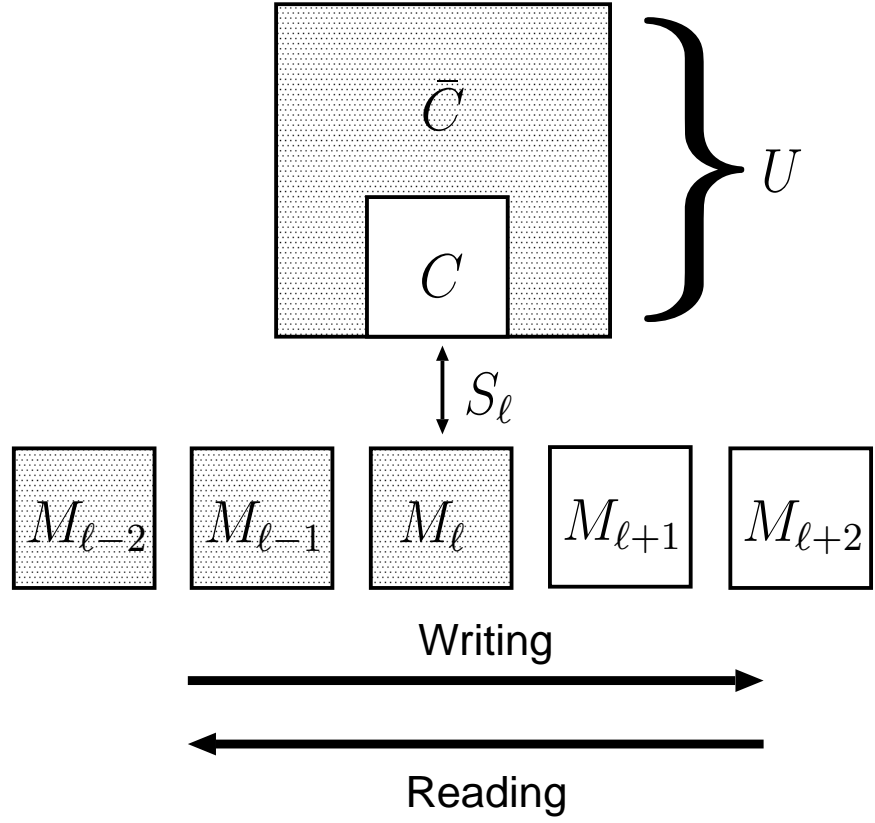}\par\end{centering}

\caption{\label{fig:memory}The system $C\bar{C}$ can only be controlled
by acting on a (small) subsystem $C.$ However system $C$ is coupled
to system $\bar{C}$ by a unitary operator $U=\exp\left\{ -iHt\right\} .$
This coupling can - in some cases - \emph{mediate} the local control
on $C$ to the full system $C\bar{C}.$ In our case, system $C$ is
controlled by performing regular swap operations $S_{\ell}$ between
it and a quantum memory $M_{\ell}.$ }
\end{figure}

We consider a tripartite finite dimensional Hilbert space given by
$\mathcal{H}=\mathcal{H}_{C}\otimes\mathcal{H}_{\bar{C}}\otimes\mathcal{H}_{M}.$
We assume that full control (the ability to prepare states and apply
unitary transformations) is possible on system $C$ and $M,$ but
no control is available on system $\bar{C}.$ However, we assume that
$C$ and $\bar{C}$ are coupled by some time-independent Hamiltonian
$H.$ We show here that under certain assumptions, if the system $C\bar{C}$
is initialised in some arbitrary state we can transfer (''read'')
this state into the system $M$ by applying some operations between
$M$ and $C.$ Likewise, by initialising the system $M$ in the correct
state, we can prepare (''write'') arbitrary states on the system
$C\bar{C.}$ The system $M$ functions as a \emph{quantum memory}\index{quantum memory}
and must be at least as large as the system $C\bar{C}.$ As sketched
in Fig.~\ref{fig:memory} we can imagine it to be split into sectors
$M_{\ell},$ I.e.. \begin{equation}
\mathcal{H}_{M}=\bigotimes_{\ell=1}^{L}\mathcal{H}_{M_{\ell}}\end{equation}
with\begin{equation}
\textrm{dim}\mathcal{H}_{M_{\ell}}=\textrm{dim}\mathcal{H}_{C}.\end{equation}
For the reading case, we assume that the memory is initialised in
the state \begin{equation}
|0\rangle_{M}\equiv\bigotimes_{\ell}|0\rangle_{M_{\ell}}\end{equation}
where $|0\rangle$ can stand for some generic state%
\footnote{Later on we will give an example where $|0\rangle$ represents a multi-qubit
state with all qubits aligned, but here we don't need to assume this.%
}. Like in the multi rail protocols considered in Chapter~\ref{cha:Multi-rail-and-Capacity},
we let the system evolve for a while, perform an operation, let it
evolve again and so forth, only that now the operation is not a measurement,
but a \emph{unitary gate}. More specifically, at step $\ell$ of the
protocol we perform a unitary swap $S_{\ell}$ between system $C$
and systems $M_{\ell}.$ After the $L$th swap operation the protocol
stops. The protocol for reading is thus represented by the unitary
operator \begin{equation}
W\equiv S_{L}US_{L-1}U\cdots S_{\ell}U\cdots S_{1}U,\end{equation}
where $U\in\mathcal{L}(\mathcal{H}_{C\bar{C}})$ is the time-evolution
operator $U=\exp\left\{ -iHt\right\} $ for some fixed time interval
$t.$ As we will see in the next section, the reduced evolution of
the system $\bar{C}$ under the protocol can be expressed in terms
of the CPT map \begin{equation}
\tau(\rho_{\bar{C}})\equiv\textrm{tr}_{C}\left[U\left(\rho_{\bar{C}}\otimes|0\rangle_{C}\langle0|\right)U^{\dag}\right],\end{equation}
where $|0\rangle_{C}$ is the state that is swapped in from the memory.
Our main assumption now is that $\tau$ is ergodic with a pure fixed
point (which we denote as $|0\rangle_{\bar{C}}$). By Theorem~\ref{thm:ergodicmixing}
this implies that $\tau$ is mixing, and therefore asymptotically
all information is transferred into the memory. %

For writing states on the system, we just make use of the unitarity
of $W.$ Roughly speaking, we initialise the memory in the state that
it \emph{would have ended up in} after applying $W$ if system $C\bar{C}$
had started in the state we want to initialise. Then we apply the
\emph{inverse} of $W$ given by\begin{equation}
W^{\dag}=U^{\dag}S_{1}\cdots U^{\dag}S_{\ell}\cdots U^{\dag}S_{L-1}U^{\dag}S_{L}.\end{equation}
We will see in Section~\ref{sec:Coding-transformation} how this
gives rise to a unitary coding transformation on the memory system,
such that arbitrary and unknown states can be initialised on the system.
The reader has probably noticed that the inverse of $W$ is generally
unphysical in the sense that it requires backward time evolution,
i.e. one has to wait \emph{negative} time steps between the swaps.
But we will see later how this can be fixed by a simple transformation.
For the moment, we just assume that $W^{\dag}$ is physical.

\section{Decomposition equations}

In this section we give a decomposition of the state after applying
the protocol which will allow us to estimate the fidelities for state
transfer in terms of the mixing properties of the map $\tau.$ Let
$|\psi\rangle_{C\bar{C}}\in\mathcal{H}_{C\bar{C}}$ be an arbitrary
state. We notice that the $C$ component of $W|\psi\rangle_{C\bar{C}}|0\rangle_{M}$
is always $|0\rangle_{C}$. Therefore we can decompose it as follows
\begin{equation}
W|\psi\rangle_{C\bar{C}}|0\rangle_{M}=|0\rangle_{C}\otimes\left[\sqrt{\eta}|0\rangle_{\bar{C}}|\phi\rangle_{M}+\sqrt{1-\eta}|\Delta\rangle_{\bar{C}M}\right]\label{eq:main}\end{equation}
 with $|\Delta\rangle_{\bar{C}M}$ being a normalised vector of $\bar{C}$
and $M$ which satisfies the identity \begin{eqnarray}
_{\bar{C}}\langle0|\Delta\rangle_{\bar{C}M}=0\;.\label{impo}\end{eqnarray}
 It is worth stressing that in the above expression $\eta$, $|\phi\rangle_{M}$
and $|\Delta\rangle_{\bar{C}M}$ are depending on $|\psi\rangle_{C\bar{C}}$.
We decompose $W^{\dag}$ acting on the first term of Eq.~(\ref{eq:main})
as 

\begin{equation}
W^{\dag}|0\rangle_{C\bar{C}}|\phi\rangle_{M}=\sqrt{\tilde{\eta}}\;|\psi\rangle_{C\bar{C}}|0\rangle_{M}+\sqrt{1-\tilde{\eta}}\;|\tilde{\Delta}\rangle_{C\bar{C}M},\label{eq:main2}\end{equation}
 where $|\tilde{\Delta}\rangle_{C\bar{C}M}$ is the orthogonal complement
of $|\psi\rangle_{C\bar{C}}|0\rangle_{M},$ i.e. \begin{equation}
_{\bar{C}C}\langle\psi|{}_{M}\langle0|\tilde{\Delta}\rangle_{C\bar{C}M}=0\;.\label{eq:o2}\end{equation}
Multiplying Eq. (\ref{eq:main2}) from the left with $_{C\bar{C}}\langle\psi|_{M}\langle0|$
and using the conjugate of Eq. (\ref{eq:main}) we find that $\eta=\tilde{\eta}.$
An expression of $\eta$ in terms of $\tau$ can be obtained by noticing
that for any vector $|\psi\rangle_{\bar{C}C}$ the following identity
applies \begin{equation}
\tau(\rho_{\bar{C}})=\textrm{tr}_{C}\left[U\left(\rho_{\bar{C}}\otimes|0\rangle_{C}\langle0|\right)U^{\dag}\right]=\textrm{tr}_{CM}\left[US_{\ell}\left(|\psi\rangle_{\bar{C}C}\langle\psi|\otimes|0\rangle_{M}\langle0|\right)S_{\ell}U^{\dag}\right]\;,\end{equation}
 with $\rho_{\bar{C}}$ being the reduced density matrix $\textrm{tr}_{C}\left[|\psi\rangle_{\bar{C}C}\langle\psi|\right]$.
Reiterating this expression one gets \begin{equation}
\textrm{tr}_{CM}\left[W(|\psi\rangle_{C\bar{C}}\langle\psi|\otimes|0\rangle_{M}\langle0|)W^{\dag}\right]=\tau^{L-1}\left(\rho_{\bar{C}}^{\prime}\right)\end{equation}
 with $\rho_{\bar{C}}^{\prime}=\textrm{tr}_{C}\left[U\left(|\psi\rangle_{\bar{C}C}\langle\psi|\right)U^{\dag}\right]$.
Therefore from Eq.~(\ref{eq:main}) and the orthogonality relation
(\ref{impo}) it follows that \begin{equation}
\eta={}_{\bar{C}}\langle0|\tau^{L-1}\left(\rho_{\bar{C}}^{\prime}\right)|0\rangle_{\bar{C}},\label{eq:fid}\end{equation}
 which, since $\tau$ is mixing, shows that $\eta\rightarrow1$ for
$L\rightarrow\infty$. Moreover we can use Eq.~(\ref{speed}) to
claim that \begin{eqnarray}
|\eta-1| & = & |{}_{\bar{C}}\langle0|\tau^{L-1}\left(\rho_{\bar{C}}^{\prime}\right)|0\rangle_{\bar{C}}-1|\nonumber \\
 & \leq & \|\tau^{L-1}\left(\rho_{\bar{C}}^{\prime}\right)-|0\rangle_{\bar{C}}\langle0|\|_{1}\leq R\;(L-1)^{d_{\bar{C}}}\;\kappa^{L-1},\label{eq:fid11}\end{eqnarray}
 where $R$ is a constant which depends upon $d_{\bar{C}}\equiv\mbox{dim}\mathcal{H}_{\bar{C}}$
and where $\kappa\in]0,1[$ is the second largest eigenvalue of $\tau.$

\section{Coding transformation\index{coding transformation}\label{sec:Coding-transformation}}

Here we derive the decoding/encoding transformation that relates states
on the memory $M$ to the states that are on the system $C\bar{C}.$
We first apply the above decompositions Eqs.~(\ref{eq:main}) and
(\ref{eq:main2}) to a fixed orthonormal basis $\left\{ |\psi_{k}\rangle_{C\bar{C}}\right\} $
of $\mathcal{H}_{C\bar{C}},$ i.e. \begin{eqnarray}
W|\psi_{k}\rangle_{C\bar{C}}|0\rangle_{M} & = & |0\rangle_{C}\otimes\left[\sqrt{\eta_{k}}|0\rangle_{\bar{C}}|\phi_{k}\rangle_{M}+\sqrt{1-\eta_{k}}|\Delta_{k}\rangle_{\bar{C}M}\right]\nonumber \\
W^{\dag}|0\rangle_{C\bar{C}}|\phi_{k}\rangle_{M} & = & \sqrt{\eta_{k}}\;|\psi_{k}\rangle_{C\bar{C}}|0\rangle_{M}+\sqrt{1-\eta_{k}}\;|\tilde{\Delta_{k}}\rangle_{C\bar{C}M}.\label{eq:maink}\end{eqnarray}
 Define a linear operator $D$ on ${\cal H}_{M}$ which performs the
following transformation \begin{eqnarray}
D|\psi_{k}\rangle_{M}=|\phi_{k}\rangle_{M}.\label{DEFD}\end{eqnarray}
Here $|\psi_{k}\rangle_{M}$ are orthonormal vectors of $M$ which
represent the states $\left\{ |\psi_{k}\rangle_{C\bar{C}}\right\} $
of $\mathcal{H}_{C\bar{C}}$ (formally they are obtained by a partial
isometry from $\bar{C}C$ to $M$). The vectors $|\phi_{k}\rangle_{M}$
are defined through Eq.~(\ref{eq:maink}) - typically they will not
be orthogonal. We first show that for large $L$ they become approximately
orthogonal.

From the unitarity of $W^{\dag}$ and from Eq. (\ref{eq:maink}) we
can establish the following identity \begin{eqnarray}
 &  & _{M}\langle\phi_{k}|\phi_{k'}\rangle_{M}=\sqrt{\eta_{k}\;\eta_{k'}}\;\delta_{kk'}+\sqrt{\eta_{k}\;(1-\eta_{k'})}\;{}_{\bar{C}CM}\langle\psi_{k}0|\tilde{\Delta}_{k'}\rangle_{\bar{C}CM}\label{QQQ}\\
 &  & +\sqrt{\eta_{k'}\;(1-\eta_{k})}\;{}_{\bar{C}CM}\langle\tilde{\Delta}_{k}|\psi_{k'}0\rangle_{\bar{C}CM}+\sqrt{(1-\tilde{\eta}_{k})(1-\tilde{\eta}_{k'})}\;{}_{C\bar{C}M}\langle\tilde{\Delta}_{k}|\tilde{\Delta}_{k'}\rangle_{C\bar{C}M}\;.\nonumber \end{eqnarray}
Defining $\eta_{0}\equiv\min_{k}\eta_{k}$ it follows for $k\neq k'$
that \begin{eqnarray}
|_{M}\langle\phi_{k}|\phi_{k'}\rangle_{M}| & \le & \sqrt{\eta_{k}\;(1-\eta_{k'})}\;|{}_{\bar{C}CM}\langle\psi_{k}0|\tilde{\Delta}_{k'}\rangle_{\bar{C}CM}|\\
 &  & +\sqrt{\eta_{k'}\;(1-\eta_{k})}\;|{}_{\bar{C}CM}\langle\tilde{\Delta}_{k}|\psi_{k'}0\rangle_{\bar{C}CM}|\nonumber \\
 &  & +\sqrt{(1-\tilde{\eta}_{k})(1-\tilde{\eta}_{k'})}\;|{}_{C\bar{C}M}\langle\tilde{\Delta}_{k}|\tilde{\Delta}_{k'}\rangle_{C\bar{C}M}|\nonumber \\
 & \leq & 2\sqrt{1-\eta_{0}}+(1-\eta_{0})\;\leq\;3\sqrt{1-\eta_{0}}.\label{IMPRT}\end{eqnarray}
Therefore for all $k,k'$ the inequality \begin{eqnarray}
|{}_{M}\langle\phi_{k}|\phi_{k'}\rangle_{M}-\delta_{k,k'}|\le3\;\sqrt{1-\eta_{0}}\label{IMPO}\end{eqnarray}
holds. It is worth noticing that, since Eq.~(\ref{eq:fid11}) applies
for all input states $|\psi\rangle_{\bar{C}C}$, we have \begin{eqnarray}
|\eta_{0}-1|\leq C\;(L-1)^{d_{\bar{C}}}\;\kappa^{L-1}\;.\label{eq:fid111}\end{eqnarray}
 Eq.~(\ref{IMPO}) allows us to make an estimation of the eigenvalues
$\lambda_{k}$ of $D^{\dag}D$ as \begin{equation}
|\lambda_{k}-1|\leq3\; d_{C\bar{C}}\;\sqrt{1-\eta_{0}},\end{equation}
with $d_{C\bar{C}}\equiv\dim\mathcal{H}_{C\bar{C}}.$ We now take
a polar decomposition $D=PV$ of $D.$ $V$ is the \emph{best unitary
approximation} to $D$ \cite[p 432]{HORNJOHNSON} and we have \begin{eqnarray}
||D-V||_{2}^{2} & = & \sum_{k}\left[\sqrt{\lambda_{k}}-1\right]^{2}\nonumber \\
 & \le & \sum_{k}\left|\lambda_{k}-1\right|\nonumber \\
 & \le & 3\; d_{C\bar{C}}^{2}\;\sqrt{1-\eta_{0}}.\label{Q0}\end{eqnarray}
 Therefore\begin{equation}
\boxed{||D-V||_{2}\le\sqrt{3}\: d_{C\bar{C}}\:(1-\eta_{0})^{1/4},}\label{Q}\end{equation}
 which, thanks to Eq.~(\ref{eq:fid111}), shows that $D$ can be
approximated arbitrary well by a unitary operator $V$ for $L\rightarrow\infty$.

\section{Fidelities for reading and writing}

In what follows we will use $V^{\dag}$ and $V$ as our reading and
writing transformation, respectively. In particular, $V^{\dag}$ will
be used to recover the input state $|\psi\rangle_{C\bar{C}}$ of the
chain after we have (partially) transferred it into $M$ through the
unitary $W$ (i.e. we first act on $|\psi\rangle_{C\bar{C}}\otimes|0\rangle_{M}$
with $W$, and then we apply $V^{\dag}$ on $M$). Vice-versa, in
order to prepare a state $|\psi\rangle_{C\bar{C}}$ on $C\bar{C}$
we first prepare $M$ into $|\psi\rangle_{M}$, then we apply to it
the unitary transformation $V$ and finally we apply $W^{\dag}$.
We now give bounds on the fidelities for both procedures.

The fidelity for reading the state $|\psi\rangle_{M}$ is given by
\begin{eqnarray}
F_{r}(\psi)\;\equiv\;{}_{M}\langle\psi|V^{\dag}\; R_{M}\; V|\psi\rangle_{M}\end{eqnarray}
 where $R_{M}$ is the state of the memory after $W$, i.e. \begin{eqnarray}
R_{M}\equiv\textrm{tr}_{C\bar{C}}\left[W(|\psi\rangle_{C\bar{C}}\langle\psi|\otimes|0\rangle_{M}\langle0|)W^{\dag}\right]=\eta\;|\phi\rangle_{M}\langle\phi|+(1-\eta)\;\sigma_{M}\;.\end{eqnarray}
 In the above expression we used Eqs.~(\ref{eq:main}) and~(\ref{impo})
and defined $\sigma_{M}=\textrm{tr}_{\bar{C}}[|\Delta\rangle_{\bar{C}M}\langle\Delta|]$.
Therefore by linearity we get \begin{eqnarray}
F_{r}(\psi)=\eta\;|{}_{M}\langle\phi|V|\psi\rangle_{M}|^{2}+(1-\eta)\;{}_{M}\langle\psi|V^{\dag}\;\sigma_{M}\; V|\psi\rangle_{M}\geq\eta\;|{}_{M}\langle\phi|V|\psi\rangle_{M}|^{2}\;.\label{fin1}\end{eqnarray}
 Notice that \begin{eqnarray}
|_{M}\langle\phi|V|\psi\rangle_{M}|=|_{M}\langle\phi|V-D+D|\psi\rangle_{M}|\geq|_{M}\langle\phi|D|\psi\rangle_{M}|-|_{M}\langle\phi|D-V|\psi\rangle_{M}|\;.\label{fin2}\end{eqnarray}
 Now we use the inequality~(\ref{Q}) to write \begin{eqnarray}
|_{M}\langle\phi|D-V|\psi\rangle_{M}|\leq||D-V||_{2}\leq\sqrt{3}\; d_{C\bar{C}}\;(1-\eta_{0})^{1/4}\;.\end{eqnarray}
If $|\psi\rangle_{M}$ was a basis state $|\psi_{k}\rangle_{M},$
then $|_{M}\langle\phi|D|\psi\rangle_{M}|=1$ by the definition Eq.~(\ref{DEFD})
of $D$. For \emph{generic} $|\psi\rangle_{M}$ we can use the linearity
to find after some algebra that \begin{eqnarray}
\sqrt{\eta}\;|_{M}\langle\phi|D|\psi\rangle_{M}|\;\geq\sqrt{\eta_{0}}\;-\;3\; d_{C\bar{C}}\;\sqrt{1-\eta_{0}}\;.\label{impo300}\end{eqnarray}
 Therefore Eq.~(\ref{fin2}) gives \begin{eqnarray}
\sqrt{\eta}\;|{}_{M}\langle\phi|V|\psi\rangle_{M}| & > & \sqrt{\eta_{0}}\;-5\; d_{C\bar{C}}\;(1-\eta_{0})^{1/4}\;.\label{fin4}\end{eqnarray}
By Eq.~(\ref{fin1}) it follows that \begin{eqnarray}
F_{r} & \geq & \eta_{0}\;-10\; d_{C\bar{C}}\;(1-\eta_{0})^{1/4}\;.\label{fin1000}\end{eqnarray}

The fidelity for writing a state $|\psi\rangle_{\bar{C}C}$ into $\bar{C}C$
is given by \begin{eqnarray}
F_{w}(\psi)\equiv{}_{C\bar{C}}\langle\psi|\textrm{tr}_{M}\left[W^{\dag}V\left(|\psi\rangle_{M}\langle\psi|\otimes|0\rangle_{\bar{C}C}\langle0|\right)V^{\dag}W\right]|\psi\rangle_{C\bar{C}}.\end{eqnarray}
 A lower bound for this quantity is obtained by replacing the trace
over $M$ with the expectation value on $|0\rangle_{M}$, i.e. \begin{eqnarray}
F_{w}(\psi) & \geq & _{C\bar{C}}\langle\psi|{}_{M}\langle0|W^{\dag}V\left(|\psi\rangle_{M}\langle\psi|\otimes|0\rangle_{\bar{C}C}\langle0|\right)V^{\dag}W|0\rangle_{M}|\psi\rangle_{C\bar{C}}\nonumber \\
 & = & \left|_{C\bar{C}}\langle0|{}_{M}\langle\psi|V^{\dag}W|0\rangle_{M}|\psi\rangle_{C\bar{C}}\right|^{2}\nonumber \\
 & = & \eta\;\left|_{M}\langle\psi|V^{\dag}|\phi\rangle_{M}\right|^{2}=\eta\;\left|_{M}\langle\phi|V|\psi\rangle_{M}\right|^{2}\label{vvv}\end{eqnarray}
 where Eqs.~(\ref{eq:main}) and the orthogonality relation~(\ref{impo})
have been employed to derive the second identity. Notice that the
last term of the inequality~(\ref{vvv}) coincides with the lower
bound~(\ref{fin1}) of the reading fidelity. Therefore, by applying
the same derivation of the previous section we can write \begin{equation}
\boxed{F\ge\eta_{0}\;-10\; d_{C\bar{C}}\;(1-\eta_{0})^{1/4},}\label{fin3000}\end{equation}
 which shows that the \index{reading and writing fidelities}reading
and writing fidelities converge to $1$ in the limit of large $L$.
Note that this lower bound can probably be largely improved.

\section{Application to spin chain communication}

We now show how the above protocol can be used to improve quantum
state transfer on a spin chain. The main advantage of using such a
memory protocol is that - opposed to all other schemes - Alice can
send arbitrary multi-qubit states with a single usage of the channel.
She needs no encoding, all the work is done by Bob. The protocol proposed
here can be used to improve the performances of any scheme mentioned
in Section~\ref{sec:Advanced-transfer-protocols}, and it works for
a large class of Hamiltonians, including Heisenberg and XY models
with arbitrary (also randomly distributed) coupling strengths. %
\begin{figure}[t]
\begin{centering}\includegraphics[width=1\columnwidth]{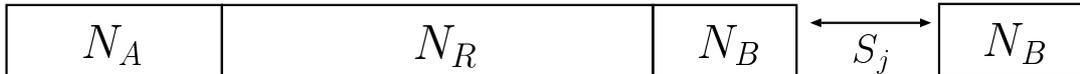}\par\end{centering}

\caption{\label{fig:swapping}Alice and Bob control the spins $N_{A}$ and
$N_{B}$ interconnected by the spins $N_{R}.$ At time $jt$ Bob performs
a swap $S_{j}$ between his spins and the memory $M_{j}$.}
\end{figure}

Consider a chain of spin-$1/2$ particles described by a Hamiltonian
$H$ which conserves the number of excitations. The chain is assumed
to be divided in three portions $A$ (Alice), $B$ (Bob) and $R$
(the remainder of the chain, connecting Alice and Bob) containing
respectively the first $N_{A}$ spins of the chain, the last $N_{B}$
spins and the intermediate $N_{R}$ spins, and the total length of
the chain is $N=N_{A}+N_{R}+N_{B}$ (see Fig \ref{fig:swapping}).
Bob has access also to a collection of quantum memories $M_{1},\cdots,M_{j}\cdots,M_{L}$
isomorphic with $B$, i.e. each having dimension equal to the dimension
$2^{N_{B}}$ of $B$. We assume that Bob's memory is initialised in
the zero excitation state $|0\rangle_{M}.$ Alice prepares an arbitrary
and unknown state $|\psi\rangle_{A}$ on her $N_{A}$ qubits. By defining
the (from Bob's perspective) controlled part of system $C=B$ and
the uncontrolled part $\bar{C}=AR,$ we can apply the results of the
last sections and get the following 

\begin{framedtheorem}
[Memory swapping]\label{thm:Let-H-be}Let $H$ be the Hamiltonian
of an open nearest-neighbour quantum chain that conserves the number
of excitations. If there is a time $t$ such that $f_{1,N}(t)\neq0$
(i.e. the Hamiltonian is capable of transport between Alice and Bob)
then the state transfer can be made arbitrarily perfect by using the
memory swapping protocol.
\end{framedtheorem}
\begin{proof}
We only have to show that the reduced dynamics on the chain is mixing
with a pure fixed point. Using the number of excitations as a conserved
additive observable, we can use the criterion of Lemma~\ref{lem:add}:
If there exists exactly one eigenstate $|E\rangle$ of factorising
form with $|0\rangle_{B}$, i.e. \begin{equation}
\exists_{1}\:|\lambda\rangle_{AR}:\quad H|\lambda\rangle_{AR}\otimes|0\rangle_{B}=E|\lambda\rangle_{AR}\otimes|0\rangle_{B},\label{eq:condition}\end{equation}
then the reduced dynamics is mixing toward $|0\rangle_{AR}.$ Assume
by contradiction that has an eigenvector $|E\rangle_{AR}\neq|0\rangle_{AR}$
which falsifies Eq.~(\ref{eq:condition}). Such an eigenstate can
be written as \begin{equation}
|E\rangle_{AR}\otimes|0\rangle_{B}=a|\mu\rangle_{AR}\otimes|0\rangle_{B}+b|\bar{\mu}\rangle_{AR}\otimes|0\rangle_{B},\label{eq:eig}\end{equation}
 where $a$ and $b$ are complex coefficients and where the spin just
before the section $B$ (with position $N_{A}+N_{R}$) is in the state
$|0\rangle$ for $|\mu\rangle_{AR}$ and in the state $|1\rangle$
for $|\bar{\mu}\rangle_{AR}.$ Since the interaction between this
spin and the first spin of section $B$ includes an exchange term
(otherwise $f_{1,N}(t)$=0 for all $t$), then the action of $H$
on the second term of~(\ref{eq:eig}) yields exactly one state which
contains an excitation in the sector $B.$ It cannot be compensated
by the action of $H$ on the first term of~(\ref{eq:eig}). But by
assumption $|E\rangle_{AR}\otimes|0\rangle_{B}$ is an eigenstate
of $H$, so we conclude that $b=0.$ This argument can be repeated
for the second last spin of section $R$, the third last spin, and
so on, to finally yield $|E\rangle_{AR}=|0\rangle_{AR}$, as long
as all the nearest neighbour interactions contain exchange parts. 
\end{proof}
\begin{remark}
Theorem~\ref{thm:Let-H-be} should be compared to Theorem \ref{thm:Let2}
for the multi rail protocol. They are indeed very similar. However
the current theorem is much stronger, since it allows to send arbitrary
multi-excitation states, and also to write states back onto the chain.
It is interesting to note that Lemma~\ref{lem:add} and Theorem~\ref{thm:Let-H-be}
indicate a connection between the dynamical controllability of a system
and its static entanglement properties. It may be interesting to obtain
a \emph{quantitative} relation between the amount of entanglement
and the convergence speed.
\end{remark}
Let us now come back to the question raised in Section~\ref{sec:Protocol}
about the operation $W^{\dag}$ being unphysical. As mentioned before,
this can be fixed using a simple transformation: if the Hamiltonian
$H$ fulfils the requirements of Lemma~\ref{lem:add}, then also
the Hamiltonian $-H$ fulfils them. Now derive the coding transformation
$\tilde{V}$ as given in Section~\ref{sec:Coding-transformation}
for the Hamiltonian $\tilde{H}=-H$. In this picture, the reading
protocol $W$ is unphysical, whereas the writing protocol becomes
physical. In the more general case where the condition of Lemma~\ref{lem:add}
is not valid, but the map \begin{equation}
\tau(\rho_{\bar{C}})\equiv\textrm{tr}_{C}\left[U\left(\rho_{\bar{C}}\otimes|0\rangle_{C}\langle0|\right)U^{\dag}\right]\end{equation}
is still ergodic with a pure fixed point, we then require the map\begin{equation}
\tilde{\tau}(\rho_{\bar{C}})\equiv\textrm{tr}_{C}\left[U^{\dag}\left(\rho_{\bar{C}}\otimes|0\rangle_{C}\langle0|\right)U\right]\end{equation}
to be also ergodic with pure fixed point to be able to use this trick.

\section{Conclusion}

We have given an explicit protocol for controlling a large permanently
coupled system by accessing a small subsystem only. In the context
of quantum chain communication this allows us to make use of the quantum
memory of the receiving party to improve the fidelity to a value limited
only by the size of the memory. We have shown that this scheme can
be applied to a Heisenberg spin chain. The main advantage of this
method is that arbitrary multi-excitation states can be transferred.
Also, our method can be applied to chains that do not conserve the
number of excitations in the system, as long as the reduced dynamic
is ergodic with a pure fixed point.

It remains an open question how much of our results remain valid if
the channel is mixing toward a \emph{mixed} state. In this case, a
part of the quantum information will in general remain in the correlations
between the system and the memory, and it cannot be expected that
the fidelity converges to one. However, by concentrating only on the
eigenstate of the fixed point density operator with the largest eigenvalue,
it should be possible to derive some bounds of the amount of information
that can be extracted.

\chapter{A valve for probability amplitude\label{cha:Single-memory}}

\section{Introduction}

We have mainly discussed two methods for quantum state transfer so
far. In the first one, multiple chains where used, and in the second
one, a single chain was used in combination with a large quantum memory.
Can we combine the best of the two schemes, i.e. is it possible to
use only a single chain and a single memory qubit? In this chapter
we will show that this is indeed the case and that the fidelity can
be improved easily by applying in certain time-intervals two-qubit
gates at the receiving end of the chain. These gates act as a \emph{valve\index{valve}}
which takes probability amplitude out of the system without ever putting
it back. The required sequence is determined \emph{a priori} by the
Hamiltonian of the system. Such a protocol is \emph{optimal} in terms
of resources, because two-qubit gates at the sending and receiving
end are required in order to connect the chain to the blocks in \emph{all}
above protocols (though often not mentioned explicitly). At the same
time, the engineering demands are not higher then for the memory swapping
protocol. Our scheme has some similarities with~\cite{Haselgrove2005},
but the gates used here are much simpler, and arbitrarily high fidelity
is guaranteed by a convergence theorem for arbitrary coupling strengths
and all non-Ising coupling types that conserve the number of excitations.
Furthermore, we show numerically that our protocol could also be realised
by a simple switchable interaction.

\section{Arbitrarily Perfect State Transfer}

\begin{figure}[htbp]
\begin{centering}\includegraphics[width=1\columnwidth]{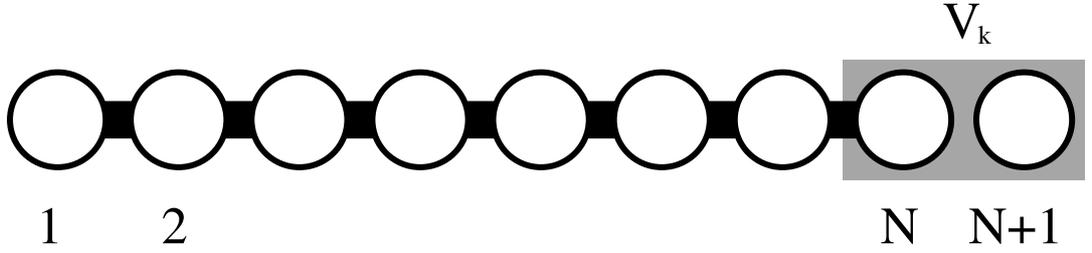}\par\end{centering}

\caption{\label{fig:setup}A quantum chain (qubits $1,2,\cdots,N$) and a
target qubit ($N+1$). By applying a sequence of two-qubit unitary
gates $V_{k}$ on the last qubit of the chain and the target qubit,
arbitrarily high fidelity can be achieved.}
\end{figure}

We now show how the receiver can improve the fidelity to an arbitrarily
high value by applying two-qubit gates between the end of the chain
and a {}``target qubit'' of the block. We label the qubits of the
chain by $1,2,\cdots,N$ and the target qubit by $N+1$ (see Fig.~\ref{fig:setup}).
The coupling of the chain is described by a Hamiltonian $H.$ We assume
that the Hamiltonian $H$ conserves the number of excitations and
that the target qubit $N+1$ is uncoupled,\begin{equation}
H|\boldsymbol{N+1}\rangle=0\label{eq:extra}\end{equation}
and set the energy of the ground state $|\boldsymbol{0}\rangle$ to
zero. For what follows we restrict all operators to the $N+2$ dimensional
Hilbert space\begin{equation}
\mathcal{H}=\textrm{span}\left\{ |\boldsymbol{n}\rangle;\; n=0,1,2,\ldots,N+1\right\} .\end{equation}
Our final assumption about the Hamiltonian of the system is that there
exists a time $t$ such that\begin{equation}
f_{N,t}(t)\equiv\langle\boldsymbol{N}|\exp\left\{ -itH\right\} |\boldsymbol{1}\rangle\neq0.\end{equation}
 Physically this means that the Hamiltonian has the capability of
transporting from the first to the last qubit of the chain. As mentioned
in the introduction, the fidelity of this transport may be very bad
in practice. %

We denote the unitary evolution operator for a given time $t_{k}$
as $U_{k}\equiv\exp\left\{ -it_{k}H\right\} $ and introduce the projector\begin{equation}
P=1-|\boldsymbol{0}\rangle\langle\boldsymbol{0}|-|\boldsymbol{N}\rangle\langle\boldsymbol{N}|-|\boldsymbol{N+1}\rangle\langle\boldsymbol{N+1}|.\end{equation}
 A crucial ingredient to our protocol is the operator\begin{eqnarray}
V(c,d) & \equiv & P+|\boldsymbol{0}\rangle\langle\boldsymbol{0}|+d|\boldsymbol{N}\rangle\langle\boldsymbol{N}|+d^{*}|\boldsymbol{N+1}\rangle\langle\boldsymbol{N+1}|\nonumber \\
 &  & +c^{*}|\boldsymbol{N+1}\rangle\langle\boldsymbol{N}|-c|\boldsymbol{N}\rangle\langle\boldsymbol{N+1}|,\end{eqnarray}
 where $c$ and $d$ are complex normalised amplitudes. It is easy
to check that\begin{equation}
VV^{\dagger}=V^{\dagger}V=1,\end{equation}
 so $V$ is a unitary operator on $\mathcal{H}.$ $V$ acts as the
identity on all but the last two qubits, and can hence be realised
by \emph{a local two-qubit gate on the qubits $N$ and $N+1$.} Furthermore
we have $VP=P$ and\begin{equation}
V(c,d)\left[\left\{ c|\boldsymbol{N}\rangle+d|\boldsymbol{N+1}\rangle\right\} \right]=|\boldsymbol{N+1}\rangle.\label{eq:wdesign}\end{equation}
 The operator $V(c,d)$ has the role of moving probability amplitude
$c$ from the $N$th qubit to target qubit, without moving amplitude
back into the system, and can be thought of as a \emph{valve.} Of
course as $V(c,d)$ is unitary, there are also states such that $V(c,d)$
acting on them would move back probability amplitude into the system,
but these do not occur in the protocol discussed here.

Using the time-evolution operator and two-qubit unitary gates on the
qubits $N$ and $N+1$ we will now develop a protocol that transforms
the state $|\boldsymbol{1}\rangle$ into |$\boldsymbol{N+1}\rangle.$
Let us first look at the action of $U_{1}$ on $|\boldsymbol{1}\rangle.$
Using the projector $P$ we can decompose this time-evolved state
as\begin{eqnarray}
U_{1}|\boldsymbol{1}\rangle & = & PU_{1}|\boldsymbol{1}\rangle+|\boldsymbol{N}\rangle\langle\boldsymbol{N}|U_{1}|\boldsymbol{1}\rangle\nonumber \\
 & \equiv & PU_{1}|\boldsymbol{1}\rangle+\sqrt{p_{1}}\left\{ c_{1}|\boldsymbol{N}\rangle+d_{1}|\boldsymbol{N+1}\rangle\right\} ,\end{eqnarray}
 where $p_{1}=\left|\langle\boldsymbol{N}|U_{1}|\boldsymbol{1}\rangle\right|^{2},$
$c_{1}=\langle\boldsymbol{N}|U_{1}|\boldsymbol{1}\rangle/\sqrt{p_{1}}$
and $d_{1}=0.$ Let us now consider the action of $V_{1}\equiv V(c_{1},d_{1})$
on the time-evolved state. By Eq. (\ref{eq:wdesign}) it follows that\begin{eqnarray}
V_{1}U_{1}|\boldsymbol{1}\rangle & = & PU_{1}|\boldsymbol{1}\rangle+\sqrt{p_{1}}|\boldsymbol{N}+1\rangle.\label{eq:firsttime}\end{eqnarray}
 Hence with a probability of $p_{1},$ the excitation is now in the
position $N+1,$ where it is {}``frozen'' (since that qubit is not
coupled to the chain. We will now show that at the next step, this
probability is increased. Applying $U_{2}$ to Eq. (\ref{eq:firsttime})
we get \begin{eqnarray}
\lefteqn{U_{2}V_{1}U_{1}|\boldsymbol{1}\rangle}\nonumber \\
 & = & PU_{2}PU_{1}|\boldsymbol{1}\rangle+\langle\boldsymbol{N}|U_{2}PU_{1}|\boldsymbol{1}\rangle|\boldsymbol{N}\rangle+\sqrt{p_{1}}|\boldsymbol{N}+1\rangle\nonumber \\
 & = & PU_{2}PU_{1}|\boldsymbol{1}\rangle+\sqrt{p_{2}}\left\{ c_{2}|\boldsymbol{N}\rangle+d_{2}|\boldsymbol{N}+1\rangle\right\} \end{eqnarray}
 with $c_{2}=\langle\boldsymbol{N}|U_{2}PU_{1}|\boldsymbol{1}\rangle/\sqrt{p_{2}},$
$d_{2}=\sqrt{p_{1}}/\sqrt{p_{2}}$ and\begin{eqnarray}
p_{2} & = & p_{1}+\left|\langle\boldsymbol{N}|U_{2}PU_{1}|\boldsymbol{1}\rangle\right|^{2}\ge p_{1}.\end{eqnarray}
 Applying $V_{2}\equiv V(c_{2},d_{2})$ we get \begin{equation}
V_{2}U_{2}V_{1}U_{1}|\boldsymbol{1}\rangle=PU_{2}PU_{1}|\boldsymbol{1}\rangle+\sqrt{p_{2}}|\boldsymbol{N}+1\rangle.\end{equation}
 Repeating this strategy $\ell$ times we get\begin{equation}
\left(\prod_{k=1}^{\ell}V_{k}U_{k}\right)|\boldsymbol{1}\rangle=\left(\prod_{k=1}^{\ell}PU_{k}\right)|\boldsymbol{1}\rangle+\sqrt{p_{\ell}}|\boldsymbol{N}+1\rangle,\label{eq:general}\end{equation}
 where the products are arranged in the time-ordered way. Using the
normalisation of the r.h.s. of Eq. (\ref{eq:general}) we get\begin{equation}
p_{\ell}=1-\left\Vert \left(\prod_{k=1}^{\ell}PU_{k}\right)|\boldsymbol{1}\rangle\right\Vert ^{2}.\end{equation}
 From Section~\ref{s:sec4}  we know that there exists a $t>0$ such
that for equal time intervals $t_{1}=t_{2}=\ldots=t_{k}=t$ we have
$\lim_{\ell\rightarrow\infty}p_{\ell}=1.$ Therefore the limit of
infinite gate operations for Eq. (\ref{eq:general}) is given by\begin{equation}
\lim_{\ell\rightarrow\infty}\left(\prod_{k=1}^{\ell}V_{k}U_{k}\right)|\boldsymbol{1}\rangle=|\boldsymbol{N+1}\rangle.\label{eq:convergence}\end{equation}
 It is also easy to see that $\lim_{k\rightarrow\infty}d_{\ell}=1,$
$\lim_{k\rightarrow\infty}c_{\ell}=0$ and hence the gates $V_{k}$
converge to the identity operator. Furthermore, since $V_{k}U_{k}|\boldsymbol{0}\rangle=|\boldsymbol{0}\rangle$
it also follows that arbitrary superpositions can be transferred.
As discussed in Theorem~\ref{speed}, this convergence is asymptotically
exponentially fast in the number of gate applied (a detailed analysis
of the relevant scaling can be found in Chapter~\ref{cha:Dual-Rail}).
Equation (\ref{eq:convergence}) is a surprising result, which shows
that \emph{any non-perfect transfer can be made arbitrarily perfect}
by only applying two-qubit gates on one end of the quantum chain.
It avoids restricting the gate times to specific times (as opposed
to the dual rail scheme) while requiring no additional memory qubit
(as opposed to the memory swapping scheme).

The sequence $V_{k}$ that needs to be applied to the end of the chain
to perform the state transfer only depends on the Hamiltonian of the
quantum chain. The relevant properties can in principle be determined
a priori by preceding measurements and tomography on the quantum chain
(as discussed in Sect.~\ref{sec:Tomography}).

\section{Practical Considerations\label{sec:Practical-Considerations}}

Motivated by the above result we now investigate how the above protocol
may be implemented in practice, well before the realisation of the
quantum computing blocks from Fig.~\ref{fig:connect2}. The two-qubit
gates $V_{k}$ are essentially rotations in the $\{|01\rangle,|10\rangle\}$
space of the qubits $N$ and $N+1.$ It is therefore to be expected
that they can be realised (up to a irrelevant phase) by a switchable
Heisenberg or $XY$ type coupling between the $Nth$ and the target
qubit. However in the above, we have assumed that the gates $V_{k}$
can be applied instantaneously, i.e. in a time-scale much smaller
than the time-scale of the dynamics of the chain. This corresponds
to a switchable coupling that is much stronger than the coupling strength
of the chain. %
\begin{figure}[htbp]
\begin{centering}\includegraphics[width=0.8\columnwidth]{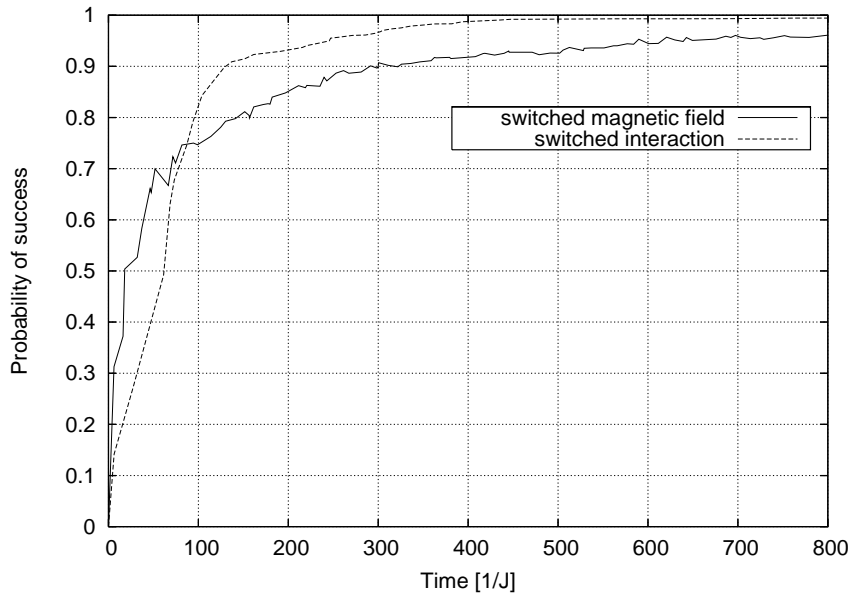}\par\end{centering}

\caption{\label{fig:numerics}Numerical example for the convergence of the
success probability. Simulated is a quantum chain of length $N=20$
with the Hamiltonian from Eq. (\ref{eq:heis}) (dashed line) and Eq.
(\ref{eq:mag}) with $B/J=20$ (solid line). Using the original protocol~\cite{Bose2003},
the same chain would only reach a success probability of $0.63$ in
the above time interval.}
\end{figure}

Here, we numerically investigate if a convergence similar to the above
results is still possible when this assumption is not valid. We \emph{do}
however assume that the switching of the interaction is still describable
by an instantaneous switching (i.e. the sudden approximation is valid).
This assumption is mainly made to keep the numerics simple. We do
not expect qualitative differences when the switching times become
finite as long as the time-dependent Hamiltonian is still conserving
the number of excitations in the chain. In fact it has recently been
shown that the finite switching time can even \emph{improve} the fidelity~\cite{Bruder}.
Intuitively, this happens because by gradually decreasing the coupling,
he not only receives the probability amplitude of the last qubit of
the chain, but can also ''swallow'' a bit of the dispersed wave-packed
(similar to the situation discussed in \cite{Haselgrove2005}).

We have investigated two types of switching. For the first type, the
coupling itself is switchable, i.e.\begin{equation}
H(t)=J\sum_{n=1}^{N-1}\sigma_{n}^{-}\sigma_{n+1}^{+}+\Delta(t)\sigma_{N}^{-}\sigma_{N+1}^{+}+\textrm{h.c.},\label{eq:heis}\end{equation}
 where $\Delta(t)$ can be $0$ or $1.$ For the second type, the
target qubit is \emph{permanently} coupled to the remainder of the
chain, but a strong magnetic field on the last qubit can be switched,
\begin{equation}
H(t)=J\sum_{n=1}^{N}\sigma_{n}^{-}\sigma_{n+1}^{+}+\textrm{h.c.}+B\Delta(t)\sigma_{N+1}^{z},\label{eq:mag}\end{equation}
 where again $\Delta(t)$ can be $0$ or $1$ and $B\gg1.$ This suppresses
the coupling between the $N$th and $N+1$th qubit due to an energy
mismatch.

In both cases, we first numerically optimise the times for unitary
evolution $t_{k}$ over a fixed time interval such that the probability
amplitude at the $N$th qubit is maximal. The algorithm then finds
the optimal time interval during which $\Delta(t)=1$ such that the
probability amplitude at the target qubit is increased. In some cases
the phases are not correct, and switching on the interaction would
result in probability amplitude floating back into the chain. In this
situation, the target qubit is left decoupled and the chain is evolved
to the next amplitude maximum at the $N$th qubit. Surprisingly, even
when the time-scale of the gates is comparable to the dynamics, near-perfect
transfer remains possible~(Fig \ref{fig:numerics}). In the case
of the switched magnetic field, the achievable fidelity depends on
the strength of the applied field. This is because the magnetic field
does not fully suppress the coupling between the two last qubits.
A small amount of probability amplitude is lost during each time evolution
$U_{k},$ and when the gain by the gate is compensated by this loss,
the total success probability no longer increases.

\section{Conclusion}

We have seen that by having a simple switchable interaction acting
as a \emph{valve} for probability amplitude, arbitrarily perfect state
transfer is possible on a single spin chain. In fact, by using the
inverse protocol, arbitrary%
\footnote{Opposed to the method for state preparation developed in the last
chapter this allows the creation of \emph{known} states only (as the
valve operations $V_{k}$ depend explicitly on the state that one
wants to prepare).%
} states in the first excitation sector can also be prepared on the
chain. Furthermore, this protocol can easily be adopted to arbitrary
graphs connecting multiple senders and receivers (as discussed for
weakly coupled systems in~\cite{Bednarska}).

\chapter{External noise\label{cha:Problems-and-Practical}}

\section{Introduction}

\begin{figure}[tbh]
\begin{centering}\includegraphics[width=0.45\paperwidth]{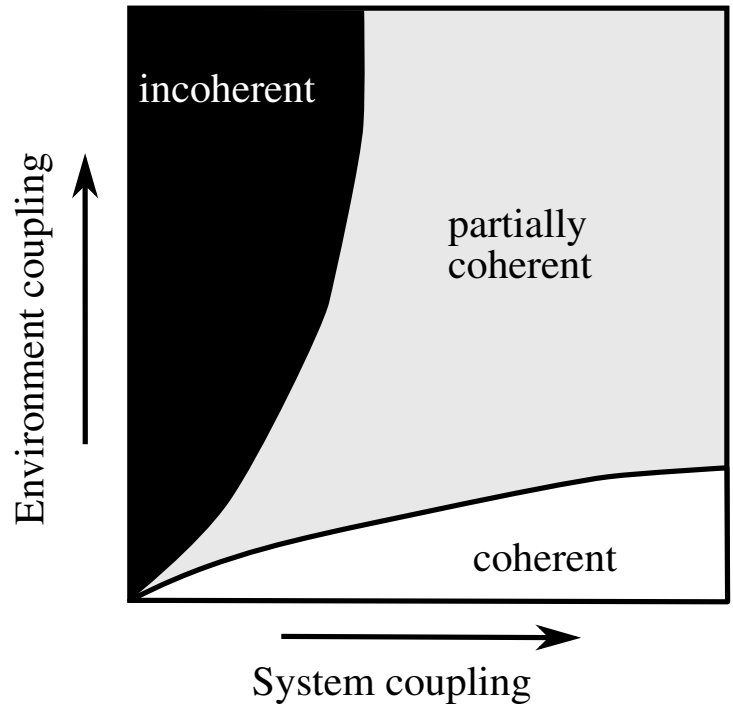}\par\end{centering}

\caption{\label{fig:Dominant}Dominant regimes of dynamics depending on the
relative strength of the system Hamiltonian and the environmental
coupling~\cite{EXCITON}.}
\end{figure}

An important question that was left open so far is what happens to
quantum state transfer in the presence of external noise. It is well
known from the theory of open quantum systems~\cite{OPENQUANTUM}
that this can lead to dissipation and decoherence, which also means
that quantum information is lost. The evolution of a closed quantum
system is described by the Schrödinger equation\index{Schrödinger equation}\begin{equation}
\partial_{t}|\psi\rangle=-iH|\psi\rangle.\end{equation}
If a system is very strongly coupled to a environment, the dynamic
is completely incoherent and described by some simple rate equations
for the occupation probabilities,\begin{equation}
\partial_{t}P_{n}=\sum_{n}k_{n\rightarrow m}P_{n}-\sum_{n}k_{m\rightarrow n}P_{m}.\end{equation}
In the more general case where the dynamic consists of coherent and
incoherent parts, the evolution can sometimes be expressed as a Lindblad
equation\index{Lindblad equation}~\cite{OPENQUANTUM}\begin{equation}
\partial_{t}\rho=\mathcal{L}\rho\end{equation}
for the reduced density matrix. These three regimes are shown in Fig.~\ref{fig:Dominant}.
 For quantum information theory, coherence is essential~\cite{NIELSEN},
and one has to try to isolate the quantum chain as much as possible
from the environment. In the partially coherent regime, typically
the quantum behaviour decays exponentially with a rate depending on
the temperature of the environment. Not surprisingly, this has also
been found in the context of quantum state transfer~\cite{Guo2006,Karimipour2005,Sun}.
From a theoretical point of view it is perhaps more interesting to
look at the low temperature and strong coupling regime, where the
dynamics is often non-Markovian~\cite{OPENQUANTUM} and can no longer
expressed as a simple Lindblad equation. This is also interesting
from a practical perspective, corresponding to effects of the environment
which cannot be avoided by cooling. Here we consider a model where
the system is coupled to a spin environment through an exchange interaction.
This coupling offers the unique opportunity of an analytic solution
of our problem without \emph{any} approximations regarding the strength
of system-environment coupling (in most treatments of the effect of
an environment on the evolution of a quantum system, the system-environment
coupling is assumed to be weak) and allows us to include inhomogeneous
interactions of the bath spins with the system. For such coupling,
decoherence is possible for mixed (thermal) initial bath states~\cite{Schmidt2005,Breuer2004}.
However if the system and bath are both initially cooled to their
ground states, is there still a non-trivial effect of the environment
on the fidelity? In this chapter we find that there are two important
effects: the spin transfer functions (Eq.~\ref{eq:spintrans}) are
\emph{slowed down} by a factor of two, and \emph{destabilised} by
a modulation of $\left|\cos Gt\right|,$ where $G$ is the mean square
coupling to the environment. This has both positive and negative implications
for the use of strongly coupled spin systems as quantum communication
channels. The spin transfer functions also occur in the charge and
energy transfer dynamics in molecular systems~\cite{EXCITON} and
in continuous time random walks~\cite{Blumen2006} to which our results
equally apply.

\section{Model}

\begin{figure}[htbp]
\begin{centering}\includegraphics[width=0.6\paperwidth]{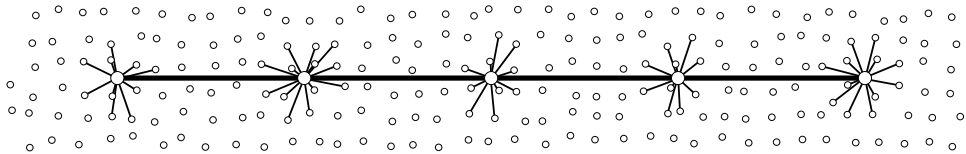}\par\end{centering}

\caption{\label{fig:spinchain}A spin chain of length $N=5$ coupled to independent
baths of spins. }
\end{figure}

We choose to start with a specific spin system, i.e. an open spin
chain of arbitrary length $N,$ with a Hamiltonian given by\begin{equation}
H_{S}=-\frac{1}{2}\sum_{\ell=1}^{N-1}J_{\ell}\left(X_{\ell}X_{\ell+1}+Y_{\ell}Y_{\ell+1}\right),\end{equation}
 where $J_{\ell}$ are some arbitrary couplings and $X_{\ell}$ and
$Y_{\ell}$ are the Pauli-X and Y matrices for the $\ell$th spin.
Toward the end of the section we will however show that our results
hold for any system where the number of excitations is conserved during
dynamical evolution. In addition to the chain Hamiltonian, each spin
$\ell$ of the chain interacts with an independent bath of $M_{\ell}$
environmental spins (see Fig \ref{fig:spinchain}) via an inhomogeneous
Hamiltonian,\begin{equation}
H_{I}^{(\ell)}=-\frac{1}{2}\sum_{k=1}^{M_{\ell}}g_{k}^{(\ell)}\left(X_{\ell}X_{k}^{(\ell)}+Y_{\ell}Y_{k}^{(\ell)}\right).\end{equation}

\begin{figure}[htbp]
\begin{centering}\includegraphics[width=0.6\paperwidth]{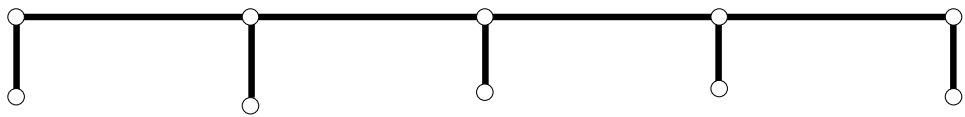}\par\end{centering}

\caption{\label{fig:spinchain_equiv}In the first excitation sector, the system
can be mapped into an effective spin model where the bath spins are
replaced by a single effective spin, as indicated here for $N=5.$}
\end{figure}

In the above expression, the Pauli matrices $X_{\ell}$ and $Y_{\ell}$
act on the $\ell$th spin of the chain, whereas $X_{k}^{(\ell)}$
and $Y_{k}^{(\ell)}$ act on the $k$th environmental spin attached
to the $\ell$th spin of the chain. We denote the total interaction
Hamiltonian by\begin{equation}
H_{I}\equiv\sum_{\ell=1}^{N}H_{I}^{(\ell)}.\end{equation}
 The total Hamiltonian is given by $H=H_{S}+H_{I},$ where it is important
to note that $\left[H_{S},H_{I}\right]\neq0.$ We assume that a homogeneous
magnetic field along the z-axis is applied. The ground state of the
system is then given by the fully polarised state $|0,0\rangle,$
with all chain and bath spins aligned along the z-axis. The above
Hamiltonian describes an extremely complex and disordered system with
a Hilbert space of dimension $2^{N+NM}.$ In the context of state
transfer however, only the dynamics of the first excitation sector
is relevant. We proceed by mapping this sector to a much simpler system~\cite{Buzek2005,Chiara2005,Hutton2004,Olaya-Castro2005,Olaya-Castro2006}.
For $\ell=1,2,\ldots,N$ we define the states\begin{equation}
|\ell,0\rangle\equiv X_{\ell}|0,0\rangle\end{equation}
 and\begin{equation}
|0,\ell\rangle\equiv\frac{1}{G_{\ell}}\sum_{k=1}^{M_{\ell}}g_{k}^{(\ell)}X_{k}^{(\ell)}|0,0\rangle\end{equation}
 with \begin{equation}
G_{\ell}=\sqrt{\sum_{k=1}^{M_{\ell}}\left(g_{k}^{(\ell)}\right)^{2}}.\label{eq:effect}\end{equation}
 It is easily verified that (setting $J_{0}=J_{N}=0$)\begin{eqnarray}
H_{S}|\ell,0\rangle & = & -J_{\ell-1}|\ell-1,0\rangle-J_{\ell}|\ell+1,0\rangle\nonumber \\
H_{S}|0,\ell\rangle & = & 0,\label{eq:hc_action}\end{eqnarray}
 and\begin{eqnarray}
H_{I}|\ell,0\rangle & = & -G_{\ell}|0,\ell\rangle\label{eq:hb_action}\\
H_{I}|0,\ell\rangle & = & -G_{\ell}|\ell,0\rangle.\label{eq:hb_action2}\end{eqnarray}
 Hence these states define a $2N-$dimensional subspace that is invariant
under the action of $H.$ This subspace is equivalent to the first
excitation sector of a system of $2N$ spin $1/2$ particles, coupled
as it is shown in Fig \ref{fig:spinchain_equiv}. %

Our main assumption is that the bath couplings are \emph{in effect}
the same, i.e. $G_{\ell}=G$ for all $\ell$. Note however that the
individual number of bath spins $M_{\ell}$ and bath couplings $g_{k}^{(\ell)}$
may still depend on $\ell$ and $k$ as long as their means square
average is the same. Also, our analytic solution given in the next
paragraph relies on this assumption, but numerics show that our main
result {[}Equation~(\ref{eq:scalingformula})] remains a good approximation
if the $G_{\ell}$ slightly vary and we take $G\equiv\left\langle G_{\ell}\right\rangle .$
Disorder in the vertical couplings is treated \emph{exactly} in the
sense that our results hold for any choice of couplings $J_{\ell}.$

\section{\label{sec:Solving-the-Schr=3D3DF6dinger}Results}

In this paragraph, we solve the Schrödinger equation for the model
outlined above and discuss the spin transfer functions. Firstly, let
us denote the orthonormal eigenstates of $H_{S}$ alone by \begin{equation}
H_{S}|\psi_{k}\rangle=\epsilon_{k}|\psi_{k}\rangle\quad(k=1,2\ldots,N)\end{equation}
 with\begin{equation}
|\psi_{k}\rangle=\sum_{\ell=1}^{N}a_{k\ell}|\ell,0\rangle.\label{eq:eigen_h_c}\end{equation}
 For what follows, it is not important whether analytic expressions
for the eigensystem of $H_{S}$ can be found. Our result holds even
for models that are not analytically solvable, such as the randomly
coupled chains considered in Section~\ref{sec:Disordered-chains}.
We now make an ansatz for the eigenstates of the full Hamiltonian,
motivated by the fact that the states \begin{equation}
|\phi_{\ell}^{n}\rangle\equiv\frac{1}{\sqrt{2}}\left(|\ell,0\rangle+\left(-1\right)^{n}|0,\ell\rangle\right)\quad(n=1,2)\end{equation}
 are eigenstates of $H_{I}^{(\ell)}$ with the corresponding eigenvalues
$\pm G$ {[}this follows directly from Eqs.~(\ref{eq:hb_action})
and~(\ref{eq:hb_action2})]. Define the vectors\begin{eqnarray}
|\Psi_{k}^{n}\rangle & \equiv & \sum_{\ell=1}^{N}a_{k\ell}|\phi_{\ell}^{n}\rangle\label{eq:ansatz}\end{eqnarray}
 with $k=1,2,\ldots,N$ and $n=0,1.$ The $|\Psi_{k}^{n}\rangle$
form an orthonormal basis in which we express the matrix elements
of the Hamiltonian. We can easily see that\begin{equation}
H_{I}|\Psi_{k}^{n}\rangle=-\left(-1\right)^{n}G|\Psi_{k}^{n}\rangle\end{equation}
 and \begin{equation}
H_{S}|\Psi_{k}^{n}\rangle=\frac{\epsilon_{k}}{\sqrt{2}}\sum_{\ell=1}^{N}a_{k\ell}|\ell,0\rangle=\frac{\epsilon_{k}}{2}\left(|\Psi_{k}^{0}\rangle+|\Psi_{k}^{1}\rangle\right).\end{equation}
 Therefore the matrix elements of the full Hamiltonian $H=H_{S}+H_{I}$
are given by \begin{equation}
\langle\Psi_{k'}^{n'}|H|\Psi_{k}^{n}\rangle=\delta_{kk'}\left(-\left(-1\right)^{n}G\delta_{nn'}+\frac{\epsilon_{k}}{2}\right).\end{equation}
 The Hamiltonian is not diagonal in the states of Eq.~(\ref{eq:ansatz}).
But $H$ is now block diagonal consisting of $N$ blocks of size $2$,
which can be easily diagonalised analytically. The orthonormal eigenstates
of the Hamiltonian are given by\begin{equation}
|E_{k}^{n}\rangle=c_{kn}^{-1}\left\{ \left(\left(-1\right)^{n}\Delta_{k}-2G\right)|\Psi_{k}^{0}\rangle+\epsilon_{k}|\Psi_{k}^{1}\rangle\right\} \label{eq:eigenstates}\end{equation}
 with the eigenvalues\begin{equation}
E_{k}^{n}=\frac{1}{2}\left(\epsilon_{k}+\left(-1\right)^{n}\Delta_{k}\right)\end{equation}
 and the normalisation\begin{equation}
c_{kn}\equiv\sqrt{\left(\left(-1\right)^{n}\Delta_{k}-2G\right)^{2}+\epsilon_{k}^{2}},\end{equation}
 where\begin{equation}
\Delta_{k}=\sqrt{4G^{2}+\epsilon_{k}^{2}}.\label{eq:delta}\end{equation}
 Note that the ansatz of Eq.~(\ref{eq:ansatz}) that put $H$ in
block diagonal form did not depend on the details of $H_{S}$ and
$H_{I}^{(\ell)}.$ The methods presented here can be applied to a
much larger class of systems, including the generalised spin star
systems (which include an interaction within the bath) discussed in~\cite{Olaya-Castro2006}.

\begin{figure}[tbh]
\begin{centering}\includegraphics[width=0.8\columnwidth]{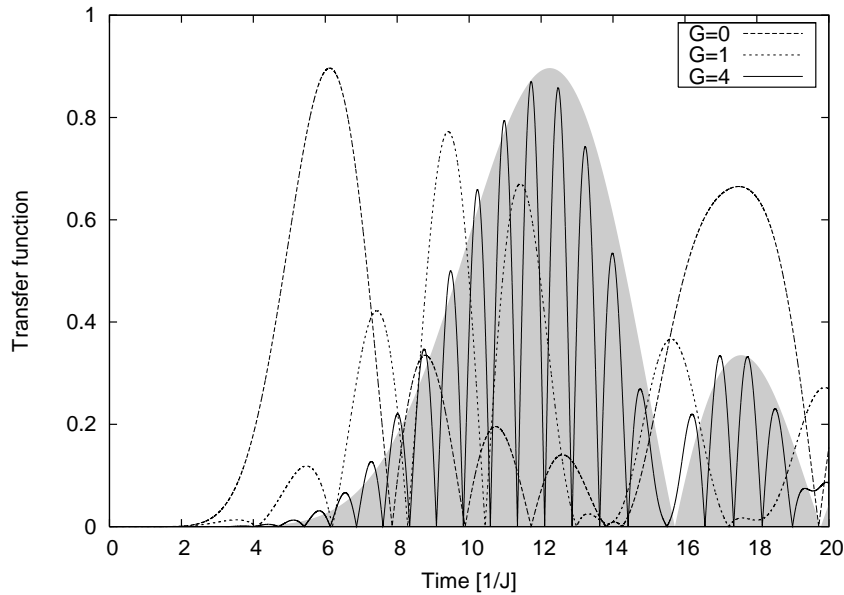}\par\end{centering}

\caption{\label{cap:Example}The absolute value of the transport function
$f_{N,1}(t)$ of an uniform spin chain (i.e. $J_{\ell}=1$) with length
$N=10$ for three different values of the bath coupling $G.$ The
filled grey curve is the envelope of the limiting function for $G\gg\epsilon_{k}/2$
given by $|f^{0}(\frac{t}{2})|.$ We can see that Eq.~(\ref{eq:scalingformula})
becomes a good approximation already at $G=4.$}
\end{figure}

\begin{figure}[tbh]
\begin{centering}\includegraphics[width=0.8\columnwidth]{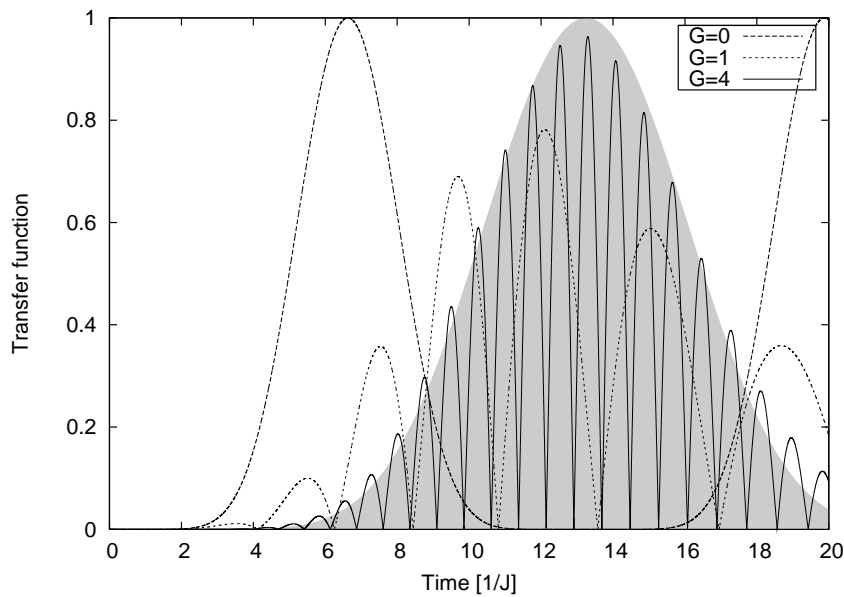}\par\end{centering}

\caption{\label{cap:Example2}The same as Fig. \ref{cap:Example}, but now
for an engineered spin chain {[}i.e. $J_{\ell}=\sqrt{\ell(N-\ell)}$]
as in~Subsection \ref{sub:Engineered-Hamiltonians}. For comparison,
we have rescaled the couplings such that $\sum_{\ell}J_{\ell}$ is
the same as in the uniform coupling case.}
\end{figure}

\begin{figure}[tbh]
\begin{centering}\includegraphics[width=0.8\columnwidth]{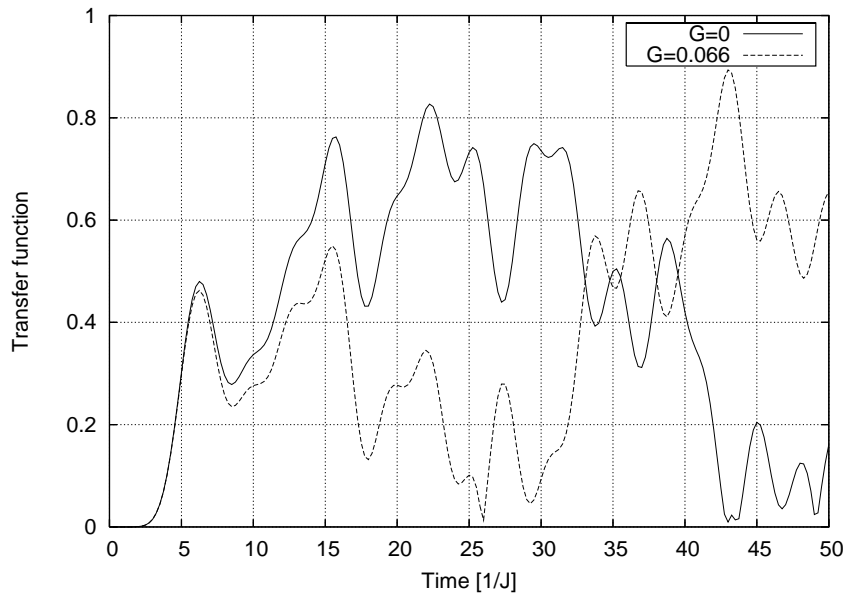}\par\end{centering}

\caption{\label{cap:Example3}A weakly coupled bath may even improve the transfer
function for some specific choices of the $J_{\ell}.$ This plot shows
the transfer function $|f_{N,1}(t)|$ for $N=10.$ The couplings $J_{\ell}$
were found numerically. }
\end{figure}

After solving the Schrödinger equation, let us now turn to quantum
state transfer. The relevant quantity~\cite{Bose2003,Haselgrove2005}
is given by the transfer function\begin{eqnarray*}
f_{N,1}(t) & \equiv & \langle N,0|\exp\left\{ -iHt\right\} |1,0\rangle\\
 & = & \sum_{k,n}\exp\left\{ -iE_{k}^{n}t\right\} \langle E_{k}^{n}|1,0\rangle\langle N,0|E_{k}^{n}\rangle.\end{eqnarray*}
 The modulus of $f_{N,1}(t)$ is between $0$ (no transfer) and $1$
(perfect transfer) and fully determines the fidelity of state transfer.
Since\begin{eqnarray*}
\langle\ell,0|E_{k}^{n}\rangle & = & c_{kn}^{-1}\left\{ \left(\left(-1\right)^{n}\Delta_{k}-2G\right)\langle\ell,0|\Psi_{k}^{0}\rangle+\epsilon_{k}\langle\ell,0|\Psi_{k}^{1}\rangle\right\} \\
 & = & \frac{c_{kn}^{-1}}{\sqrt{2}}\left(\left(-1\right)^{n}\Delta_{k}-2G+\epsilon_{k}\right)a_{k\ell}\end{eqnarray*}
 we get\begin{eqnarray}
\lefteqn{f_{N,1}(t)=}\label{eq:transfer}\\
 &  & \frac{1}{2}\sum_{k,n}e^{\frac{-it}{2}\left(\epsilon_{k}+\left(-1\right)^{n}\Delta_{k}\right)}\frac{\left(\left(-1\right)^{n}\Delta_{k}-2G+\epsilon_{k}\right)^{2}}{\left(\left(-1\right)^{n}\Delta_{k}-2G\right)^{2}+\epsilon_{k}^{2}}a_{k1}a_{kN}^{*}.\nonumber \end{eqnarray}
 Eq.~(\ref{eq:transfer}) is the main result of this section, fully
determining the transfer of quantum information and entanglement in
the presence of the environments. In the limit $G\rightarrow0,$ we
have $\Delta_{k}\approx\epsilon_{k}$ and $f_{N,1}(t)$ approaches
the usual result without an environment,\begin{equation}
f_{N,1}^{0}(t)\equiv\sum_{k}\exp\left\{ -it\epsilon_{k}\right\} a_{k1}a_{kN}^{*}.\end{equation}
 In fact, a series expansion of Eq.~(\ref{eq:transfer}) yields that
the first modification of the transfer function is of the order of
$G^{2},$\begin{equation}
G^{2}\sum_{k}a_{k1}a_{kN}^{*}\left[\exp\left\{ -it\epsilon_{k}\right\} \left(-\frac{1}{\epsilon_{k}^{2}}-\frac{it}{\epsilon_{k}}\right)+\frac{1}{\epsilon_{k}^{2}}\right].\end{equation}
 Hence the effect is small for very weakly coupled baths. However,
as the chains get longer, the lowest lying energy $\epsilon_{1}$
usually approaches zero, so the changes become more significant (scaling
as $1/\epsilon_{k}$). For intermediate $G,$ we evaluated Eq.~(\ref{eq:transfer})
numerically and found that the first peak of the transfer function
generally becomes slightly lower, and gets shifted to higher times
(Figures \ref{cap:Example} and \ref{cap:Example2}). A numeric search
in the coupling space $\left\{ J_{\ell},\ell=1,\ldots,N-1\right\} $
however also revealed some rare examples where an environment can
also slightly improve the peak of the transfer function (Fig \ref{cap:Example3}).

In the strong coupling regime $G\gg\epsilon_{k}/2,$ we can approximate
Eq.~(\ref{eq:delta}) by $\Delta_{k}\approx2G.$ Inserting it in
Eq.~(\ref{eq:transfer}) then becomes\begin{eqnarray}
f_{N,1}(t) & \approx & \frac{1}{2}e^{-iGt}\sum_{k}\exp\left\{ -it\epsilon_{k}\frac{1}{2}\right\} a_{k1}a_{kN}^{*}+\nonumber \\
 &  & +\frac{1}{2}e^{iGt}\sum_{k}\exp\left\{ -it\epsilon_{k}\frac{1}{2}\right\} a_{k1}a_{kN}^{*}\nonumber \\
 & = & \cos(Gt)f_{N,1}^{0}(\frac{t}{2}).\end{eqnarray}
 This surprisingly simple result consists of the normal transfer function,
slowed down by a factor of $1/2,$ and modulated by a quickly oscillating
term (Figures \ref{cap:Example} and \ref{cap:Example2}). We call
this effect \emph{destabilisation\index{destabilisation}.} Our derivation
actually did not depend on the indexes of $f(t)$ and we get for the
transfer from the $n$th to the $m$th spin of the chain that \begin{equation}
\boxed{f_{n,m}(t)\approx\cos(Gt)f_{n,m}^{0}(\frac{t}{2}).}\label{eq:scalingformula}\end{equation}
 It may look surprising that the matrix $f_{n,m}$ is no longer unitary.
This is because we are considering the dynamics of the chain only,
which is an open quantum system~\cite{OPENQUANTUM}. A heuristic
interpretation of Eq.~(\ref{eq:scalingformula}) is that the excitation
oscillates back and forth between the chain and the bath (hence the
modulation), and spends half of the time trapped in the bath (hence
the slowing). If the time of the maximum of the transfer function
$|f_{n,m}^{0}(t)|$ for $G=0$ is a multiple of $\pi/2G$ then this
maximum is also reached in the presence of the bath. We remark that
this behaviour is strongly non-Markovian\index{non-Markovian}~\cite{OPENQUANTUM}.

Finally, we want to stress that Eq.~(\ref{eq:scalingformula}) is
\emph{universal} for any spin Hamiltonian that conserves the number
of excitations, i.e. with $\left[H_{S},\sum_{\ell}Z_{\ell}\right]=0$.
Thus our restriction to chain-like topology and exchange couplings
for $H_{S}$ is not necessary. In fact the only difference in the
whole derivation of Eq.~(\ref{eq:scalingformula}) for a more general
Hamiltonian is that Eq.~(\ref{eq:hc_action}) is replaced by\begin{eqnarray}
H_{S}|\ell,0\rangle & = & \sum_{\ell'}h_{\ell'}|\ell',0\rangle.\label{eq:hc_action2}\end{eqnarray}
 The Hamiltonian can still be formally diagonalised in the first excitation
sector as in Eq.~(\ref{eq:eigen_h_c}), and the states of Eq.~(\ref{eq:eigenstates})
will still diagonalise the total Hamiltonian $H_{S}+H_{I}.$ Also,
rather than considering an exchange Hamiltonian for the interaction
with the bath, we could have considered a Heisenberg interaction~\cite{Khaetskii2003},
but only for the special case where all bath couplings $g_{k}^{(\ell)}$
are all the same~\cite{Rao2006}. Up to some irrelevant phases, this
leads to the same results as for the exchange interaction.

\section{Conclusion}

We found a surprisingly simple and universal scaling law for the spin
transfer functions in the presence of spin environments. In the context
of quantum state transfer this result is double-edged: on one hand,
it shows that even for very strongly coupled baths quantum state transfer
is possible, with the same fidelity and only reasonable slowing. On
the other hand, it also shows that the fidelity as a function of time
becomes destabilised with a quickly oscillating modulation factor.
In practice, this factor will restrict the time-scale in which one
has to be able to read the state from the system. The results here
are very specific to the simple bath model and do not hold in more
general models (such as these discussed in~\cite{Guo2006,Sun}, where
true decoherence and dissipation takes place). What we intended to
demonstrate is that even though a bath coupling need not introduce
decoherence or dissipation to the system, it can cause other dynamical
processes that can be problematic for quantum information processing.
Because the effects observed here cannot be avoided by cooling the
bath, they may become relevant in some systems as a low temperature
limit.

\chapter{Conclusion and outlook}

Our research on quantum state transfer with spin chains has taken
us on a journey from a very practical motivation to quite fundamental
issues and back again. On one hand, our results are quite abstract
and fundamental, and have related state transfer to number theory,
topology and quantum convergence. On the other hand, we have developed
schemes which are simple and practical, taking into account experimental
hurdles such as disorder and restricted control. While the multi rail
scheme and the memory swapping scheme will probably become useful
only after much further progress in experimental QIT, the dual rail
scheme and in particular the valve scheme have some good chances to
be realised in the near future.

State transfer with quantum chains has become an area of large interest,
with more than seventy articles on the subject over the last three
years. The most important goal now is an experiment that demonstrates
coherent transfer on a short chain (say of length $N\ge5$). Such
an experiment is not only useful building a quantum computer, but
also from a fundamental perspective. For instance, the violation of
a Bell-inequality between distant entangled solid state qubits would
be a milestone in the field. Since this requires a very high transfer
fidelity, the design of such an experiment would probably require
system dependent theoretical research on how to overcome specific
types of noise and how to improve the fidelity for specific Hamiltonians.

\listoffigures

\listoftables



\printindex{}

\begin{theindex}

  \item amplitude damping, 41
  \item amplitude delaying channel, 33
  \item Anderson localisation, 43
  \item arbitrary and unknown qubit, 11
  \item asymptotic deformation, 81

  \indexspace

  \item black box, 12, 49, 53

  \indexspace

  \item chocolate, 11
  \item classical averaged fidelity, 27
  \item coding transformation, 96
  \item conclusively perfect state transfer., 33
  \item cooling protocol, 17
  \item coupled chains, 53
  \item CPT, 83
  \item criteria for quantum state transfer, 31

  \indexspace

  \item decoherence-free subspace, 41
  \item destabilisation, 116
  \item dispersion, 21
  \item distillation, 27
  \item dual rail, 35

  \indexspace

  \item efficiency, 58
  \item engineered couplings, 29
  \item entanglement distillation, 37
  \item entanglement of distillation, 27, 63
  \item entanglement of formation, 27
  \item entanglement transfer, 27
  \item ergodic, 76
  \item experiments, 15

  \indexspace

  \item fidelity, 16
  \item fix-point, 76
  \item flux qubits, 15

  \indexspace

  \item generalised Lyapunov function, 77

  \indexspace

  \item Heisenberg Hamiltonian, 20
  \item homogenisation, 92

  \indexspace

  \item Lindblad equation, 109

  \indexspace

  \item maximal peak, 23
  \item minimal fidelity, 16, 20
  \item mixing, 76
  \item multi rail, 68

  \indexspace

  \item non-expansive map, 80
  \item non-Markovian, 117

  \indexspace

  \item peak width, 27
  \item peripheral eigenvalues, 86
  \item phase noise, 41
  \item pure fix-points, 87

  \indexspace

  \item quantum capacity, 28
  \item quantum chain, 12
  \item quantum channel, 16, 83
  \item quantum computer, 10
  \item quantum erasure channel, 37
  \item quantum gates, 11
  \item quantum memory, 93
  \item quantum relative entropy, 84
  \item quantum-jump approach, 41
  \item qutrits, 55

  \indexspace

  \item reading and writing fidelities, 100

  \indexspace

  \item scalability, 12
  \item Schr\"odinger equation, 109
  \item Shor's algorithm, 10
  \item spectral radius, 67
  \item spin chain, 12
  \item strict contraction, 80
  \item swap gates, 26

  \indexspace

  \item time-scale, 38
  \item tomography, 49
  \item topological space, 74
  \item transfer functions, 18

  \indexspace

  \item valve, 103

  \indexspace

  \item weak contraction, 80

\end{theindex}
\end{document}